\documentclass{ws-ijmpa}
\usepackage[super,compress]{cite}
\usepackage{graphicx}
\usepackage{here}
\def\beq{\begin{equation}}
\def\eeq{\end{equation}}
\def\bea{\begin{eqnarray}}
\def\eea{\end{eqnarray}}
\def\bq{\begin{quote}}
\def\eq{\end{quote}}

\def\nnb{\nonumber}
\def\ga{\left(}
\def\dr{\right)}

\def\lrar{\Longrightarrow}
		
\def\nnb{\nonumber}
\def\la{\langle}
\def\ra{\rangle}
\def\nin{\noindent}
\def\ba{\vspace*{-0.2cm}\begin{array}}
\def\ea{\end{array}\vspace*{-0.2cm}}

\def\b{$\bullet~$}
\def\als{\alpha_s}

\def\gg2{ \la\alpha_s G^2 \ra}
\def\gg3{g^3f_{abc}\la G^aG^bG^c \ra}
\def\ggg4{\la\als^2G^4\ra}

\def\beq{\begin{equation}}
\def\enq{\end{equation}}
\def\beqa{\begin{eqnarray}}
\def\enqa{\end{eqnarray}}
\def\nnb{\nonumber}

\def\qq{\lag\bar{q}q\rag}

\def\qqs{\lag\bar{s}s\rag}

\def\GGG{\lag g^3G^3\rag}

\def\lb{\label}
\def\nn{\nonumber}

\def\sGs{\lag\bar{s}Gs\rag}

\newcommand{\rag}{\rangle}
\newcommand{\lag}{\langle}

\def\lv{\mathcal{L}_v}
\def\lp{\mathcal{L}_+}
\def\lid{\mbox{Li}_2}
\def\ln{\mbox{Log}}
\def\gg{\lag g^{2}_{s} G^2 \rag}
\def\ggg{\lag g^{3}_{s}G^3\rag}

\def\GG{\lag G^2 \rag}
\def\gGG{\lag g_s^2 G^2 \rag}
\def\GGG{\lag G^3 \rag}
\def\gGGG{\lag g_s^3G^3\rag}

\begin{document}
\markboth{R. Albuquerque et al.}
{XYZ-like spectra...}
\title{XYZ-SU3 Breakings from  Laplace Sum Rules at Higher Orders }
\author{R. Albuquerque}
\address{Faculty of Technology,Rio de Janeiro State University (FAT,UERJ), Brazil\\
Email address:  raphael.albuquerque@uerj.br}
\author{S. Narison$^{*}$}
\address{Laboratoire
Univers et Particules de Montpellier (LUPM), CNRS-IN2P3, \\
Case 070, Place Eug\'ene
Bataillon, 34095 - Montpellier, France.\\
Email address: snarison@gmail.com}

\author{
D. Rabetiarivony$^{**}$ and G. Randriamanatrika$^{**}$}

\address{Institute of High-Energy Physics of Madagascar (iHEPMAD)\\
University of Ankatso,
Antananarivo 101, Madagascar}


\maketitle

\begin{history}
\end{history}
\vspace*{-0.5cm}
\begin{abstract}
\nin
We  present new compact integrated expressions of SU3 breaking corrections to QCD spectral functions of heavy-light molecules and four-quark $XYZ$-like states at lowest order (LO) of perturbative (PT) QCD and up to $d=8$ condensates of the Operator Product Expansion (OPE). 
Including next-to-next-to-leading order (N2LO)  PT corrections in the chiral limit and next-to-leading order (NLO) SU3 PT corrections, which we have estimated by  assuming the factorization of the four-quark spectral functions, we improve previous  LO results for the $XYZ$-like masses and decay constants from QCD spectral sum rules (QSSR). Systematic errors are estimated 
from a geometric growth of the higher order PT corrections  and 
from some  partially known  $d=8$ non-perturbative contributions. 
Our optimal results, based on stability criteria, are summarized in Tables\,\ref{tab:resultc} to \ref{tab:4q-resultb} while the $0^{++}$ and $1^{++}$ channels are compared  with some existing LO results in Table\,\ref{tab:theory}. One can notice that , in most channels, the SU3 corrections on the meson masses are tiny: $\leq 10\%$ (resp. $\leq 3\%$)  for the $c$ (resp. $b$)-quark channel but can be large for the couplings ($\leq 20\%$). Within the lowest dimension currents, most of the $0^{++}$ and $1^{++}$ states are below the physical thresholds while our predictions cannot discriminate a molecule from a four-quark state.  
A comparison with the masses of some experimental candidates indicates that the $0^{++}$ 
X(4500) might have a large $\bar D^*_{s0}D^*_{s0}$ molecule component while an interpretation of the $0^{++}$ candidates as four-quark ground states is not supported by our findings.  
The $1^{++}$ X(4147) and X(4273) 
  are compatible with the $\bar D^*_{s}D_{s}$, $\bar D^*_{s0}D_{s1}$ molecules and/or with the axial-vector $A_c$ four-quark ground state. 
  Our results for the $0^{-\pm}$, $1^{-\pm}$ and for different beauty  states can be tested in the future data. Finally, we revisit our previous estimates\,\cite{ChXYZ} for the $\bar D^*_{0}D^*_{0}$ and $\bar D^*_{0}D_{1}$ and present new results for the $\bar D_1D_1$.
  

\ccode{Pac numbers: 11.55.Hx, 12.38.Lg, 13.20-v}  
\vspace*{0.25cm}
\nin
* ICTP-Trieste consultant for Madagascar. \\
** PhD Students.
\end{abstract}

\newpage
\section{Introduction and Experimental Facts}
\nin


In recent papers\,\cite{ChXYZ,SNX,X3C,X3D}, we have used QCD spectral (Laplace\,\cite{SVZa,SVZb,SNRAF,BELLa,BELLb,BELLc} and FESR\,\cite{FESRa,FESRb})  sum rules \,\cite{SNB1,SNB2,SNB3a,SNB3b,SNB3c,RRY,CK} to 
 improve our previous results for the masses and decay constants (couplings) of the $XYZ$ exotic heavy-light and charmonium-like mesons  obtained in the chiral limit $m_q=0$\,\cite{X1A,X1B,X2,X3A,X3B}\,\footnote{For reviews, see e.g.\,\cite{MOLE5,MOLE11,MOLE7}.}. In so doing,  we include up to next-to-next-leading (N2LO) PT factorizable corrections to the heavy-light exotic (molecule and four-quark) correlators which are necessary for giving more sense on the input values of the heavy quark masses which play an important r\^ole in the analysis. 
 
 In this paper
 \,\footnote{A compact form of the paper can be found in\,\cite{SU3-PROC}}, we pursue our investigation by including analogous N2LO PT corrections in the chiral limit and adding to these the SU3 NLO PT corrections to the heavy-light exotic correlators. 
To these new higher order (HO) PT contributions, we add the contributions of  condensates having a dimension ($d\leq 6$) already available in the literature but rederived in this paper. Due to the uncertainties on the size (violation of the factorization assumption\,\cite{SNTAU,LNT,JAMI2a,JAMI2b,JAMI2c}, mixing of operators\,\cite{SNT})  and incomplete contributions (only one class of contributions are only computed in the literature) of higher dimension $(d\geq 8)$, we do not include them into the analysis but only consider their effects as a source of systematic errors due to the truncation of the Operator Product Expansion (OPE). The contributions of the unknown order $\alpha_s^3$ N2LO contribution 
are estimated from a geometric growth of the PT series\,\cite{SNZ} and are
added as a source of systematic uncertainties in the truncation of the PT series.
In this sense, we consider this work as an improvement of previous related works in the literature and, in particular, our works in\,\cite{X3A,X3B}. 

Recent measurements of the $J/\psi\phi$ invariant masses   from $B^+\to J/\psi\phi K^+$ decays by the LHCb collaboration\,\cite{LHCb1} confirmed the existence of the $X(4147)$ and $X(4273)$ with quantum numbers $1^{++}$ within 8.4 and 6.0$ \sigma$ significance found earlier by the CDF\,\cite{CDFX,CDFX1}, the CMS\,\cite{CMS} and the D0\,\cite{D0} collaborations. In the same time, the LHCb collaboration has reported the existence of the $0^{++}$ states in the analogous  $J/\psi\phi$ invariant masses.
On the other, the BELLE collaboration\,\cite{BELLES} has found a $J/\psi\phi$ narrow structure within a 3.2 $\sigma$, the Y(4351) with a width of 13 MeV [but not the $X(4140)$ found earlier by the CDF collaboration\,\cite{CDFX}]  from $\gamma\gamma$ scattering which can be a $0^{++}$ or a $2^{++}$ state. These experimental results are summarized in Table\,\ref{tab:exp}.

{\scriptsize
\begin{center}
\begin{table}[hbt]
\setlength{\tabcolsep}{0.8pc}
\newlength{\digitwidth} \settowidth{\digitwidth}{\rm 0}
\catcode`?=\active \def?{\kern\digitwidth}

 \tbl{
Experimental $1^{++}$ and $0^{++}$ observed states from the $J/\psi\phi$ invariant masses of the $B^\pm\to J/\psi\phi K^\pm$ decays and $\gamma\gamma\to J/\psi\phi$ scattering process. 
}
    {\footnotesize
\begin{tabular}{lllll}

&\\
\hline
\\
States& $J^{PC}$&Mass [MeV] & Total width [MeV]  & Refs.   \\
\\
\hline
\\
&$\bf 1^{++}$& \\
X(4147)  & &$4146.5\pm 4.5^{+4.6}_{-2.8}$&  $83\pm 21^{+21}_{-14}$ & \cite{LHCb1,CDFX,CDFX1,CMS,D0}\\
X(4273)  & &$4273.3\pm 8.3^{+17.2}_{-3.6}$&  $56\pm 11^{+8}_{-11}$ & \cite{LHCb1,CDFX1}\\
&\bf $\bf 0^{++}$ or $\bf 2^{++}$& \\
X(4350)&& $4350.6^{+4.6}_{-5.1}\pm 0.7$&$13^{+18}_{-9}\pm 4$&\cite{BELLES}\\
&$\bf 0^{++}$& \\
X(4500)  & &$4506\pm 11^{+12}_{-15}$&  $92\pm 21^{+21}_{-21}$ & \cite{LHCb1}\\
X(4700)  & &$4704\pm 10^{+14}_{-24}$&  $120\pm 31^{+42}_{-33}$ & \cite{LHCb1}\\
\\
\hline
\end{tabular}
}
\label{tab:exp}
\end{table}
\end{center}
}
Our results are sumarized in Tables\,\ref{tab:resultc} to \ref{tab:4q-resultb}  and in the last section : Summary and Conclusions. A comparison with some other lowest order (LO) QCD spectral sum rules  for the $0^{++}$ and $1^{++}$ states is given in Table\,\ref{tab:theory} of  Section\,\ref{sec:confront}.  
 A confrontation with different experimental candidates is also given in this section.
 
\section{QCD expressions of the Spectral Functions}
\nin

Compared to previous LO QCD expressions of the spectral functions given in the literature, we provide new integrated compact expressions which are more elegant and easier to handle for the numerical analysis. These expressions are tabulated in the Appendices. 

 The PT expression of the spectral function obtained using on-shell renormalization has been transformed into the  $\overline{MS}$-scheme by using the relation between the  $\overline{MS}$ running mass $\overline{m}_Q(\mu)$ and the on-shell mass (pole) ~$M_Q$ , to order $\alpha_s^2$ \cite{TAR,COQUEa,COQUEb,SNPOLEa,SNPOLEb,BROAD2a,AVDEEV,BROAD2b,CHET2a,CHET2b}:
\bea
M_Q &=& \overline{m}_Q(\mu)\Big{[}
1+\frac{4}{3} a_s+ (16.2163 -1.0414 n_l)a_s^2\nnb\\
&&+\ln{\ga\frac{\mu}{ M_Q}\dr^2} \ga a_s+(8.8472 -0.3611 n_l) a_s^2\dr\nnb\\
&&+\ln^2{\ga\frac{\mu}{ M_Q}\dr^2} \ga 1.7917 -0.0833 n_l\dr a_s^2...\Big{]},
\label{eq:pole}
\eea
for $n_l$ light flavours where $\mu$ is the arbitrary subtraction point and $a_s\equiv \alpha_s/\pi$.

 Higher order PT corrections are obtained using the factorization assumption of the four-quark correlators into a convolution of bilinear current correlators
as we shall discuss later on.
\section{QSSR analysis of the  Heavy-Light Molecules}
\subsection{Molecule  currents and the QCD two-point function}
\nin
For describing these molecule states, we shall consider the usual lowest dimension local interpolating currents where each bilinear current has the quantum number of the corresponding open $D_s(0^-)$, $D^*_{s0}(0^+)$, $D^*_s(1^-)~,D_{s1}(1^+)$  states and the analogous states in the $b$-quark channel\,\footnote{For convenience, we shall not consider colored and more general combinations of interpolating operators discussed e.g in \cite{NIELSEN3,STEELE} as well as higher dimension ones involving derivatives. }. 
The previous  assignment is consistent with the definition of a molecule to be a weakly bound state of two mesons within a Van der Vaals force other than a gluon exchange. This feature can justify the approximate  use (up to order $1/N_c$) of the factorization of the four-quark currents as a convolution of two bilinear quark-antiquark currents when estimating the HO PT corrections.
These states and the corresponding interpolating currents are given in Table \ref{tab:current}.  

{\scriptsize
\begin{center}
\begin{table}[hbt]
\setlength{\tabcolsep}{1.8pc}

 \tbl{
Interpolating currents with a definite $C$-parity describing the molecule-like 
states. $Q\equiv$ $c$ (resp. $b$) for the $\bar D_sD_s$ (resp. $\bar B_sB_s$)-like molecules.  
}
    {\footnotesize
\begin{tabular}{lcl}

&\\
\hline
\\
States& $J^{PC}$&Molecule Currents  $\equiv{\cal O}_{mol}(x)$  \\
\\
\hline
\\
\bf Scalar &$\bf 0^{++}$& \\
$\bar D_sD_s,~\bar B_sB_s$  &
&$( \bar{s} \gamma_5 Q ) (\bar{Q} \gamma_5 s)$ \\
%
$\bar D_s^*D_s^*,\bar B_s^*B_s^*$ & 
&$( \bar{s} \gamma_\mu Q ) (\bar{Q} \gamma^\mu s)$ \\ 
$\bar D^*_{s0}D^*_{s0},~\bar B_{s0}^*B^*_{s0}$&  & $( \bar{s} Q ) (\bar{Q} s)$\\ 
$\bar D_{s1}D_{s1},\bar B_{s1}B_{s1}$ & 
&$( \bar{s} \gamma_\mu\gamma_5 Q ) (\bar{Q} \gamma^\mu\gamma_5 s)$ \\ 
\\
\bf Axial-vector &$\bf 1^{++}$& \\
$\bar D_s^*D_s,~\bar B_s^*B_s$& 
&$ \frac{i}{\sqrt{2}} \Big[ (\bar{Q} \gamma_\mu s) ( \bar{s} \gamma_5 Q ) 
     - (\bar{s} \gamma_\mu Q) ( \bar{Q} \gamma_5 s ) \Big]$\\ 
 $\bar D^*_{s0}D_{s1},~\bar B^*_{s0}B_{s1}$  & 
&$ \frac{1}{\sqrt{2}} \Big[ ( \bar{s} Q ) (\bar{Q} \gamma_\mu \gamma_5 s)
     + ( \bar{Q} s ) (\bar{s} \gamma_\mu \gamma_5 Q) \Big]$\\ 
\\
\bf Pseudoscalar &$\bf 0^{-\pm}$& \\
     $\bar D^*_{s0}D_s,~\bar B_{s0}^*B_s$ & & $ \frac{1}{\sqrt{2}} \Big[ ( \bar{s} Q ) (\bar{Q} \gamma_5 s) 
     \pm ( \bar{Q} s ) (\bar{q} \gamma_5 Q) \Big]$\\
$\bar D^*_sD_{s1},~\bar B^*_sB_{s1}$ & 
&$ \frac{1}{\sqrt{2}} \Big[ ( \bar{Q} \gamma_\mu s ) (\bar{s} \gamma^\mu \gamma_5 Q)
     \mp ( \bar{Q} \gamma_\mu \gamma_5 s ) (\bar{s} \gamma^\mu Q) \Big]$\\
     \\
\bf Vector&$\bf 1^{-\pm}$& \\
$\bar D^*_{s0}D^*_s,~\bar B^*_{s0}B^*_s$  & 
&$ \frac{1}{\sqrt{2}} \Big[ ( \bar{s} Q ) (\bar{Q} \gamma_\mu s) 
     \mp ( \bar{Q} s ) (\bar{q} \gamma_\mu Q) \Big]$\\ 
     $\bar D_sD_{s1},~\bar B_sB_{s1}$  &
 &    $ \frac{i}{\sqrt{2}} \Big[ ( \bar{Q} \gamma_\mu \gamma_5 s ) (\bar{s} \gamma_5 Q)
     \pm ( \bar{s} \gamma_\mu \gamma_5 Q ) (\bar{Q} \gamma_5 s) \Big]$\\

     \\
\hline
\end{tabular}
}
\label{tab:current}
\end{table}
\end{center}
}
\nin

 The  two-point correlators associated to the (axial)-vector interpolating operators  are:

\bea
\Pi^{\mu\nu}_{mol}(q)&\equiv&i\int d^4x ~e^{iq.x}\lag 0
|T[{\cal O}^\mu_{mol}(x){\cal O}_{mol}^{\nu\dagger}(0)]
|0\rag\nnb\\
&=&-\Pi^{(1)}_{mol}(q^2)(g^{\mu\nu}-\frac{q^\mu q^\nu}{q^2})+\Pi^{(0)}_{mol}(q^2)\frac{q^\mu
q^\nu}{ q^2}~,
\lb{2po}
\eea
The two invariants, $\Pi^{(1)}_{mol}$ and $\Pi^{(0)}_{mol}$, appearing in 
Eq.~(\ref{2po}) are independent and have respectively the quantum numbers 
of the spin 1 and 0 mesons. 

$\Pi^{(0)}_{mol}$ is related via  Ward identities\,\cite{SNB1,SNB2} to the (pseudo)scalar two-point functions $\psi^{(s,p)}(q^2)$ 
built directly from the (pseudo)scalar currents given in Table\,\ref{tab:current}:
\beq 
\psi^{(s,p)}_{mol}(q^2)=i\int d^4x ~e^{iq.x}\lag 0
|T[{\cal O}^{(s,p)}_{mol}(x){\cal O}^{(s,p)}_{mol}(0)]
|0\rag~,
\label{2po5}
\eeq
with which we shall work in the following.

Thanks  to their analyticity properties, the invariant functions $\Pi^{(1,0)}_{mol}(q^2)$ in Eq.~(\ref{2po}) and the two-point correlator $ \psi^{(s,p)}_{mol}(q^2)$ in Eq.\,\ref{2po5} obey the dispersion relation:
\beq
\Pi^{(1,0)}_{mol}(q^2), ~\psi^{(s,p)}_{mol}(q^2)=\frac{1}{\pi}\int_{4M_Q^2}^\infty dt \frac{{\rm Im}\{\Pi^{(1,0)}_{mol}(t),~ \psi^{(s,p)}_{mol}(t)\}}{ t-q^2-i\epsilon}+\cdots \;,
\lb{ope}
\enq
where $\mbox{Im}\:\Pi^{(1,0)}_{mol}(t),~ {\rm Im}\:\psi^{(s,p)}_{mol}(t)$ are the spectral functions and $\cdots$ indicate subtraction points which are polynomial in $q^2$. 
\subsection{LO PT and NP corrections to the molecule spectral functions}
The new different LO integrated expressions including non-perturbative (NP) corrections up to dimension $d$=6-8 used in the analysis are tabulated in Appendix~A. 

Compared to the ones in the literature, the expressions of the spectral functions are
in integrated and compact forms which are more easier to handle for the numerical phenomenological analysis. 

However, one should note that some of the expressions given in the literature do not agree each others. Due to the 
few informations given by the authors on their derivation, it is difficult to trace back the origin of such discrepancies\,\footnote{A numerical comparison in some specific channels is given in Section 10.}. 
Hopefully, within  the accuracy of the approach, such discrepancies affect only slightly the final results if the errors are taken properly. 

In the chiral limit $m_q=0$ and $\la\bar uu\ra=\la\bar dd\ra$, we have checked that the orthogonal combinations of $\bar D^*D, \bar B^*B (1^{++}), \bar D_0^* D_1,\bar B_0^* B_1 (0^{--})$ and $\bar D^*D_1, \bar B^*B_1 (0^{--} )$ molecules give the same results up to the $d=6$ contributions. This is due to the presence of one $\gamma_5$ matrix in the current which neutralizes the different traces appearing in each pair. This is not the case of the $\bar D^*_0D^*, \bar B^*_0B^*$ (without $\gamma_5$) and $\bar DD_1, \bar BB_1$ (with two  $\gamma_5$). 

\subsection{N2LO PT corrections using factorization}
\nin
Assuming a factorization of the four-quark interpolating current as a natural consequence of the molecule
definition of the state, we can write the corresponding spectral function as a convolution of the
spectral functions associated to quark bilinear current\,\footnote{It is called properly sesquilinear instead of bilinear  current as it is a formed by a quark field and its anti-particle.} as illustrated by the Feynman diagrams in Fig.\,\ref{fig:factor}. In this way, we obtain \cite{PICH}\,\footnote{For some applications to the $\bar BB$ mixing, see e.g. \cite{BBAR1,BBAR2,BBAR3}.} for the  $\bar DD^*$- and $\bar D^*_0D^*$-like spin 1 states:
\bea
\frac{1}{ \pi}{\rm Im} \Pi^{(1)}_{mol}(t)&=& \theta (t-4M_Q^2)\ga \frac{1}{ 4\pi}\dr^2 t^2 \int_{M_Q^2}^{(\sqrt{t}-M_Q)^2}\hspace*{-0.5cm}dt_1\int_{M_Q^2}^{(\sqrt{t}-\sqrt{t_1})^2} \hspace*{-1cm}dt_2\nnb\\
&&\times~\lambda^{3/2}\frac{1}{ \pi}{\rm Im} \Pi^{(1)}(t_1) \frac{1}{ \pi}{\rm Im} \psi^{(s,p)}(t_2)~.
\label{eq:convolution}
\eea
For the  $\bar DD$-like  spin 0 state, one has:
\bea
\frac{1}{ \pi}{\rm Im} \psi^{(s)}_{mol}(t)&=& \theta (t-4M_Q^2)\ga \frac{1}{4\pi}\dr^2 t^2 \int_{M_Q^2}^{(\sqrt{t}-M_Q)^2}\hspace*{-0.5cm}dt_1\int_{M_Q^2}^{(\sqrt{t}-\sqrt{t_1})^2}  \hspace*{-1cm}dt_2~\nnb\\
&&\times~\lambda^{1/2}\ga \frac{t_1}{ t}+ \frac{t_2}{ t}-1\dr^2\nnb\\
&&\times ~\frac{1}{ \pi}{\rm Im}\psi^{(p)}(t_1) \frac{1}{ \pi} {\rm Im} \psi^{(p)}(t_2),
\eea
and for the $\bar D^*D^*$-like spin 0 state:
\bea
\frac{1}{ \pi}{\rm Im} \psi^{(s)}_{mol}(t)&=& \theta (t-4M_Q^2)\ga \frac{1}{4\pi}\dr^2 t^2 \int_{M_Q^2}^{(\sqrt{t}-M_Q)^2}\hspace*{-0.5cm}dt_1\int_{M_Q^2}^{(\sqrt{t}-\sqrt{t_1})^2}  \hspace*{-1cm}dt_2~\nnb\\
&&\times~\lambda^{1/2}\Big{[}\ga \frac{t_1}{ t}+ \frac{t_2}{ t}-1\dr^2
+\frac{8t_1t_2}{ t^2}\Big{]}\nnb\\
&&\times ~\frac{1}{ \pi}{\rm Im} \Pi^{(1)}(t_1) \frac{1}{ \pi} {\rm Im} \Pi^{(1)}(t_2),
\eea
where:
\beq
\lambda=\ga 1-\frac{\ga \sqrt{t_1}- \sqrt{t_2}\dr^2}{ t}\dr \ga 1-\frac{\ga \sqrt{t_1}+ \sqrt{t_2}\dr^2}{ t}\dr~,
\eeq
is the phase space factor and $M_Q$ is the on-shell heavy quark mass. 
Im $ \Pi^{(1)}(t)$ is the spectral function associated to the bilinear $\bar c\gamma_\mu (\gamma_5)q$  vector or axial-vector current, while Im $\psi^{(5)}(t)$ is associated to the 
$\bar c(\gamma_5)q$  scalar or pseudoscalar current\,\footnote{In the limit where the light quark mass $m_q=0$, the PT expressions of the vector (resp. scalar) and axial-vector (resp. pseudoscalar) spectral functions are the same.}. 
This representation simplifies the evaluation of the PT $\alpha_s^n$-corrections as we can use the PT expression of the spectral functions for heavy-light bilinear currents known to order $\alpha_s$ (NLO) from \cite{BROAD}\,\footnote{Within the above procedure, we have checked that we reproduce the factorized PT LO contributions obtained using for example the PT expressions of $\bar D^*_0D^*_0$ and $\bar D^*_0D^*$ given in Appendix A.}. Order $\alpha_s^2$ (N2LO) corrections are known in the chiral limit $m_q=0$ from \cite{CHETa,CHETb} which are available as a Mathematica Program named  Rvs\,\footnote{We have seen in\,\cite{ChXYZ} that these N2LO corrections are relatively
small which demonstrates the good convergence of the PT series.}. We shall use the NLO SU3 breaking PT corrections obtained in\,\cite{PIVO2} from the two-point function formed by bilinear currents.
From the above representation, the anomalous dimensions of the molecule correlators come from  the (pseudo)scalar current.  Therefore,  the corresponding renormalization group invariant interpolating current reads to NLO\,\footnote{The spin 0 current built from two (axial)-vector currents has no anomalous dimension. We have introduced the super-indices $(v,a)$ for denoting the vector and axial-vector spin 1 channels.}:
\beq
 \bar {\cal O}^{(s,p)}_{mol}(\mu)= a_s(\mu)^{4/\beta_1} {\cal O}^{(s,p)}_{mol}~,~~~~~~~~  \bar {\cal O}^{(v,a)}_{mol}(\mu)= a_s(\mu)^{2/\beta_1} {\cal O}^{(v,a)}_{mol}~,
 \eeq
 with $-\beta_1=(1/2)(11-2n_f/3)$ is the first coefficient of the QCD $\beta$-function for $n_f$ flavours and  $a_s\equiv (\alpha_s/\pi)$. 
\subsection{$1/q^2$ tachyonic gluon mass and large order PT corrections}
\nin
The $1/q^2$ corrections due to a tachyonic gluon  mass  discussed in \cite{CNZ1,CNZ2} (for reviews see: \,\cite{ZAK1,ZAK2}) will not be included here. Instead, we shall consider the fact that they are dual to the sum of the large order PT series\,\cite{SNZ} such that, with the inclusion of the N3LO terms estimated from the geometric growth of the QCD PT series \cite{SNZ} as a source of the PT errors,  we expect to give a good approximation of these uncalculated higher order terms. The estimate of these errors is given in Tables\,\ref{tab:errorc} to \,\ref{tab:error-4b}.

\subsection{Parametrization of the Spectral Function within MDA}
\nin

 We shall  use the Minimal Duality Ansatz (MDA) given in Eq. \ref{eq:duality} for parametrizing the spectral function (generic notation):
\beq
\frac{1}{\pi}\mbox{ Im}\Pi_{mol}(t)\simeq f^2_{mol}M^8_{mol}\delta(t-M_{mol}^2)
  \ + \
  ``\mbox{QCD continuum}" \theta (t-t_c),
\label{eq:duality}
\eeq
where $f_{mol}$ is the decay constant defined as:
\beq
\la 0| {\cal O}^{(s,p)}_{mol}|mol\ra=f^{(s,p)}_{mol}M^4_{mol}~,~~~~~~~~~~~~\la 0| {\cal O}^\mu_{mol}|mol\ra=f^{(v,a)}_{mol}M^5_{mol}\epsilon_\mu~,
\label{eq:coupling}
\eeq
respectively for spin 0 and 1 molecule states with  $\epsilon_\mu$ the vector polarization.
The higher states contributions are smeared by the ``QCD continuum" coming from the discontinuity of the QCD diagrams and starting from a constant threshold $t_c$.  This {\it simple model} has been tested successfully in the $e^+e^-\to hadrons$  ($\rho$ and $\phi$ mesons) channels where complete data are available\,\cite{SNB1,SNB2}. Finite width corrections to this simple model has been e.g studied in\,\cite{VENEZIA,SNGLUE1,SNGLUE2} and have been found to be negligible. We then expect that such results also hold here. 

 Noting that, in the previous definition in Table \ref{tab:current},  the bilinear (pseudo)scalar current acquires an anomalous dimension due to its normalization, thus the decay constants run  to order $\alpha_s^2$ as\,\footnote{The coupling of the (pseudo)scalar molecule built from two (axial)-vector currents has no anomalous dimension and does not run.}:
\beq
f^{(s,p)}_{mol}(\mu)=\hat f^{(s,p)}_{mol} \ga -\beta_1a_s\dr^{4/\beta_1}/r_m^2~,~~~~f^{(v,a)}_{mol}(\mu)=\hat f^{(v,a)}_{mol} \ga  -\beta_1a_s\dr^{2/\beta_1}/r_m~,
\label{eq:fhat}
\eeq
where we have introduced the renormalization group invariant coupling $\hat f_{mol}$.  
 The QCD corrections numerically read to N2LO:
\beq
r_m(n_f=4)=1+1.014 a_s +1.389a_s^2~
,~~~~
r_m(n_f=5)=1+1.176a_s +1.501a_s^2~.  
\eeq
This coupling is the analogue of the pion decay constant $f_\pi=132$ MeV where the corresponding hadronic coupling behaves as $1/f_{mol}$. This behaviour can be understood from a three-point sum rule analysis (for reviews, see e.g \cite{SNB1,SNB2,MOLE5}), \`a la Golberger-Treiman like-relation\,\cite{FURLAN} or from some low-energy theorems\,\cite{VENEZIA,SNG1,SNG2}. 
\subsection{The inverse Laplace  transform sum rule (LSR)}
\nin
The exponential sum rules firstly derived by SVZ\,\cite{SVZa,SVZb} have been called Borel sum rules due to the factorial suppression factor of the condensate contributions in the OPE. Their quantum mechanics version have been studied by Bell-Bertlmann in \cite{BELLa,BELLb,BELLc} through the harmonic oscillator where $\tau$ has the property of an imaginary time. The derivation of their radiative corrections has been firstly shown by Narison-de Rafael\, \cite{SNRAF} to have the properties of the inverse Laplace sum rule (LSR). The LSR and its ratio read\,\footnote{The last equality in Eq.\,\ref{eq:ratioLSR} is obtained when one uses MDA in Eq.\,\ref{eq:duality} for parametrizing the spectral function.}:
\beq
{\cal L}_{mol}(\tau,t_c,\mu)=\frac{1}{\pi}\int_{4M_Q^2}^{t_c}dt~e^{-t\tau} \mbox{Im}\{\Pi^{(v,a)}_{mol},~\psi^{(s,p)}_{mol}\}(t,\mu)~,
\label{eq:LSR}
\eeq
\beq\label{eq:ratioLSR}
{\cal R}_{mol}(\tau,t_c,\mu) = \frac{\int_{4M_Q^2}^{t_c} dt~t~ e^{-t\tau}\mbox{Im}\{\Pi^{(v,a)}_{mol},~\psi^{(s,p)}_{mol}\}(t,\mu)}
{\int_{4M_Q^2}^{t_c} dt~ e^{-t\tau}\mbox{Im}\{\Pi^{(v,a)}_{mol},~\psi^{(s,p)}_{mol}\}(t,\mu)}\simeq M_R^2~,
\eeq
where $\mu$ is the subtraction point which appears in the approximate QCD series when radiative corrections are included and $\tau$ is the sum rule variable replacing $q^2$. The variables $\tau,\mu$ and $t_c$ are, in principle, free parameters. 
We shall use stability criteria (if any), with respect to these free 3 parameters,  for extracting the optimal results. 
\subsection{Double ratios of inverse Laplace  transform sum rules (DRSR)}
Double ratios of sum rules (DRSR)\, \cite{DRSR} are also useful for extracting the SU3 breaking effects on couplings and mass ratios. They read:
\beq
f^{sd}_{mol}\equiv \frac{{\cal L}^s_{mol}(\tau,t_c,\mu)}{ {\cal L}^d_{mol}(\tau,t_c,\mu)}~,~~~~~~~~~
r^{sd}_{mol}\equiv {{\cal R}^s_{mol}(\tau,t_c,\mu)\over {\cal R}^d_{mol}(\tau,t_c,\mu)}~.
\label{eq:rsd}
\eeq
The upper indices $s,d$ indicate the $s$ and $d$ quark channels. 
These DRSR can be used when each sum rule optimizes at the same values of the parameters $(\tau,t_c,\mu)$. In this case, they lead to a more precise determination of the SU3 breaking effects on the couplings and mass ratios as obtained in some other channels\, \cite{X2,SNFORM1,SNGh3,SNGh1,SNGh5,SNmassa,SNmassb,SNhl,SNFORM2,HBARYONa,HBARYONb}.  
\subsection{Tests of MDA and Stability Criteria}
\nin

 In the standard Minimal Duality Ansatz (MDA) given in Eq. \ref{eq:duality} for parametrizing the spectral function,
 the ``QCD continuum" threshold $t_c$ is constant and is independent on the subtraction point $\mu$ \,\footnote{Some model with a $\mu$-dependence of $t_c$ has been discussed e.g in\,\cite{LUCHA}.}. One should notice that this standard MDA with constant $t_c$ describes quite well the properties of the lowest ground state as explicitly demonstrated in \cite{SNFB12a,SNFB12b} and in various examples\,\cite{SNB1,SNB2}, while it has been also successfully tested in the large $N_c$ limit of QCD in \cite{PERISa,PERISb}. 

 Refs. \cite{SNFB12a,SNFB12b}  have explicitly tested  this simple model by confronting the predictions of the integrated spectral function within this simple parametrization with the full data measurements. One can notice  
in Figs. 1 and 2 of Ref.  \cite{SNFB12a,SNFB12b} the remarkable agreement of the model predictions and of the measured data of the $J/\psi$ charmonium and $\Upsilon$ bottomium systems for a large range of the inverse sum rule variable $\tau$. Though it is difficult to estimate with a good precision the systematic error related to this simple model, this feature indicates the ability of the model for reproducing accurately the data. We expect that the same feature is reproduced for the case of the XYZ discussed here where complete data are still lacking.

In order to extract an optimal information for the lowest resonance parameters from this rather crude  description of the spectral function and from the approximate QCD expression, one often applies the stability criteria at which an optimal result can be extracted. This stability is signaled by the existence of a stability plateau, an extremum or an inflexion point (so-called ``sum rule window") versus the changes of the external sum rule variables $\tau$ and $t_c$ where the simultaneous  requirement on the dominance over the continuum contribution and on the convergence of the OPE is automatically satisfied. This optimization criterion demonstrated in series of papers by Bell-Bertlmann \cite{BELLa,BELLb,BELLc}, in the case of the $\tau$-variable, by taking the examples of harmonic oscillator and charmonium sum rules and extended to the case of the $t_c$-parameter in \cite{SNB1,SNB2} gives a more precise meaning of  the so-called ``sum rule window" originally discussed by SVZ \cite{SVZa,SVZb} and used in the sum rules literature. Similar applications of the optimization method to the pseudoscalar $D$ and $B$ open meson states have been successful when comparing these results with the ones from some other determinations as discussed in Refs.\,\cite{SNFB12a,SNFB12b} and reviewed in\,\cite{SNB1,SNB2,SNFB01,SNREV14,SNREV15} and in some other recent reviews\,\cite{ROSNERb,LATT13}. 

 In this paper, we shall add to the previous well-known $\tau$- and $t_c$-stability criteria, the one associated  to the requirement of stability  versus the arbitrary subtraction constant $\mu$ often put by hand  in the current literature  and which is often the source of large errors from the PT series in the sum rule analysis.  The $\mu$-stability procedure has been applied recently in\,\cite{SNFB12a,SNFB12b,SNFB13,SNLIGHT,SNREV14,SNREV15,SNFB14,SNCOREL18}\,\footnote{Some other alternative approaches for optimizing the PT series can be found in \cite{STEVENSONa,STEVENSONb,STEVENSONc,STEVENSONd,STEVENSONe}.} which gives a much better meaning on the choice of $\mu$-value at which the observable is extracted, while the errors  in the determinations of the results have been reduced due to a better control of the $\mu$ region of variation which is not the case in the existing literature.

{\scriptsize
\begin{table}[hbt]
\setlength{\tabcolsep}{2pc}
 \tbl{QCD input parameters:
the original errors for 
$\la\alpha_s G^2\ra$, $\la g^3  G^3\ra$ and $\rho \la \bar qq\ra^2$ have been multiplied by about a factor 3 for a conservative estimate of the errors (see also the text). }  
    {\small
 {\begin{tabular}{@{}lll@{}} \toprule
&\\
\hline
\\
Parameters&Values& Ref.    \\
\\
\hline
\\
$\alpha_s(M_\tau)$& $0.325(8)$&\cite{SNTAU,BNPa,BNPb}\\
$\hat m_s$&$(0.114\pm0.006)$ GeV &\cite{SNB1,SNTAU,SNmassa,SNmassb,SNmass98a,SNmass98b,SNLIGHT}\\
$\overline{m}_c({m}_c)$&$1261(12)$ MeV &average \cite{SNmass02,SNH10a,SNH10b,SNH10c,PDG,IOFFEa,IOFFEb}\\
$\overline{m}_b(m_b)$&$4177(11)$ MeV&average \cite{SNmass02,SNH10a,SNH10b,SNH10c,PDG}\\
$\hat \mu_q$&$(253\pm 6)$ MeV&\cite{SNB1,SNmassa,SNmassb,SNmass98a,SNmass98b,SNLIGHT}\\
$\kappa\equiv \la \bar ss\ra/\la\bar dd\ra$& $(0.74^{+0.34}_{- 0.12})$&\cite{HBARYONa,HBARYONb,SNB1}\\
$M_0^2$&$(0.8 \pm 0.2)$ GeV$^2$&\cite{JAMI2a,JAMI2b,JAMI2c,HEIDa,HEIDb,HEIDc,SNhl}\\
$\la\alpha_s G^2\ra$& $(7\pm 3)\times 10^{-2}$ GeV$^4$&
\cite{SNTAU,LNT,SNIa,SNIb,YNDU,BELLa,BELLb,BELLc,SNH10a,SNH10b,SNH10c,SNG1,SNG2,SNGH}\\
$\la g^3  G^3\ra$& $(8.2\pm 2.0)$ GeV$^2\times\la\alpha_s G^2\ra$&
\cite{SNH10a,SNH10b,SNH10c}\\
$\rho \alpha_s\la \bar qq\ra^2$&$(5.8\pm 1.8)\times 10^{-4}$ GeV$^6$&\cite{SNTAU,LNT,JAMI2a,JAMI2b,JAMI2c}\\
\\
\hline
\end{tabular}}
}
\label{tab:param}
\end{table}
} 
\subsection{QCD Input Parameters}
\hspace*{0.5cm} 
The QCD parameters which shall appear in the following analysis will be the charm and bottom quark masses $m_{c,b}$, the strange quark $m_s$ (we shall neglect  the light quark masses $m_{u,d}$),
the light quark condensates $\qq$ ($q\equiv u,d,s$),  the gluon condensates $ \lag
\alpha_sG^2\rag
\equiv \la \alpha_s G^a_{\mu\nu}G_a^{\mu\nu}\ra$ 
and $ \la g^3G^3\ra
\equiv \la g^3f_{abc}G^a_{\mu\nu}G^{b,\nu}_{\rho}G^{c,\rho\mu}\ra$, 
the mixed quark condensate $\la\bar qGq\ra
\equiv {\la\bar qg\sigma^{\mu\nu} (\lambda_a/2) G^a_{\mu\nu}q\ra}=M_0^2\la \bar qq\ra$ 
and the four-quark 
 condensate $\rho\alpha_s\la\bar qq\ra^2$, where
 $\rho\simeq 3-4$ indicates the deviation from the four-quark vacuum 
saturation. Their values are given in Table \ref{tab:param}
\,\footnote{A recent analysis\,\cite{SNCOREL18} confirms the quoted values of $m_c,b,~\alpha_s$ and improve the one of  $ \lag
\alpha_sG^2\rag$.}

We shall work with the running
light quark condensates, which read to leading order in $\alpha_s$: 
\beq
{\la\bar qq\ra}(\tau)=-{\hat \mu_q^3  \ga-\beta_1a_s\dr^{2/{
\beta_1}}},~~~~~~~~
{\la\bar q Gq\ra}(\tau)=-{M_0^2{\hat \mu_q^3} \ga-\beta_1a_s\dr^{1/{3\beta_1}}}~,
\label{d4g}
\eeq
and the running strange quark mass to NLO (for the number of flavours $n_f=3$):
\beq
\overline{m}_s(\tau)={\hat m_s\ga-\beta_1a_s\dr^{-2/{
\beta_1}}}(1+0.8951a_s)~,
\label{ms}
\eeq
where $\beta_1=-(1/2)(11-2n_f/3)$ is the first coefficient of the $\beta$ function 
for $n_f$ flavours; $a_s\equiv \alpha_s(\tau)/\pi$; 
$\hat\mu_q$ and $ \hat m_s$ are the spontaneous RGI light quark condensate and strange quark mass\,\cite{FNR}. 
We shall use:
\beq   
\alpha_s(M_\tau)=0.325(8) \lrar  \alpha_s(M_Z)=0.1192(10)
\label{eq:alphas}
\eeq
from $\tau$-decays \cite{SNTAU,BNPa,BNPb}\,\footnote{A recent update is done in\,\cite{PICHTAU} where the same central value is obtained and where more complete references can be found.}
 which agree with the 2016 world average\,\cite{BETHKE}: 
\beq
\alpha_s(M_Z)=0.1181(11)~. 
\eeq
The value of the running $\la \bar qq\ra$ condensate is deduced from  the well-known GMOR relation: 
\beq
(m_u+m_d)\la \bar uu+\bar dd\ra=-m_\pi^2f_\pi^2~,
\eeq
where $f_\pi=130.4(2)$ MeV \cite{ROSNERb}. The value of $(\overline{m}_u+\overline{m}_d)(2)=(7.9\pm 0.6)$ MeV obtained in  \cite{SNmassa,SNmassb} agrees with the PDG  in \cite{PDG}  and lattice averages in \cite{LATT13}. Then, we deduce the RGI light quark spontaneous mass $\hat\mu_q$ given  in Table~\ref{tab:param}. 

 For the heavy quarks, we shall use the running mass\,\footnote{This choice is not justified if one works at lowest order (LO) like in the existing literature due to the ill-defined mass definition at LO. Effects of the use of the running or the on-shell mass at LO has been explicitly shown in\,\cite{X3A,X3B,ChXYZ}.} and the corresponding value of $\alpha_s$ evaluated at the scale $\mu$. These sets of correlated parameters are given in Table \ref{tab:alfa} for different values of $\mu$ and for a given number of flavours $n_f$. The value of $\mu$ used here corresponds to the optimal one obtained in\,\cite{ChXYZ}. 

 For the $\la \alpha_s G^2\ra$ condensate, we have enlarged the original error by a factor about 3 in order to have
a conservative result for recovering the original SVZ estimate and the alternative extraction in \cite{IOFFEa,IOFFEb} from charmonium sum rules.
However, a direct naive comparison of this  range of values obtained within short QCD series (few terms) with the one from lattice calculations \cite{BALIa} obtained within a long QCD series\,\cite{BALIb} can be misleading. 


We shall see later on that the effects of the gluon and four-quark condensates on the values of the decay constants and masses are relatively small though they play an important r\^ole in the stability analysis. 

{\scriptsize
\begin{table}[hbt]
\setlength{\tabcolsep}{1.1pc}
 \tbl{
$\alpha_s(\mu)$ and correlated values of $\overline{m}_Q(\mu)$ used in the analysis for different values of the subtraction scale $\mu$. The error in $\overline{m}_Q(\mu)$ has been induced by the one of $\alpha_s(\mu)$ to which one has added the error on their determination given in Table\,\ref{tab:param}. }
    {\small
{\begin{tabular}{@{}llll@{}} \toprule
&\\
\hline
\\
Input for $\bar D_sD_s,...,~[cs\bar c\bar s],$ : $n_f=4$\\
\\
\hline
\\
$\mu$[GeV]&$\alpha_s(\mu)$&& $\overline{m}_c(\mu)$[GeV]\\
\\
\hline
Input: $\overline{m}_c({m}_c)$&0.4084(144)&&1.26\\
1.5&0.3649(110)&&1.176(5)\\
2&0.3120(77)&&1.069(9)\\
2.5&0.2812(61)&&1.005(10)\\
3.0&0.2606(51)&&0.961(10)\\
3.5&0.2455(45)&&0.929(11)\\
4.0&0.2339(41)&&0.903(11) \\
4.5&0.2246(37)&&0.882(11)\\
5.0&0.2169(35)&&0.865(11)\\
5.5&0.2104(33)&&0.851(12)\\
6.0&0.2049(30)&&0.838(12)\\
\hline
\\
Input for $\bar B_sB_s,...,~[bs\bar b\bar s]$ : $n_f=5$\\
\\
\hline
\\
$\mu$[GeV]&$\alpha_s(\mu)$&& $\overline{m}_b(\mu)$[GeV]\\
\\
\hline
3&0.2590(26)&&4.474(4)\\
3.5&0.2460(20)&&4.328(2)\\
Input: $\overline{m}_b(m_b)$&0.2320(20)&&4.177\\
4.5&0.2267(20)&&4.119(1)\\
5.0&0.2197(18)&&4.040(1)\\
5.5&0.2137(17)&&3.973(2)\\
6.0&0.2085(16)&&3.914(2)\\
6.5&0.2040(15)&&3.862(2)\\
7.0&0.2000(15)&&3.816(3)\\
\hline
\hline
\end{tabular}}
}
\label{tab:alfa}
\end{table}
} 
\nin
\section{Accuracy of the Factorization Assumption}\label{sec:factor}
\subsection{PT Lowest order tests}
\begin{figure}[hbt] 
\begin{center}
{\includegraphics[width=6.29cm  ]{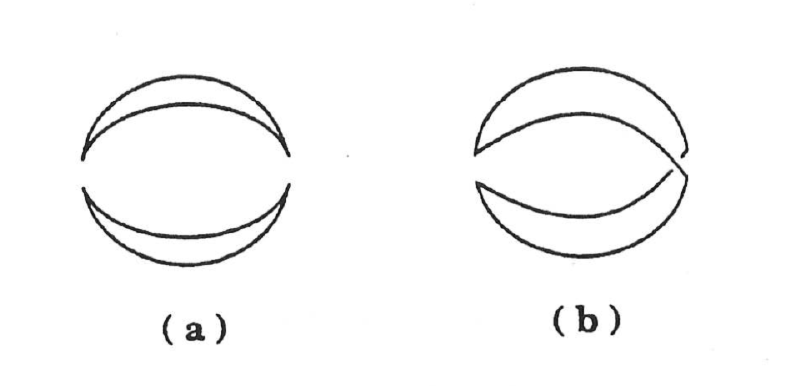}}
\caption{
\scriptsize 
{\bf (a)} Factorized contribution to the four-quark correlator at lowest order of PT; {\bf (b)} Non-factorized contribution at lowest order of PT (the figure comes from\,\cite{PICH}).
}
\label{fig:factor} 
\end{center}
\end{figure} 
\nin
\begin{figure}[hbt] 
\begin{center}
{\includegraphics[width=6.29cm  ]{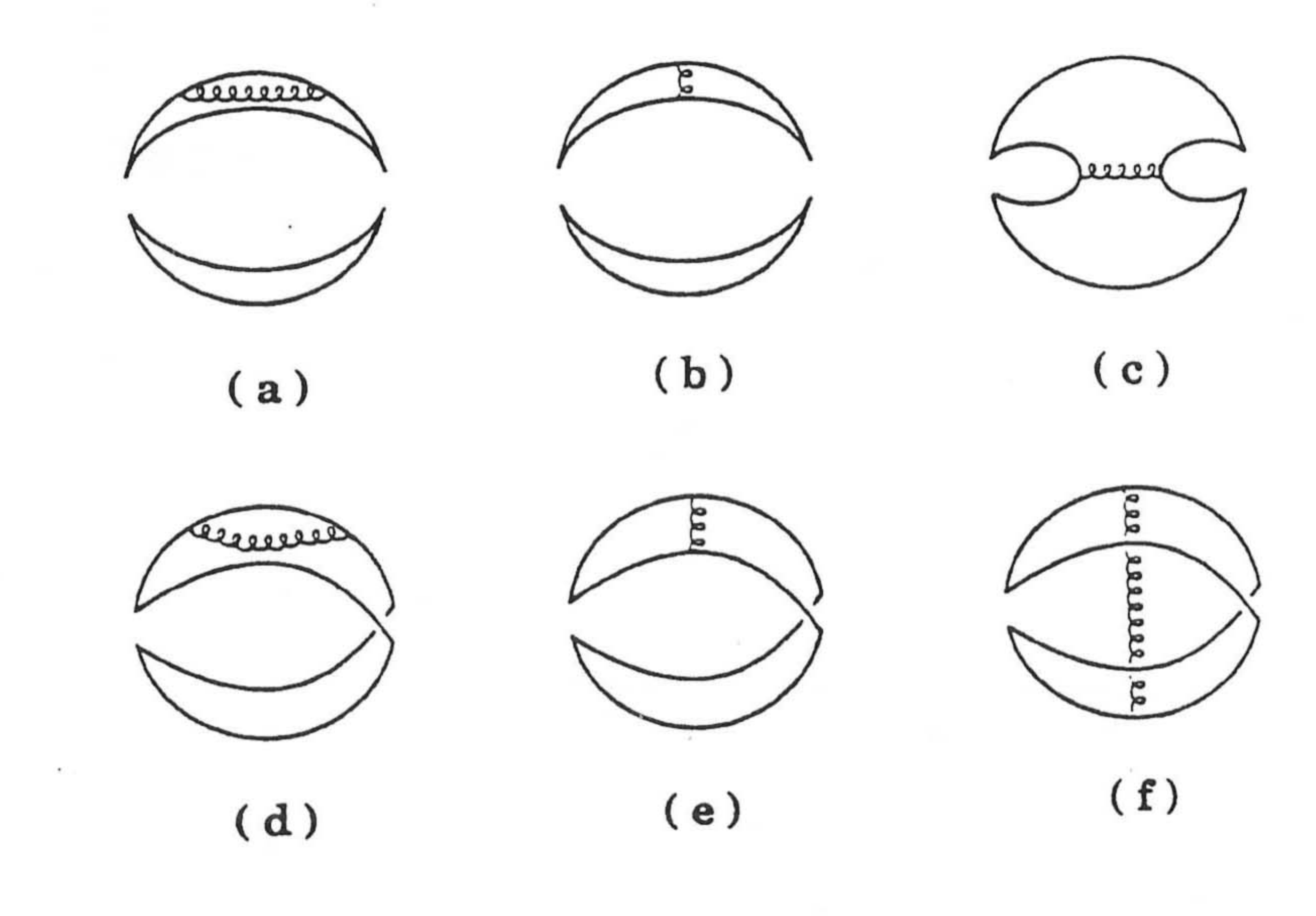}}
\caption{
\scriptsize 
{\bf (a,b)} Factorized contributions to the four-quark correlator at NLO of PT; {\bf (c to f)} Non-factorized contributions  at NLO of PT (the figure comes from\,\cite{PICH}).
}
\label{fig:factoras} 
\end{center}
\end{figure} 
\nin
To lowest order of PT QCD, the four-quark correlator can be subdivided  into its factorized (Fig.\,\ref{fig:factor}a) and its non-factorized (Fig.\,\ref{fig:factor}b) parts.  

We have tested in\,\cite{ChXYZ} the factorization assumption at LO by taking the example of the  $\bar M^*_0M^*(1^-)$  molecule states where $M\equiv D$ (resp. $B$) meson in the charm (resp. bottom) quark channels. We concluded from the previous two examples that assuming a factorization of the PT contributions
at LO induces an almost negligible effect on the decay constant ($\simeq 1.5\%$) and mass ($\simeq 7\times 10^{-4}$) determinations for the $\bar D^*_0D^*$ and $\bar B^*_0B^*$ vector molecules. 
\subsection{Factorization tests for PT$\oplus$NP contributions at LO}
We have  noticed in\,\cite{ChXYZ} that the effect of factorization of the PT$\oplus$NP at LO is about 2.2\% for the 
decay constant and 0.5\% for the mass which is quite tiny. However, to avoid this (small) effect, we shall work in the following with the full non-factorized PT$\oplus$NP of the  LO expressions. 
\subsection{Test at NLO of PT from the $B^0\bar B^0$ four-quark correlator}
For extracting the PT $\alpha_s^n$ corrections to the correlator and due to the technical complexity of the calculations, we shall assume that these radiative corrections are dominated by the ones from the factorized diagrams (Fig.\,\ref{fig:factoras}a,b) while we neglect the ones from non-factorized diagrams (Fig.\,\ref{fig:factoras}c to f). This fact has been proven explicitly by \,\cite{BBAR2,BBAR3}  in the case of the $\bar B^0B^0$ systems (very similar correlator as the ones discussed in the following) where the non-factorized $\alpha_s$ corrections do not exceed 10\% of the total 
$\alpha_s$ contributions. 
\subsection{Conclusions of the factorization tests} 
\nin
We expect from the previous LO examples that the masses of the molecules are known with a good accuracy while, for the coupling, we shall have in mind the systematics induced by the radiative corrections estimated by keeping only the factorized diagrams. The contributions of the factorized diagrams will be extracted from the convolution integrals given in Eq.\,\ref{eq:convolution}. Here, the suppression of the NLO corrections will be more pronounced for the extraction of the meson masses from the ratio of sum rules compared to the case of the $\bar B^0B^0$ systems. 
\section{The  Heavy-light Charm  Molecule States}
They are described by the interpolating currents given in Table\,\ref{tab:current}.
The corresponding spectral functions are given to LO of PT QCD in \ref{app.a}. 
The different sources of the errors from the analysis are given in Table\,\ref{tab:errorcplus} to \ref{tab:error-rap-cmoins}. 
\subsection{The  $(0^{++})$ Charm  Scalar Molecule States}
We shall study the $\bar D_sD_s,~\bar D^*_sD^*_s$, $\bar D^*_{s0}D^*_{s0}$ 
and their beauty analogue. Noticing that the qualitative behaviours of the curves in these channels are very similar, we shall illustrate
the analysis in the case of $\bar D_sD_s$ and $\bar B_sB_s$ molecule states by working with the SU3 ratios $f^{sd}_{DD}$ and $r^{sd}_{DD}$ of couplings and masses defined in Eq.\ref{eq:rsd}.  
   \subsection*{\b The $\bar D_sD_s$ molecule state}
\subsubsection*{$-$ $\bar D_sD_s$  coupling and mass}
\begin{figure}[hbt] 
\begin{center}
{\includegraphics[width=6.29cm  ]{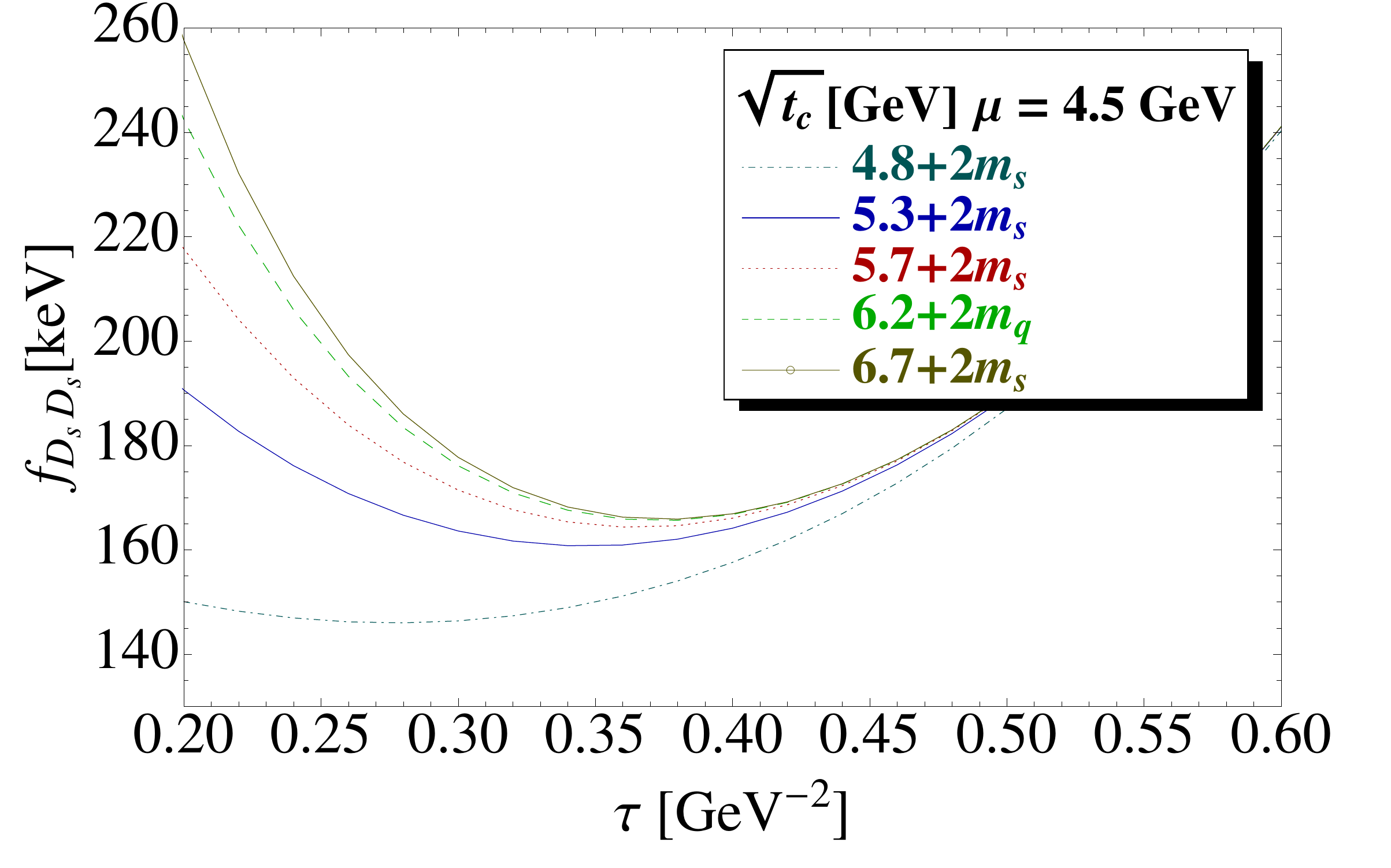}}
{\includegraphics[width=6.29cm  ]{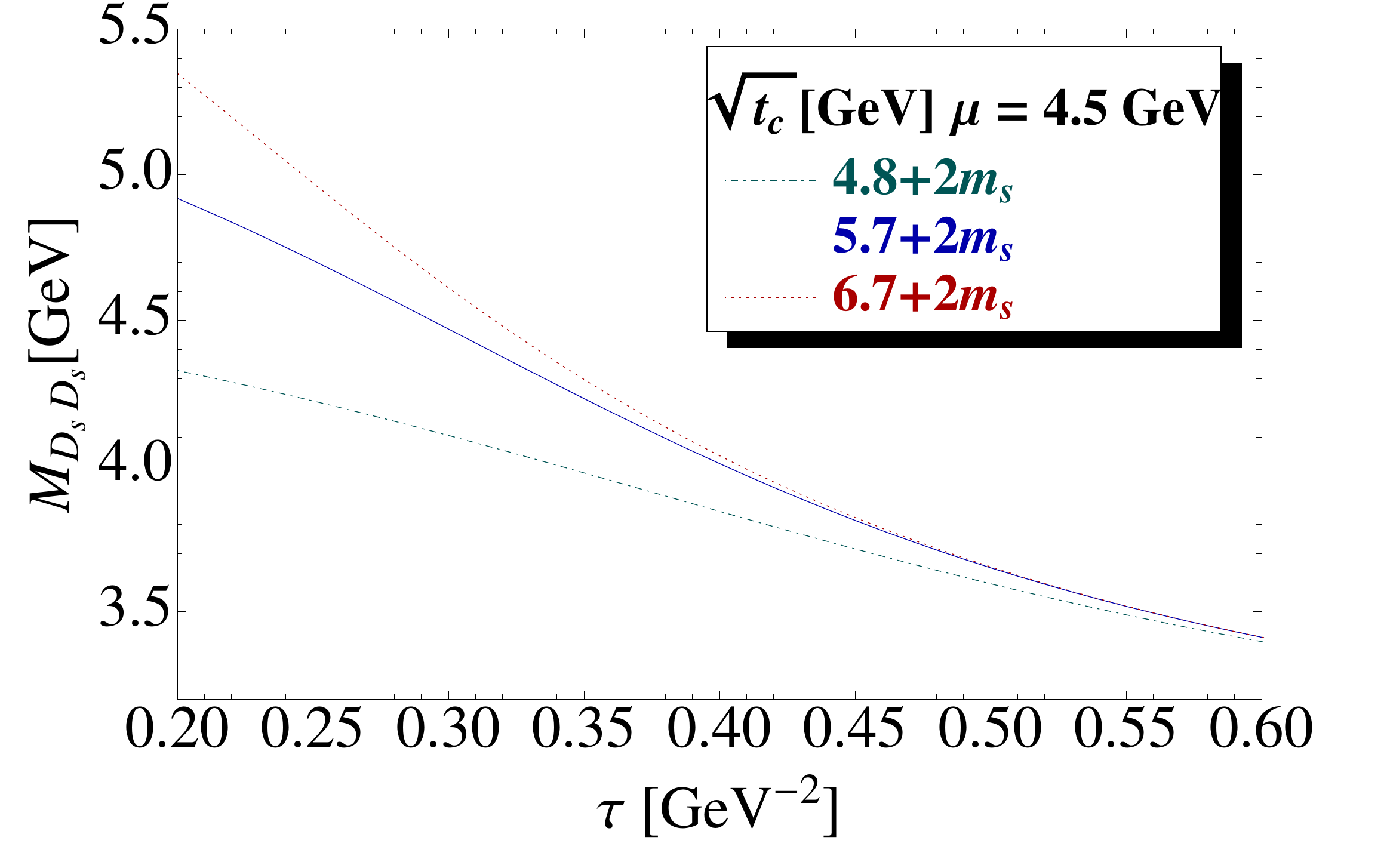}}
\centerline {\hspace*{-3cm} a)\hspace*{6cm} b) }
\caption{
\scriptsize 
{\bf a)} The coupling $f_{D_sD_s}$ at NLO  as function of $\tau$ for different values of $t_c$, for $\mu=4.5$ GeV  and for the QCD parameters in Tables\,\ref{tab:param} and \ref{tab:alfa}; {\bf b)} The same as a) but for the mass $M_{D_sD_s}$.
}
\label{fig:dsds} 
\end{center}
\end{figure}

\nin
The analysis of the $\mu$ subtraction point behaviour of the $\bar D_sD_s$ coupling and mass is very similar to the chiral limit case discussed in detail in\,\cite{ChXYZ} and will not be repeated here. We use the optimal choice obtained there:
\beq
\mu= (4.5\pm 0.5)~{\rm GeV}~.
\eeq
Taking the previous value of $\mu$, we study, in Fig.\,\ref{fig:dsds}a), the behaviour of the $\bar D_sD_s$ coupling\,\footnote{Here and in the following ``decay constant" is the same as ``coupling".} and in Fig.\,\ref{fig:dsds}b) its mass in terms of the LSR variable $\tau$ at different values of $t_c$  at NLO of PT QCD by including the contributions of condensates up to dimension 6. Higher dimension $(d= 8)$ though partially known (see \ref{app.a}) will not be included in the OPE but serves as an estimate of the errors induced by unknown higher dimension condensates. We shall use, for the estimate of the coupling, the input $\bar D_sD_s$ mass value obtained iteratively from the sum rule. 
\nin

\subsubsection*{$-$ $\bar D_sD_s/\bar DD$  SU3 ratios of couplings and masses}
\nin
Taking the previous value of input parameters, we study, in Fig.\,\ref{fig:dds}a), the behaviour of the SU3 ratio of couplings  $f^{sd}_{DD}$  and in Fig.\,\ref{fig:dds}b) the ratio of masses $r^{sd}_{DD}$ (see Eq.\,\ref{eq:rsd}) in terms of the LSR variable $\tau$ at different values of $t_c$.  
Noticing from Fig.\,\ref{fig:dds} that the value of $\tau$ at which the decay constant $f_{D_sD_s}$ reaches a minimum is about the same as the one of $f_{DD}$ (Fig. 6 of Ref. \,\cite{ChXYZ}), then, it is legitimate to use the ratio or double ratio of sum rules (DRSR)  for extracting with a good accuracy the SU3 breaking corrections to 
the coupling and mass in this channel.  We  show this ratio in Fig. \,\ref{fig:dds} where only the mass ratio presents $\tau$-stability.
\subsubsection*{$-$ Results}
\begin{figure}[hbt] 
\begin{center}
{\includegraphics[width=6.29cm  ]{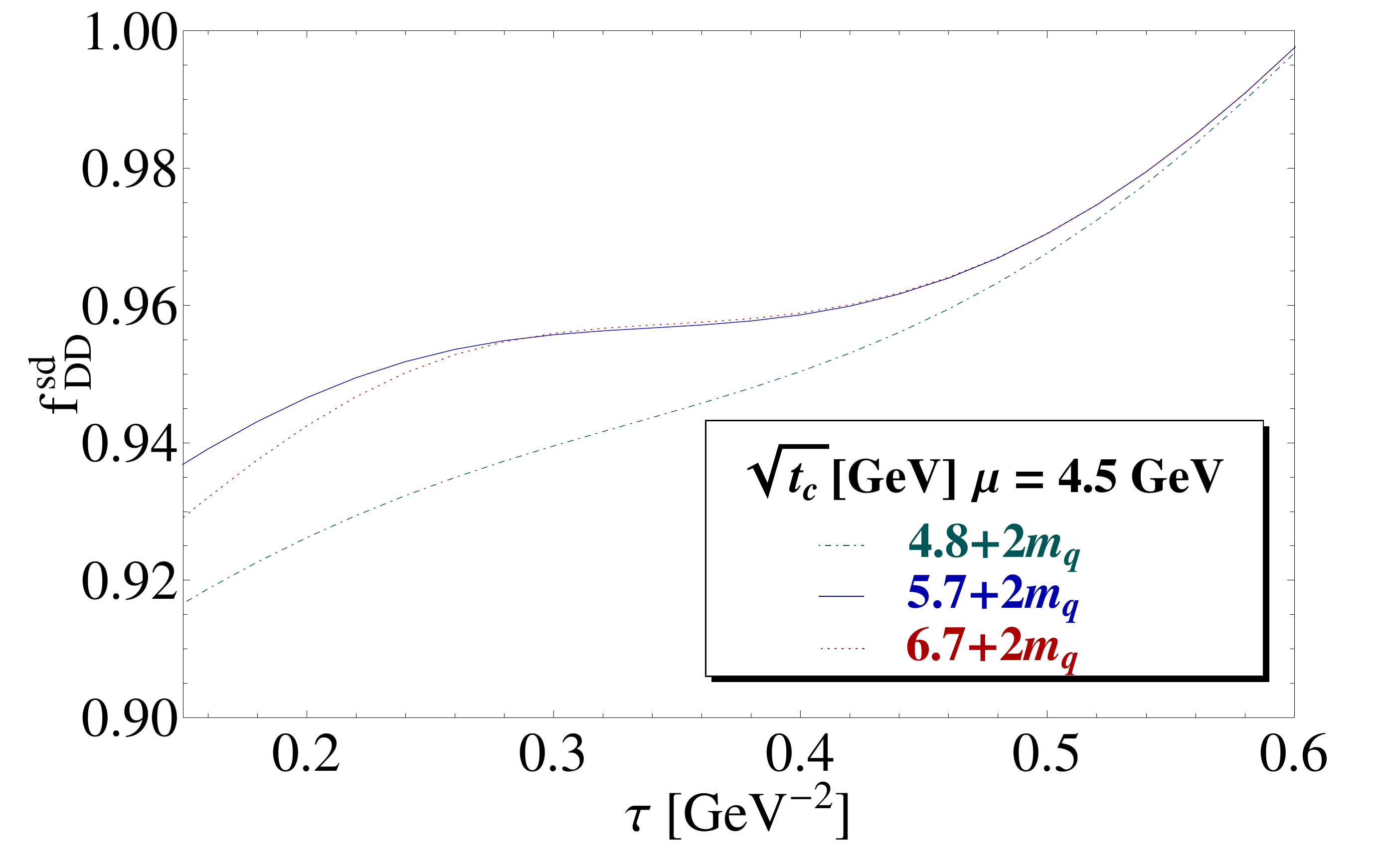}}
{\includegraphics[width=6.29cm  ]{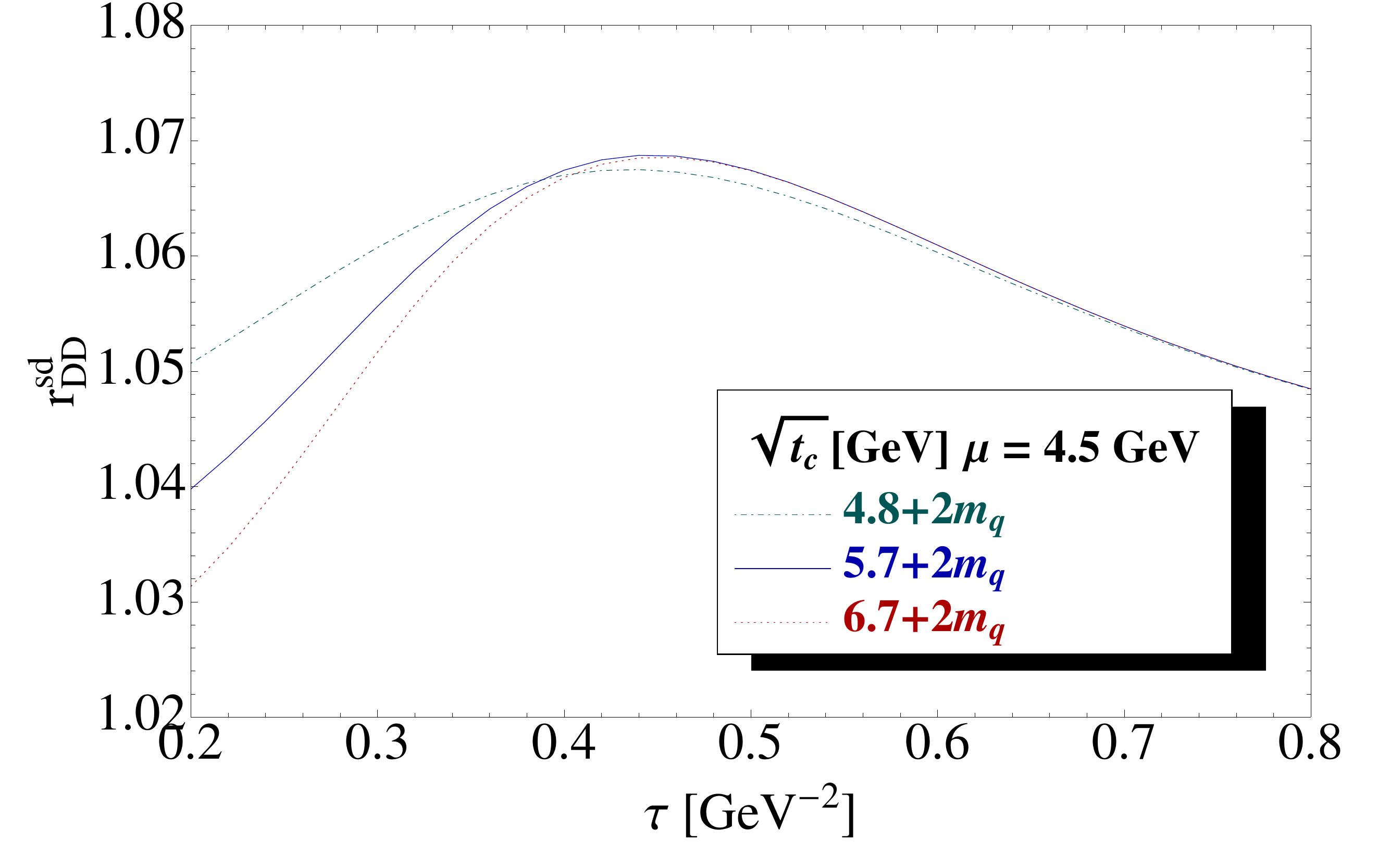}}
\centerline {\hspace*{-3cm} a)\hspace*{6cm} b) }
\caption{
\scriptsize 
{\bf a)} SU3 ratio of couplings $f^{sd}_{DD}$ at NLO  as function of $\tau$ for different values of $t_c$, for $\mu=4.5$ GeV  and for the QCD parameters in Tables\,\ref{tab:param} and \ref{tab:alfa}; {\bf b)} The same as a) but for the SU3 ratio of masses $r^{sd}_{DD}$.
}
\label{fig:dds} 
\end{center}
\end{figure}

\nin
From the previous analysis, we consider as an optimal estimate the mean value of the coupling , mass and their SU3 ratios obtained at the minimum or inflexion point for the common range  of $t_c$-values ($\sqrt{t_c}\simeq 4.8+2\overline m_s$ GeV) corresponding to the starting of the $\tau$-stability f and the one where (almost) $t_c$-stability ($\sqrt{t_c}\simeq 6.7+2\overline m_s$ GeV) is reached
 for $\tau\simeq (0.38\pm 0.02)$ GeV$^{-2}$. 
In these stability regions, the requirement that the pole contribution is larger than the one of the continuum  is automatically satisfied (see e.g.\,\cite{MOLE5}).
In this way, we obtain from a direct determination of the mass and coupling in Fig.\,\ref{fig:dsds} and for $\mu=4.5$ GeV at NLO:
\beq
f_{D_sD_s}(4.5)\simeq 156(10)_{t_c}(1)_{\tau}\cdots{\rm keV},~~~~
M_{D_sD_s}\simeq 4144(10)_{t_c}(38)_{\tau}\cdots{\rm MeV}~,
\label{eq:ddsa}
\eeq
at $\tau \simeq$ 0.28 (resp. 0.38) GeV$^{-2}$ for $\sqrt{t_c}\simeq$ 4.8+$2m_s$ (resp. 6.7+$2m_s$) GeV. $\cdots$ correspond to errors given in Table\,\ref{tab:errorcplus} induced by the QCD input parameters. 
Using the input values of $f_{DD}$=164(8) keV and $M_{DD}$ =3901(6) MeV at NLO from Ref.\,\cite{ChXYZ}\,\footnote{In order to avoid double counting, we retain only the error due to $(\tau,t_c)$ for these inputs.}, we deduce:
\beq
f^{sd}_{DD}\equiv\frac{f_{D_sD_s}}{f_{DD}}\simeq 0.95(4)_f(6)_{t_c}(0)_{\tau}\cdots,
r^{sd}_{DD}\equiv\frac{M_{D_sD_s}}{M_{DD}}\simeq 1.062(2)_M(3)_{t_c}(10)_{\tau}\cdots,
\label{eq:dds}
\eeq
where the first errors come from the determination of the $\bar DD$ coupling and mass\,\cite{ChXYZ}.
A direct determination of the SU3 ratios  of couplings and masses from Fig.\,\ref{fig:dds} shows $\tau$ and $t_c$-stabilities from which we deduce a more accurate determination:
\beq
f^{sd}_{DD}\simeq 0.950(8)_{t_c}(1)_{\tau}\cdots,~~~~~r^{sd}_{DD}\simeq 1.069(1)_{t_c}(0)_{\tau}\cdots,
\label{eq:rdd}
\eeq
in perfect agreement within the errors with the ones in Eq.\,\ref{eq:dds}.   Using the input values of $M_{DD}$ =3901(6) MeV and $f_{DD}$=164(8) keV at NLO from Ref.\,\cite{ChXYZ}, we can deduce:
\beq
f_{D_sD_s}\simeq 156(8)_f(1)_{t_c}(0)_{\tau}\cdots{\rm keV}~,~~~M_{D_sD_s}\simeq 4170(6)_M(4)_{t_c}(0)_{\tau}\cdots{\rm MeV}~,
\eeq
which agree with the direct determination in Eq.\,\ref{eq:ddsa}. We take as a final estimate the mean:
\bea
f^{sd}_{DD}&\simeq& 0.950(5)\cdots,~~~~~f_{D_sD_s}\simeq 156(8)\cdots{\rm keV}~,\nnb\\
r^{sd}_{DD}&\simeq& 1.069(1)\cdots,~~~~~M_{D_sD_s}\simeq 4169(7)...{\rm MeV}~,
\eea
\label{eq:dd}
\nin
   \subsection*{\b The $\bar D^*_sD^*_s$ molecule state}
  \subsubsection*{$-$ $\bar D^*_sD^*_s$  coupling and mass}
\nin
  The qualitative behaviour of different curves are very similar to the case of the $\bar D_sD_s$  state. From these
  curves, one can see that the ones of the coupling present stabilities, while the ones of the mass  have inflexion points which cannot be precisely located. We deduce from the analysis a direct determination of the coupling: 
  \beq
  f_{D^*_{s}D^*_{s}}(4.5)\simeq  259(23)_{t_c}(0)_{\tau}\cdots~{\rm keV},
 \eeq
where one has stability at $\tau\simeq$ 0.24 (resp. 0.36) GeV$^{-2}$ for $\sqrt{t_c}\simeq$ 4.8+$2\overline m_s$ (resp. 6.7+$2\overline m_s$) GeV.
\subsubsection*{$-$ $\bar D^*_sD^*_s/\bar D^*D^*$  SU3 ratios of couplings and masses}
\nin
\begin{figure}[hbt] 
\begin{center}
{\includegraphics[width=6.29cm  ]{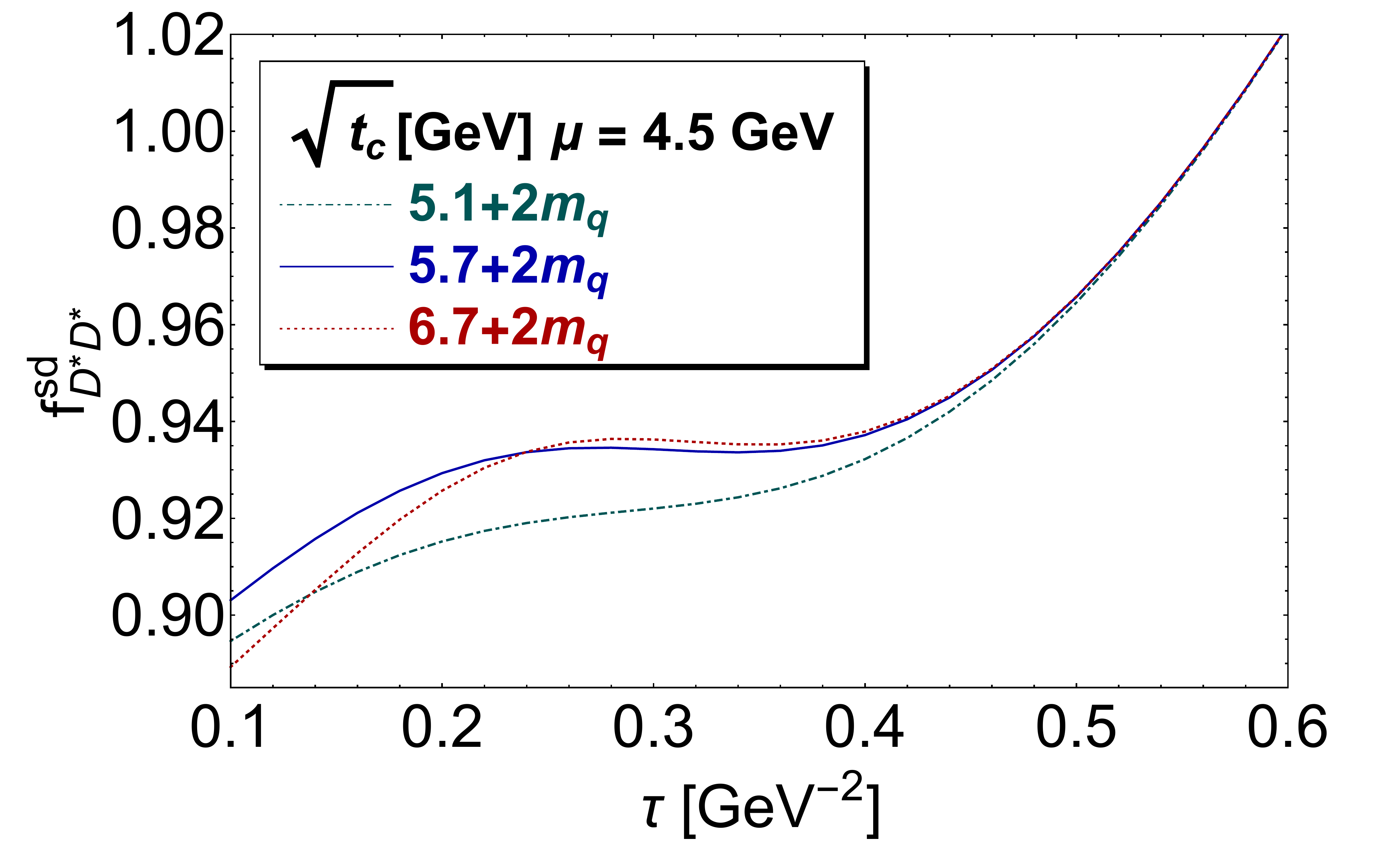}}
{\includegraphics[width=6.29cm  ]{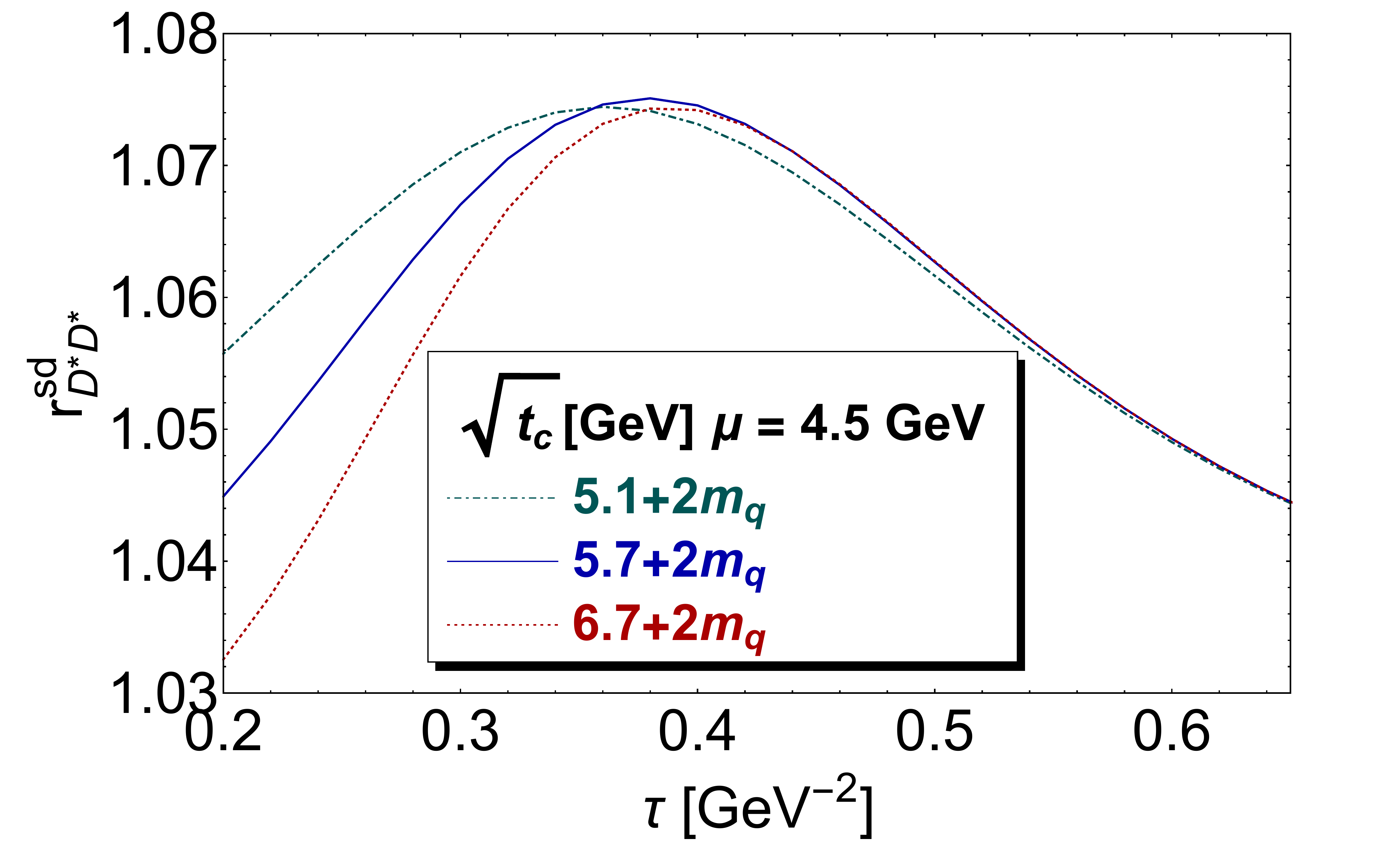}}
\centerline {\hspace*{-3cm} a)\hspace*{6cm} b) }
\caption{
\scriptsize 
{\bf a)} SU3 ratio of couplings $f^{sd}_{D^*D^*}$ at NLO  as function of $\tau$ for different values of $t_c$, for $\mu=4.5$ GeV  and for the QCD parameters in Tables\,\ref{tab:param} and \ref{tab:alfa}; {\bf b)} The same as a) but for the SU3 ratio of masses $r^{sd}_{D^*D^*}$.
}
\label{fig:dstdst} 
\end{center}
\end{figure} 
\nin
In this channel, the ratios of masses present extrema at  $\tau\simeq$ 0.36 (resp. 0.38) GeV$^{-2}$ like in the case of $D_sD_s$ but more pronounced while the ratio of  couplings presents net  inflexion points at  $\tau\simeq$ 0.28 (resp. 0.36) GeV$^{-2}$as shown in Fig.\,\ref{fig:dstdst}. We deduce:
 \beq
f^{sd}_{D^*D^*}\equiv\frac{f_{D^*_{s}D^*_{s}}}{f_{D^*D^*}}\simeq 0.929(8)_{t_c}(0)_\tau\cdots~,~~~ r^{sd}_{D^*D^*}\equiv\frac{M_{D^*_{s}D^*_{s}}}{M_{D^*D^*}}\simeq 1.074(1)_{t_c}(0)_{\tau}\cdots~.
\label{eq:fsdst}
\eeq
\subsubsection*{$-$ Results at NLO}
\nin
Combining the previous value of  $f_{D^*_{s}D^*_{s}}$ and the ratio $ r^{sd}_{D^*D^*}$ with the ones at NLO from \cite{ChXYZ}: $f_{D^*D^*}=288(9)$ keV and $M_{D^*D^*}=3903(17)$ MeV, one can deduce:
\beq
f^{sd}_{D^*D^*}\equiv\frac{f_{D^*_{s}D^*_{s}}}{f_{D^*D^*}}\simeq 0.90(3)_{f}(8)_{t_c}(0)_\tau\cdots,~~M_{D^*_{s}D^*_{s}}\simeq 4192(18)_{M}(4)_{t_c}(0)_{\tau}\cdots{\rm MeV},
\label{eq:fsdst2}
\eeq   
where the first errors come from the determination of the $\bar D^*D^*$ coupling and mass\,\cite{ChXYZ}. Taking the mean of the ratio of couplings and re-using the value $f_{D^*D^*}=288(9)$ keV, we deduce our final estimate:
\beq
f^{sd}_{D^*D^*}\simeq 0.929(8)\cdots~\lrar~ f_{D^*_{s}D^*_{s}}(4.5)\simeq  265(8.3)(2.9)~{\rm keV}~.
\eeq
   \subsection*{\b The $\bar D^*_{s0}D^*_{s0}$  and $\bar D^*_{0}D^*_{0}$ molecule states}
   \subsubsection*{$-$ The $\bar D^*_{s0}D^*_{s0}$ molecule }
 The analysis is very similar to the above (see Fig.\,\ref{fig:dds}), where the SU3 ratio of masses presents maxima
 at  $\tau\simeq$ 0.28 (resp. 0.32) GeV$^{-2}$ for $\sqrt{t_c}\simeq$ 5.3 (resp. 7) +$2\overline m_s$ GeV while the decay constant presents $\tau$ and $t_c$-stabilities (see Fig.\,\ref{fig:dsds}) 
for $\tau\simeq$ 0.18 (resp. 0.28) GeV$^{-2}$ for the previous range of $t_c$-values. 
In this way, we obtain at NLO:
 \beq
  f_{D^*_{s0}D^*_{s0}}(4.5)\simeq  86(8)_{t_c}(0)_{\tau}\cdots~{\rm keV}~,~~~
  r^{sd}_{D^*_0D^*_0}\equiv\frac{M_{D^*_{s0}D^*_{s0}}}{M_{D^*_0D^*_0}}\simeq 1.069(68)_{t_c}(13)_{\tau}\cdots~.
\eeq 
A direct determination of the ratio of coupling shows net inflexion points as in Fig.\,\ref{fig:dds} at $\tau\simeq$ 0.24 (resp. 0.30) GeV$^{-2}$ for $\sqrt{t_c}\simeq$ 5.3+2$\overline m_q$ (resp. 7+2$\overline m_q$) GeV. In this way, we obtain:
\beq
f^{sd}_{D^*_0D^*_0}\simeq 0.875(6)_{t_c}(0)_\tau\cdots~
\label{eq:dstfsd}
\eeq
The analysis of the mass presents an inflexion point as in Fig.\,\ref{fig:dsds} which is senstive to the $\tau$-values. We fix this range as the one from 
$ r^{sd}_{D^*_0D^*_0}$ which is $\tau\simeq$ 0.28 (resp. 0.32) GeV$^{-2}$ for $\sqrt{t_c}\simeq$ 5.3 (resp. 7) +$2\overline m_s$ GeV. In this way, we obtain:
\beq
M_{D^*_{s0}D^*_{s0}}\simeq 4277(102)_{t_c}(92)_{\tau}\cdots{\rm MeV}~.
\eeq
   \subsubsection*{$-$ Revisiting the $\bar D^*_{0}D^*_{0}$  molecule}
Here, we cross-check our results obtained in the chiral limit in\,\cite{ChXYZ} and we notice an error as the $\tau$-stability of the coupling $  f_{D^*_{0}D^*_{0}}$ starts earlier  from $\sqrt{t_c}=5.3$ GeV and for $\tau\simeq 0.24$ GeV$^{-2}$ than the one used in\,\cite{ChXYZ}. Taking this larger range of $\sqrt{t_c}$ values from 5.3 to 6.8 GeV, we deduce at NLO:
\beq
f_{D^*_{0}D^*_{0}}\simeq 96(10)\cdots~{\rm keV}~,
\label{eq:dst0f}
\eeq
instead of 116 keV obtained in\,\cite{ChXYZ}.  This change of the low-$t_c$ values also affects the direct mass determination.
The curves present  inflexion points around $\tau\simeq 0.29$ GeV$^{-2}$ which leads at NLO to:
\beq
M_{D^*_{0}D^*_{0}}\simeq 4008(411)_{t_c}(35)_{\tau}\cdots{\rm MeV}~,
\eeq
with a large error instead of 4402(54) MeV quoted in\,\cite{ChXYZ}.    
   \subsubsection*{$-$ Final Results}
We also deduce from the SU3 ratios and the result from $\bar D^*_{s0}D^*_{s0}$, the value of the coupling:
\beq
  f_{D^*_{s0}D^*_{s0}}(4.5)\simeq  84(9)\cdots~{\rm keV}~,~~~~
  M_{D^*_{0}D^*_{0}}\simeq 4001(95)_{t_c}(255)_f\cdots{\rm MeV}~,
\eeq
Taking the  mean of the two values of couplings and masses, we deduce the final value at NLO:
\bea
 f_{D^*_{s0}D^*_{s0}}(4.5)&\simeq& 85(6)\cdots~{\rm keV}~~\lrar~~  f_{D^*_{0}D^*_{0}}(4.5)\simeq  97(7)\cdots~{\rm keV},\nnb\\
  M_{D^*_{0}D^*_{0}}&\simeq& 4003(227)\cdots{\rm MeV}.
  \label{eq:d*0d*0}
 \eea 
   \subsection*{\b The $\bar D_{s1}D_{s1}$ and $\bar D_1D_1$ molecule states}
We perform a similar analysis. The behaviours of the different curves are very similar to the case of the $\bar D_sD_s$ molecule states and will not be shown here. They present stabilites for $\sqrt{t_c}\simeq$ 5.1+$2\overline m_s$ to  6.7+$2\overline m_s$ GeV for $\tau\simeq 0.28-0.34$ (resp. 0.32--0.34) (resp. 0.28--0.34) GeV$^{-2}$ for the coupling $f_{D_{s1}D_{s1}}$ (resp. SU3 ratio of couplings $f^{sd}_{D_1D_1}$) (resp. SU3 ratio of masses  $r^{sd}_{D_1D_1}$) leading to the values at NLO:
\bea
f_{D_{s1}D_{s1}}\simeq 209(14)\cdots~{\rm keV},~f^{sd}_{D_1D_1}\simeq 0.906(9)\cdots~,r^{sd}_{D_1D_1}\simeq 1.097(6)\cdots~,
\eea
where the quoted errors come from the correlated values of $(t_c,\tau)$ and $\cdots$ are QCD corrections given in Table\,\ref{tab:errorcplus}. The mass presents an inflexion point which is difficult to localize.  To fix the $\tau$-values, we take the  range where the SU3 ratio of masses optimizes, which corresponds to $\tau\simeq 0.32-0.34$ GeV$^{-2}$. In this way, we obtain:
\beq
M_{D_{s1}D_{s1}}\simeq 4187(34)\cdots~{\rm MeV}~.
\eeq
Using the previous values of the SU3 ratio, we can deduce for the $\bar D_1D_1$ molecule at NLO:
\beq
f_{D_{1}D_{1}}\simeq 231(16)\cdots~{\rm keV}~,~~~~M_{D_{1}D_{1}}\simeq 3838(37)\cdots~{\rm MeV}~.
\label{eq:d1d1}
\eeq
\subsection{The  $(1^{+\pm})$ Charm Axial-Vector Molecule States}				     %
Here, within our choice of interpolating currents, the $(1^{++})$ and $(1^{+-})$ are degenerate in masses and have the same
couplings like in the case of the pseudoscalar molecules. 
   \subsection*{\b The $\bar D^*_sD_{s}$  molecule state}
The curves are very similar to the case of the scalar  molecules where the coupling presents a minimum for $\tau\simeq$ 0.26 (resp. 0.36) GeV$^{-2}$ for $\sqrt{t_c}\simeq$ 4.8+$2\overline m_s$ (resp. 6.7+$2\overline m_s$) GeV, while the SU3 ratio of masses presents a maximum both for $\tau=0.42$ GeV$^{-2}$ and for $\sqrt{t_c}\simeq$ 5.7+$2\overline m_q$ GeV $,~(q\equiv d,s)$.  
We obtain:
 \beq
  f_{D^*_{s}D_{s}}(4.5)\simeq  145(10)_{t_c}(0)_{\tau}\cdots~{\rm keV}~,~~~
  r^{sd}_{D^*D}\equiv\frac{M_{D^*_{s}D_{s}}}{M_{D^*D}}\simeq1.070(1)_{t_c}(0)_{\tau}\cdots~.
\eeq
Using the values: $M_{D^*D}=3901(3.4)$ MeV and $ f_{D^*D}=$ 154(7.6) keV from Ref. 1, we deduce:
\beq
f^{sd}_{D^*D}\simeq 0.94(5)_f(7)_{t_c}(0)_\tau\cdots,~~
M_{D^*_{s}D_{s}}\simeq 4174(3)_M(4)_{t_c}(0)_{\tau}\cdots{\rm MeV}.
\eeq   
One can improve the determination of the ratio of couplings by its direct determination. At the inflexion points for $\tau\simeq$ 0.30 (resp. 0.34) GeV$^{-2}$ for $\sqrt{t_c}\simeq$ 4.8+$2\overline m_q$ (resp. 6.7+$2\overline m_q$) GeV, one deduces:
\beq
f^{sd}_{D^*D}\simeq 0.934(10)_{t_c}(0)_\tau\cdots.
\eeq
Taking the mean value of the SU3 ratio of coupling, we deduce at NLO:
\beq
f^{sd}_{D^*D}\simeq 0.930(7)\cdots~\lrar~ f_{D^*_{s}D_{s}}(4.5)\simeq  143(7)_f(1.1)_{t_c}(0)_{\tau}\cdots~{\rm keV}~
\eeq
   \subsection*{\b The $\bar D^*_{s0}D_{s1}$ and  $\bar D^*_{0}D_{1}$ molecule states}
\subsubsection*{$-$ The $\bar D^*_{s0}D_{s1}$  molecule}
 The coupling and SU3 ratio of masses stabilizes for $\tau\simeq$ 0.26 (resp. 0.32) GeV$^{-2}$ for $\sqrt{t_c}\simeq$ 5.3+$2\overline  m_q$ (resp. 6.7+$2\overline m_q$) GeV.  The SU3 ratio of coupling stabilizes for $\tau\simeq$ 0.27(resp. 0.28) GeV$^{-2}$. 
 We obtain at NLO:
 \bea
  f_{D^*_{s0}D_{s1}}(4.5)&\simeq&  87(7)_{t_c}(0)_{\tau}\cdots~{\rm keV},~~f^{sd}_{D^*_0D_1}\simeq 0.904(9)_{t_c}(0)_\tau~,\nnb\\
  r^{sd}_{D^*_0D_1}&\simeq&1.119(18)_{t_c}(0)_{\tau}\cdots
\eea
Using the previous range of values of $\tau\simeq$ 0.26 (resp. 0.32) GeV$^{-2}$,  we deduce at NLO:
\beq
M_{D^*_{s0}D_{s1}}\simeq 4269(7)_{t_c}(0)_{\tau}\cdots~{\rm MeV}.
\eeq 
 \subsubsection*{$-$ Revisiting the $\bar D^*_{0}D_{1}$ molecule}
 Here we revise our previous result in Ref.\,\cite{ChXYZ} by correcting the range  of $t_c$ and of $\tau$ used there. The coupling stabilizes at $\tau\simeq$ 0.26 (resp. 0.32) GeV$^{-2}$ for $\sqrt{t_c}\simeq$ 5.3+$2\overline  m_q$ (resp. 6.7+$2\overline m_q$) GeV. Within these ranges of values,  a direct determination gives:
\beq
  f_{D^*_{0}D_{1}}(4.5)\simeq 96(7)_{t_c}(0)_{\tau}\cdots~{\rm keV},~~
 M_{D^*_{0}D_{1}}\simeq 3857(29)_{t_c}(0)_{\tau}\cdots~{\rm MeV}.
\eeq 
Combining the previous values of $M_{D^*_{s0}D_{s1}},~f_{D^*_{s0}D_{s1}}$ with $ r^{sd}_{D^*_0D_1}$ and  $ f^{sd}_{D^*_0D_1}$, one can also deduce:
\beq
  f_{D^*_{0}D_{1}}(4.5)\simeq  96(7)_{t_c}(0)_{\tau}\cdots~{\rm keV},~~
 M_{D^*_{0}D_{1}}\simeq 3815(61)_{t_c}(0)_{\tau}\cdots~{\rm MeV}.
\eeq 
Taking the mean of the two determinations lead to our final estimate at NLO:
\beq
  f_{D^*_{0}D_{1}}(4.5)\simeq  96(7)\cdots~{\rm keV},~~
 M_{D^*_{0}D_{1}}\simeq 3849(26)\cdots~{\rm MeV}.
\eeq 
 These corrected values replace the ones obtained in Ref.\,\cite{ChXYZ} at NLO :
 \beq
  f_{D^*_{0}D_{1}}(4.5)\simeq  118(16)~{\rm keV},~~
 M_{D^*_{0}D_{1}}\simeq 4395(164)~{\rm MeV}.
\eeq 
These revisited values together with the ones of $D^*_0D^*_0$ given in Eq.\,\ref{eq:d*0d*0} and the new value of the ones of $D_1D_1$ in Eq.\,\ref{eq:d1d1} are quoted in Table\,\ref{tab:resultc}  to NLO and N2LO.
{\scriptsize
\begin{table}[hbt]
\setlength{\tabcolsep}{0.25pc}
\tbl{Different sources of errors for the estimate of the $0^{+}$ and $1^{+}$ $\bar D_sD_s$-like molecule masses (in units of MeV) and couplings $f_{M_sM_s}(\mu)$ (in units of keV). We use $\mu=4.5(5)$ GeV.}
   {\scriptsize
 {\begin{tabular}{@{}llllllllllllllllllll@{}} 
&\\
\hline
\hline
 &\multicolumn{2}{c}{$D_sD_s$}
					&\multicolumn{2}{c}{$D^{*}_{s}D^{*}_{s}$}
					&\multicolumn{2}{c}{$D^{*}_{s0}D^{*}_{s0}$}
					&\multicolumn{2}{c}{$D_{s1}D_{s1}$}
					&\multicolumn{2}{c}{$D^{*}_{s}D_{s}$}
					&\multicolumn{2}{c}{$D^{*}_{s0}D_{s1}$}
					\\
\cline{2-3} \cline{4-5}\cline{6-7}\cline{8-9}\cline{10-11}\cline{12-13}
                 & \multicolumn{1}{l}{$\Delta M$} 
                 & \multicolumn{1}{l }{$\Delta f$} 
                 & \multicolumn{1}{l}{$\Delta M$}
                 & \multicolumn{1}{l }{$\Delta f$} 
                 & \multicolumn{1}{l}{$\Delta M$}
                 & \multicolumn{1}{l}{$\Delta f$} 
		    & \multicolumn{1}{l}{$\Delta M$} 
                 & \multicolumn{1}{l}{$\Delta f$} 
                 & \multicolumn{1}{l}{$\Delta M$}
                 & \multicolumn{1}{l}{$\Delta f$} 
		    & \multicolumn{1}{l}{$\Delta M$} 
                 & \multicolumn{1}{l}{$\Delta f$} 
                  \\
\hline

\hline 
{\bf Inputs}\\
{\it LSR parameters}&\\
$(t_c,\tau)$&7&10&18&23&59&4&34&14&5&10&7&7\\
$\mu$&24.56&0.16&24.14&0.60&28.31&0.61&25.50&8.65&27.89&0.63&30.33&2.87\\
{\it QCD inputs }&\\
$\bar M_Q$&11.19&4.82&7.98&7.67&8.63&2.83&4.61&5.45&11.07&3.17&5.05&1.99\\
$\alpha_s$&12.45&3.50&12.30&5.79&15.65&1.33&11.76&3.68&12.63&4.26&18.68&1.21\\
$N3LO$&0.0&0.77&0.28&1.33&3.64&1.19&4.41&1.40&0.98&0.91&0.42&1.61\\
$\la\bar qq\ra$&11.20&1.93&9.32&1.68&38.38&5.16&4.58&2.26&8.51&1.70&32.09&3.09\\
$\la\alpha_s G^2\ra$&5.66&1.04&1.32&0.58&0.06&2.54&4.31&1.82&0.46&0.11&2.02&0.17\\
$M_0^2$&9.93&0.20&8.28&2.59&15.18&1.00&14.28&2.81&6.32&1.17&13.19&1.67\\
$\la\bar qq\ra^2$&11.01&7.47&9.95&18.08&51.01&13.29&27.77&20.46&10.24&7.41&38.73&12.90\\
$\la g^3G^3\ra$&0.08&0.11&0.12&0.25&0.57&0.23&0.16&0.31&0.30&0.10&1.54&0.16\\
$d\geq8$&32.0&9.62&196.5&4.80&95.5&2.74&29.9&6.39&56.0&8.10&194.8&1.22\\
{\it Total errors}&48.31&17.01&199.97&31.33&134.32&12.22&62.41&28.14&66.62&15.95&204.94&13.97\\
\hline\hline
\end{tabular}}
\label{tab:errorcplus}
}
\end{table}
} 
{\scriptsize
\begin{table}[H]
\setlength{\tabcolsep}{0.25pc}
\tbl{Different sources of errors for the estimate of the $0^{+}$ and $1^{+}$  $\bar DD$-like molecule SU(3) ratios of masses $r^{sd}_{MM}$ and SU(3) ratios of couplngs $f^{sd}_{MM}$. We use $\mu=4.5(5)$ GeV.}
   {\scriptsize
 {\begin{tabular}{@{}llllllllllllllllllll@{}} 
&\\
\hline
\hline
 &\multicolumn{2}{c}{$D_sD_s$}
					&\multicolumn{2}{c}{$D^{*}_{s}D^{*}_{s}$}
					&\multicolumn{2}{c}{$D^{*}_{s0}D^{*}_{s0}$}
					&\multicolumn{2}{c}{$D_{s1}D_{s1}$}
					&\multicolumn{2}{c}{$D^{*}_{s}D_{s}$}
					&\multicolumn{2}{c}{$D^{*}_{s0}D_{s1}$}
					\\
\cline{2-3} \cline{4-5}\cline{6-7}\cline{8-9}\cline{10-11}\cline{12-13}
                 & \multicolumn{1}{l}{$r^{sd}$} 
                 & \multicolumn{1}{l }{$f^{sd}$} 
                 & \multicolumn{1}{l}{$r^{sd}$}
                 & \multicolumn{1}{l }{$f^{sd}$} 
                 & \multicolumn{1}{l}{$r^{sd}$}
                 & \multicolumn{1}{l}{$f^{sd}$} 
		    & \multicolumn{1}{l}{$r^{sd}$} 
                 & \multicolumn{1}{l}{$f^{sd}$} 
                 & \multicolumn{1}{l}{$r^{sd}$}
                 & \multicolumn{1}{l}{$f^{sd}$} 
		    & \multicolumn{1}{l}{$r^{sd}$} 
                 & \multicolumn{1}{l}{$f^{sd}$} 
                  \\
\hline

\hline 
{\bf Inputs}\\
{\it LSR parameters}&\\
$(t_c,\tau)$&0.001&0.005&0.001&0.008&0.013&0.036&0.006&0.009&0.001&0.007&0.018&0.009\\
$\mu$&0.0&0.001&0.0&0.001&0.0&0.002&0.001&0.001&0.0&0.001&0.0&0.002\\
{\it QCD inputs}&\\
$\bar M_Q$&0.0&0.002&0.0&0.003&0.0&0.003&0.0&0.003&0.0&0.002&0.001&0.002\\
$\alpha_s$&0.0&0.001&0.0&0.0&0.0&0.0&0.0&0.0&0.0&0.0&0.001&0.00\\
$N3LO$&0.00&0.00&0.00&0.00&0.00&0.001&0.001&0.002&0.00&0.002&0.01&0.00\\
$\la\bar qq\ra$&0.001&0.0&0.001&0.0&0.008&0.004&0.002&0.001&0.001&0.0&0.006&0.00\\
$\la\alpha_s G^2\ra$&0.001&0.002&0.0&0.001&0.0&0.007&0.001&0.002&0.0&0.0&0.0&0.0\\
$M_0^2$&0.001&0.0&0.002&0.0&0.0&0.0&0.002&0.0&0.002&0.0&0.003&0.0\\
$\la\bar qq\ra^2$&0.002&0.025&0.001&0.025&0.010&0.031&0.002&0.030&0.002&0.024&0.007&0.030\\
$\la g^3G^3\ra$&0.0&0.0&0.0&0.0&0.0&0.001&0.0&0.0&0.0&0.0&0.0&0.0\\
$d\geq8$&0.003&0.009&0.002&0.010&0.010&0.010&0.002&0.010&0.002&0.007&0.012&0.002\\
{\it Total errors}&0.004&0.028&0.003&0.028&0.021&0.050&0.007&0.033&0.004&0.026&0.024&0.031\\
\hline\hline
\end{tabular}}
\label{tab:error-rap-cplus}
}
\end{table}
} 

\subsection{ The  $(0^{-\pm})$ Charm Pseudoscalar Molecule States}				     %
Here, within our choice of interpolating currents, the $(0^{--})$ and $(0^{-+})$ are degenerate in masses and have the same
couplings. 

   \subsection*{\b The $\bar D^*_{s0}D_{s}$  molecule state}
\begin{figure}[hbt] 
\begin{center}
{\includegraphics[width=6.29cm  ]{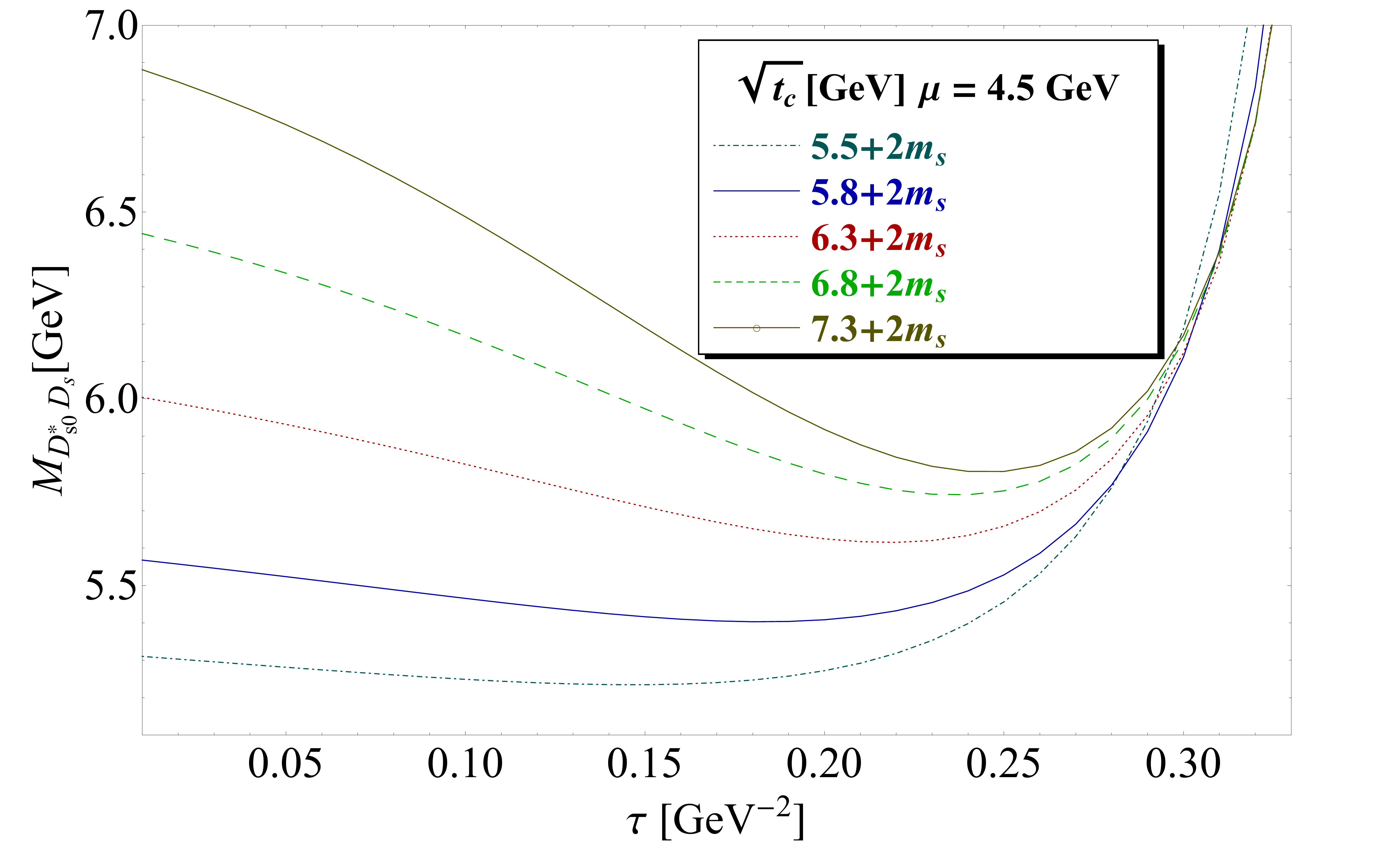}}
{\includegraphics[width=6.29cm  ]{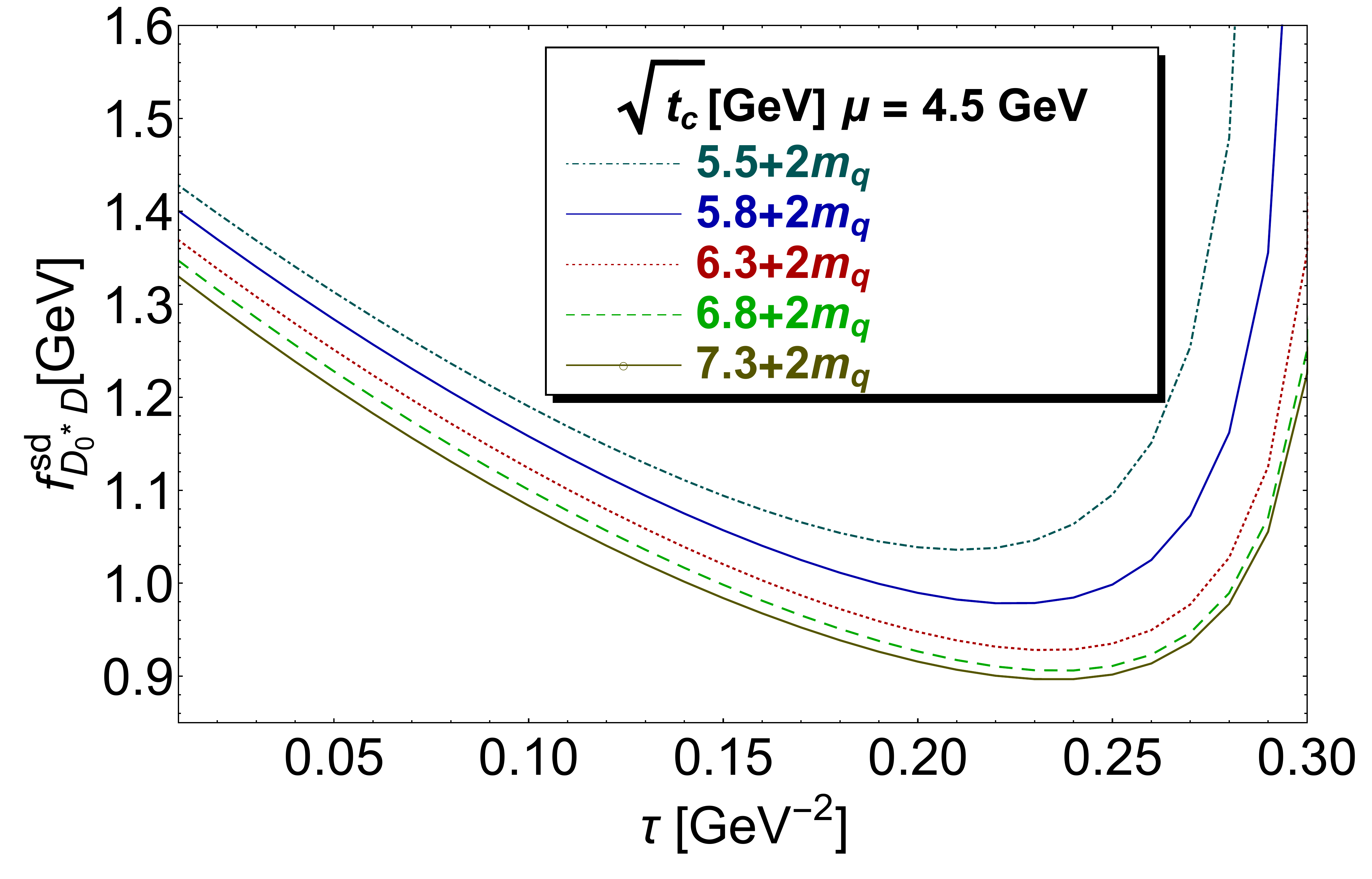}}
\centerline {\hspace*{-3cm} a)\hspace*{6cm} b) }
\caption{
\scriptsize 
{\bf a)} The mass $M_{D^*_0D}$ at NLO  as function of $\tau$ for different values of $t_c$, for $\mu=4.5$ GeV  and for the QCD parameters in Tables\,\ref{tab:param} and \ref{tab:alfa}; {\bf b)} The same as a) but for the SU3 ratio of couplings $f^{sd}_{D^*_0D}$.
}
\label{fig:dst0d} 
\end{center}
\end{figure} 
\nin
The mass presents minima (but not the coupling) for $\tau$=0.18 (resp. 0.25) GeV$^{-2}$ corresponding to $\sqrt{t_c}=5.8+2\overline m_q$ (resp. $\sqrt{t_c}=7.3+2\overline m_q$) GeV as shown in Fig.\,\ref{fig:dst0d}. Within the same range of $t_c$, the ratio of couplings has minima for
$\tau$=0.22 (resp. 0.24) GeV$^{-2}$ . In these regions, we deduce:
\beq
f^{sd}_{D^*_0D}\simeq 0.938(41)_{t_c}(2)_\tau\cdots,~
M_{D^*_{s0}D_{s}}\simeq 5604(201)_{t_c}(17)_{\tau}\cdots~{\rm MeV}.
\eeq  
Using the values: $M_{D^*_0D}=5800(115)$ MeV and $ f_{D^*_{0}D}=$ 240(16) keV from\,\cite{ChXYZ}, we deduce at NLO:
\beq
r^{sd}_{D^*_0D}\simeq 0.97(2)_M(5)_{t_c}(0)_\tau\cdots,~
f_{D^*_{s0}D_{s}}\simeq 225(15)_f(10)_{t_c}(1)_{\tau}\cdots~{\rm keV}.
\eeq 
   \subsection*{\b The $\bar D^*_{s}D_{s1}$  molecule state}
\begin{figure}[hbt] 
\begin{center}
{\includegraphics[width=6.29cm  ]{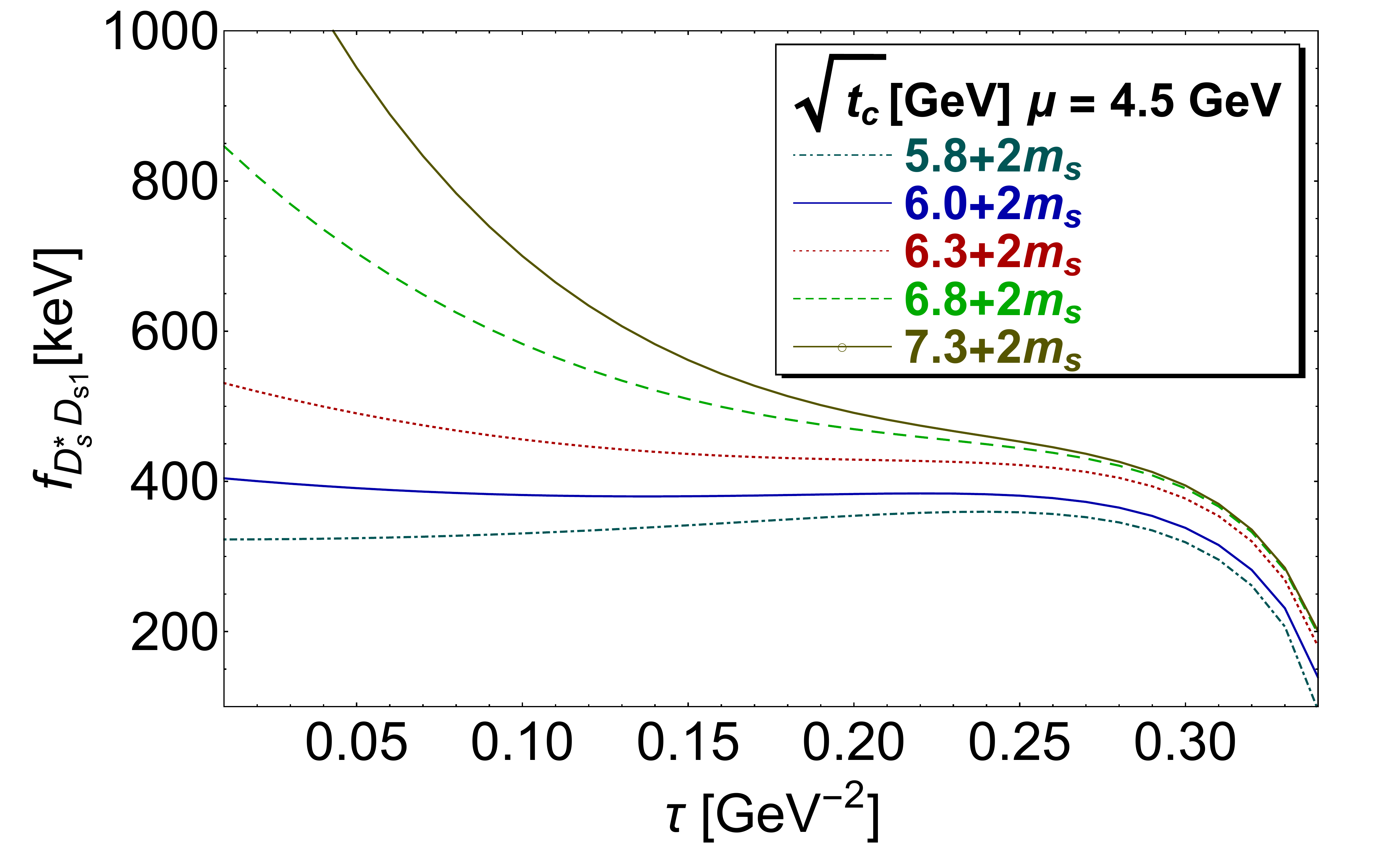}}
\caption{
\scriptsize 
The coupling $f_{D^*_sD_{s1}}$ at NLO  as function of $\tau$ for different values of $t_c$, for $\mu=4.5$ GeV  and for the QCD parameters in Tables\,\ref{tab:param} and \ref{tab:alfa}.
}
\label{fig:dstd1} 
\end{center}
\end{figure}

The shapes of different curves for the mass and SU3 ratio of couplings are very similar to the case of the $\bar D^*_{s0}D_{s}$ and will not be shown here. Unlike the previous case of  $D^*_{s0}D_s$, the coupling presents $\tau$-stabilities as shown in Fig.\,\ref{fig:dstd1} from $\sqrt{t_c}=6.0+2\overline m_q$ (resp. $\sqrt{t_c}=7.3+2\overline m_q$) GeV and for $\tau$=0.14 (resp. 0.21) GeV$^{-2}$.
\nin
Within the above range of $t_c$, 
 the ratio of couplings presents stability for $\tau$=0.22 (resp. 0.24) GeV$^{-2}$ from $\sqrt{t_c}=6+2\overline m_q$ (resp. $\sqrt{t_c}=7.3+2\overline m_q$) GeV, while the minima for the mass occur at  $\tau$=0.19 (resp. 0.25) GeV$^{-2}$. In these regions, we deduce:
\bea
f_{D^*_{s}D_{s1}}&\simeq&431(51)_{t_c}(8)_\tau\cdots~{\rm keV},\nnb\\
f^{sd}_{D^*D_1}&\simeq& 0.94(1)_{t_c}(0)_\tau\cdots,~
M_{D^*_{s}D_{s1}}\simeq 5724(176)_{t_c}(14)_{\tau}\cdots~{\rm MeV}.
\eea  
Using the values: $M_{D^*D_1}=5898(89)$ MeV and $ f_{D^*D_1}=$ 490(25) keV from\,\cite{ChXYZ}, we deduce at NLO:
\beq
r^{sd}_{D^*D_1}\simeq 0.97(1.5)_M(5)_{t_c}(0)_\tau\cdots,~
f_{D^*_{s}D_{s1}}\simeq 460(23.5)_f(5)_{t_c}(0)_{\tau}\cdots~{\rm keV}~,
\eeq 
where the coupling agrees within the errors with the previous direct determination. Taking the mean of the couplings and re-using $ f_{D^*D_1}=$ 490(25) keV, we deduce the final estimate:
\beq
f_{D^*_{s}D_{s1}}\simeq 455(22)\cdots~{\rm keV}~,\lrar f^{sd}_{D^*D_1}\simeq 0.93(1)\cdots,
\eeq 
where we have taken the error on the ratio from the direct determination. 
\subsection{ The  $(1^{--})$ Charm Vector Molecule States}				     %
   \subsection*{\b The $(1^{--})~\bar D^*_{s0}D^*_{s}$ molecule state}
The results are similar to the previous ones. 
The coupling presents $\tau$-stabilities from $\sqrt{t_c}=6.0+2\overline m_q$ to $\sqrt{t_c}=7.3+2\overline m_q$ GeV and for $\tau$=0.21-0.24 (resp. 0.24) GeV$^{-2}$.
   Within these range of $t_c$-values, the mass stabilizes for $\tau\simeq$  0.18 (resp. 0.24) GeV$^{-2}$ and the SU3 ratio of couplings for $\tau\simeq$  0.23 (resp. 0.24) GeV$^{-2}$. 
We obtain:
\bea
f_{D^*_{s0}D^*_{s}}&\simeq& 201(13)_{t_c}(2)_{\tau}\cdots~{\rm keV}~,\nnb\\
f^{sd}_{D^*_0D^*}&\simeq&0.90(3)_{t_c}(0)_\tau\cdots,~M_{D^*_{s0}D^*_{s}}\simeq 5708(170)_{t_c}(12)_{\tau}~{\rm MeV}.
\eea   
 Using the values: $M_{D^*_0D^*}=5861(83.4)$ MeV and $ f_{D^*_0D^*}=$ 238(11.4) keV from\,\cite{ChXYZ}, we can deduce:
\beq
r^{sd}_{D^*_0D^*}\simeq 0.98(1.4)_M(3)_{t_c}(0)_\tau\cdots,~
f_{D^*_{s0}D^*_{s}}\simeq 214(10)_f(7)_{t_c}(0)_{\tau}\cdots~{\rm keV}~.
\eeq   
Taking the mean value of the couplings and re-using $ f_{D^*_0D^*}=$ 238(11.4) keV, we deduce the final estimate:
\beq
f_{D^*_{s0}D^*_{s}}\simeq 208(9)\cdots~{\rm keV}~\lrar~f^{sd}_{D^*_0D^*}\simeq 0.87(3)\cdots~,
\eeq
where again the error of the SU3 ratio comes from the precise direct determination. 
   \subsection*{\b The $(1^{--})~\bar D_sD_{s1}$ molecule state}
   The results of the analysis are  similar to the previous $D^*_sD_{s1}$ pseudoscalar case and the figures will not be shown.  
  The coupling presents $\tau$-stabilities from $\sqrt{t_c}=6.0+2\overline m_q$ to $\sqrt{t_c}=7.3+2\overline m_q$ GeV and for $\tau$=0.14 (resp. 0.24) GeV$^{-2}$.
   Within these range of $t_c$-values, the mass stabilizes for $\tau\simeq$  0.28 GeV$^{-2}$ and the ratio of couplings for $\tau\simeq$  0.28 GeV$^{-2}$. 
We obtain:
\bea
f_{D_{s}D_{s1}}&\simeq& 202(8)_{t_c}(0)_{\tau}\cdots~{\rm keV}~,\nnb\\
f^{sd}_{DD_1}&\simeq&0.96(2)_{t_c}(0)_\tau\cdots,~M_{D_{s}D_{s1}}\simeq 5459(100)_{t_c}(0)_{\tau}~{\rm MeV}.
\eea   
 Using the values: $M_{DD_1}=5639(150)$ MeV and $ f_{DD_1}=$ 209(19) keV from\,\cite{ChXYZ}, we can deduce:
\beq
r^{sd}_{DD_1}\simeq 0.97(2.6)_M(2)_{t_c}(0)_\tau\cdots,~
f_{D_{s}D_{s1}}\simeq 201(18)_f(4)_{t_c}(0)_{\tau}\cdots~{\rm keV}~.
\eeq   
Taking the mean of the couplings, we deduce our final result:
\beq
f_{D_{s}D_{s1}}\simeq 202(7)\cdots~{\rm keV}~,\lrar f^{sd}_{DD_1}\simeq 0.97(2)\cdots~.
\eeq
\subsection{ The  $(1^{-+})$ Charm Vector Molecule States}				     %
   \subsection*{\b The $(1^{-+})~\bar D^*_{s0}D^*_{s}$ molecule state}
The results are similar to the previous ones. 
The coupling presents $\tau$-stabilities from $\sqrt{t_c}=6.0+2\overline m_q$ to $\sqrt{t_c}=7.2+2\overline m_q$ GeV and for $\tau$=0.12(resp. 0.21) GeV$^{-2}$.
   Within these range of $t_c$-values, the mass stabilizes for $\tau\simeq$  0.18 (resp. 0.23) GeV$^{-2}$ and the ratio of couplings for $\tau\simeq$  0.18 (resp. 0.19) GeV$^{-2}$. 
We obtain:
\bea
f_{D^*_{s0}D^*_{s}}&\simeq& 203(20)_{t_c}(2)_{\tau}\cdots~{\rm keV}~,\nnb\\
f^{sd}_{D^*_0D^*}&\simeq&1.021(44)_{t_c}(12)_\tau\cdots,~M_{D^*_{s0}D^*_{s}}\simeq 5699(169)_{t_c}(3)_{\tau}~{\rm MeV}.
\eea   
 Using the values: $M_{D^*_0D^*}=5920(83.4)$ MeV and $ f_{D^*_0D^*}=$ 224(11.4) keV from\,\cite{ChXYZ}, we can deduce:
\beq
r^{sd}_{D^*_0D^*}\simeq 0.963(14)_M(29)_{t_c}(0)_\tau\cdots,~
f_{D^*_{s0}D^*_{s}}\simeq 229(12)_f(10)_{t_c}(3)_{\tau}\cdots~{\rm keV}~.
\eeq   
Taking the mean value of the couplings and re-using $ f_{D^*_0D^*}=$ 224(11.4) keV, we deduce the final estimate:
\beq
f_{D^*_{s0}D^*_{s}}\simeq 219(13)\cdots~{\rm keV}~\lrar~f^{sd}_{D^*_0D^*}\simeq 0.98(5)\cdots~,
\eeq
where again the error of the ratio comes from the precise direct determination. 
   \subsection*{\b The $(1^{-+})~\bar D_sD_{s1}$ molecule state}
   The results of the analysis are  similar to the previous $D^*_sD_{s1}$ pseudoscalar case and the figures will not be shown.  
  The coupling presents $\tau$-stabilities from $\sqrt{t_c}=6.0+2\overline m_q$ to $\sqrt{t_c}=7.3+2\overline m_q$ GeV and for $\tau$=0.14 (resp. 0.24) GeV$^{-2}$.
   Within these range of $t_c$-values, the mass stabilizes for $\tau\simeq$  0.28 GeV$^{-2}$ and the ratio of couplings for $\tau\simeq$  0.28 GeV$^{-2}$. 
We obtain:
\bea
f_{D_{s}D_{s1}}&\simeq& 193.4(55)_{t_c}(3)_{\tau}\cdots~{\rm keV}~,\nnb\\
f^{sd}_{DD_1}&\simeq&1.054(123)_{t_c}(0)_\tau\cdots,~M_{D_{s}D_{s1}}\simeq 5599(139)_{t_c}(5)_{\tau}\cdots~{\rm MeV}.
\eea   
 Using the values: $M_{DD_1}=5840(150)$ MeV and $ f_{DD_1}=$ 213(19) keV from\,\cite{ChXYZ}, we can deduce:
\beq
r^{sd}_{DD_1}\simeq 0.959(34)_M(2)_{t_c}(1)_\tau\cdots,~
f_{D_{s}D_{s1}}\simeq 224.5(20.0)_f(26.2)_{t_c}(0.)_{\tau}\cdots~{\rm keV}~.
\eeq   
Taking the mean of the couplings, we deduce our final result:
\beq
f_{D_{s}D_{s1}}\simeq 194(5)\cdots~{\rm keV}~,\lrar f^{sd}_{DD_1}\simeq 0.911(106)\cdots~,
\eeq
where we have taken the error from the direct determination of $f^{sd}_{DD_1}$.
{\scriptsize
\begin{table}[hbt]
\setlength{\tabcolsep}{0.1pc}
\tbl{Different sources of errors for the estimate of the $0^{-}$ and $1^{- }$ $\bar D_sD_s$-like molecule masses (in units of MeV) and couplings $f_{M_sM_s}$ (in units of keV). We use $\mu=4.5(5)$ GeV.}
   {\scriptsize
 {\begin{tabular}{@{}llllllllll@{}} 
&\\
\hline
\hline
\bf Inputs&$\Delta M_{D^*_{s0}D_s}$&$\Delta f_{D^*_{s0}D_s}$&$\Delta M_{D^*_{s}D_{s1}}$&$\Delta f_{D^*_{s}D_{s1}}$&$\Delta M_{D^*_{s0}D^*_s}$&$\Delta f_{D^*_{s0}D^*_s}$&$\Delta M_{D_{s}D_{s1}}$&$\Delta f_{D_{s}D_{s1}}$\\
\hline 
{\it LSR parameters}&\\
$(t_c,\tau)$&202&18&177 &22&170&9&100&7 \\
$\mu$&10&4.07&8&10.33&11&3.36&12&4.39\\
{\it QCD inputs}&\\
$\bar M_Q$&26.51&5.73 &26.46&12.23&27.09&5.36&30.28&5.11 \\
$\alpha_s$&5.51&2.20&5.20&4.68&7.1&2.14&6.51&2.16\\
$N3LO$&15.33&0.49 &6.44 & 3.71& 9.59 & 0.56&13.09& 0.77  \\
$\la\bar qq\ra$ &6.65&0.48&5.72&0.98&11.92&0.21&0.0&0.0\\
$\la\alpha_s G^2\ra$&15.0&1.75&6.25&0.98&3.94&0.19&0.0 &0.0 \\
$M_0^2$&1.02&0.0&3.75&0.49&6.33&0.76&10.78&1.72\\
$\la\bar qq\ra^2$&83.12&12.61&74.5&25.47&61.18&8.94&59.11&8.11\\
$\la g^3G^3\ra$&2.36&0.11&2.45&0.24&2.24&0.09&2.47&0.12 \\
$d\geq8$&17.05&5.54&15.84&8.53&2.8&0.36&9.4&1.96\\
{\it Total errors}&223.12&23.57&195.06&33.88&183.97&10.93&122.38&11.86\\
\hline\hline
\end{tabular}}
\label{tab:errorc}
}
\end{table}
} 
{\scriptsize
\begin{table}[hbt]
\setlength{\tabcolsep}{1.7pc}
\tbl{Different sources of errors for the direct estimate of the $0^{-}$ and $1^{- }$ $\bar D_sD_s$-like molecule SU(3) ratio of couplings $f^{sd}_{MM}$ (the SU3 ratio of masses are not determined directly). We use $\mu=4.5(5)$ GeV.}
   {\scriptsize
 {\begin{tabular}{@{}llllllllll@{}} 
&\\
\hline
\hline
\bf Inputs &$\Delta f^{sd}_{D^*_{0}D}$&$\Delta f^{sd}_{D^*D_1}$&$\Delta f^{sd}_{D^*_{0}D^*}$&$\Delta f^{sd}_{DD_1}$\\
\hline 
{\it LSR parameters}&\\
$(t_c,\tau)$&0.04&0.01&0.03&0.02\\
$\mu$&0.001&0.002&0.001&0.004\\
{\it QCD inputs}&\\
$\bar M_Q$&0.003&0.003&0.002&0.003  \\
$\alpha_s$&0.0&0.0&0.0&0.0 \\
$N3LO$&0.003&0.006& 0.001  &0.003  \\
$\la\bar qq\ra$ &0.002&0.002&0.001&0.0\\
$\la\alpha_s G^2\ra$&0.001&0.0&0.0&0.0\\
$M_0^2$&0.0&0.001&0.003&0.003  \\
$\la\bar qq\ra^2$&0.033&0.032&0.025&0.024  \\
$\la g^3G^3\ra$&0.0&0.0&0.0&0.0   \\
$d\geq8$&0.007&0.004&0.0&0.006 \\
{\it Total errors}&0.053&0.036&0.039&0.032\\
\hline\hline
\end{tabular}}
\label{tab:error-rap-cmoins}
}
\end{table}
} 

\section{The  Heavy-light Charm  Four-Quark States}
\subsection{Interpolating currents}
The four-quark states are described by the interpolating currents given in Table\,\ref{tab:current4}. 
The corresponding spectral functions are given to LO of PT QCD in \ref{app.b}
The different sources of the errors are given in Table\,\ref{tab:4q-errorc} and Table\,\ref{tab:error-rap-4c}.
{\scriptsize
\begin{center}
\begin{table}[hbt]
\setlength{\tabcolsep}{0.8pc}

 \tbl{
Interpolating currents describing the four-quark 
states. $Q\equiv$ $c$ (resp. $b$). $k$ is an arbitrary current mixing where the optimal value is found to be $k=0$ from\,\cite{X3A,X3B}.   
}
    {\footnotesize
\begin{tabular}{lcl}

&\\
\hline
\\
States& $J^{P}$&Four-Quark  Currents  $\equiv{\cal O}_{4q}(x)$  \\
\\
\hline
\\
\bf Scalar &$\bf 0^{+}$&$ \epsilon_{abc}\epsilon_{dec}  \bigg[
		\big( s^T_a \: C\gamma_5 \: Q_b \big) \big( \bar{s}_d \: \gamma_5 C \: \bar{Q}^T_e \big) 
		+ k \big( s^T_a \: C \:Q_b \big) \big( \bar{s}_d \: C \: \bar{Q}^T_e \big) \bigg] $\\
\\
\bf Axial-vector &$\bf 1^{+}$& $\epsilon_{abc}\epsilon_{dec}  \bigg[
		\big( s^T_a \: C\gamma_5 \:Q_b \big) \big( \bar{s}_d \: \gamma_\mu C \: \bar{Q}^T_e \big) 
		+ k \big( s^T_a \: C \:Q_b \big) \big( \bar{s}_d \:\gamma_\mu \gamma_5 C \: \bar{Q}^T_e \big) \bigg] $ \\
\\
\bf Pseudoscalar &$\bf 0^{-}$& $\epsilon_{abc}\epsilon_{dec}  \bigg[
		\big( s^T_a \: C\gamma_5 \:Q_b \big) \big( \bar{s}_d \: C \: \bar{Q}^T_e \big) 
		+ k \big( s^T_a \: C \:Q_b \big) \big( \bar{s}_d \:\gamma_5 C \: \bar{Q}^T_e \big) \bigg] $ \\
     \\
\bf Vector&$\bf 1^{-}$& $ \epsilon_{abc}\epsilon_{dec}  \bigg[
		\big( s^T_a \: C\gamma_5 \:Q_b \big) \big( \bar{s}_d \: \gamma_\mu \gamma_5 C \: \bar{Q}^T_e \big) 
		+ k \big( s^T_a \: C \:Q_b \big) \big( \bar{s}_d \:\gamma_\mu C \: \bar{Q}^T_e \big) \bigg] $
\\
\hline
\end{tabular}
}
\label{tab:current4}
\end{table}
\end{center}
}
  \subsection{The  $S_{sc}(0^{+})$ Charm Scalar Four-Quark State}
\begin{figure}[hbt] 
\begin{center}
{\includegraphics[width=6.29cm  ]{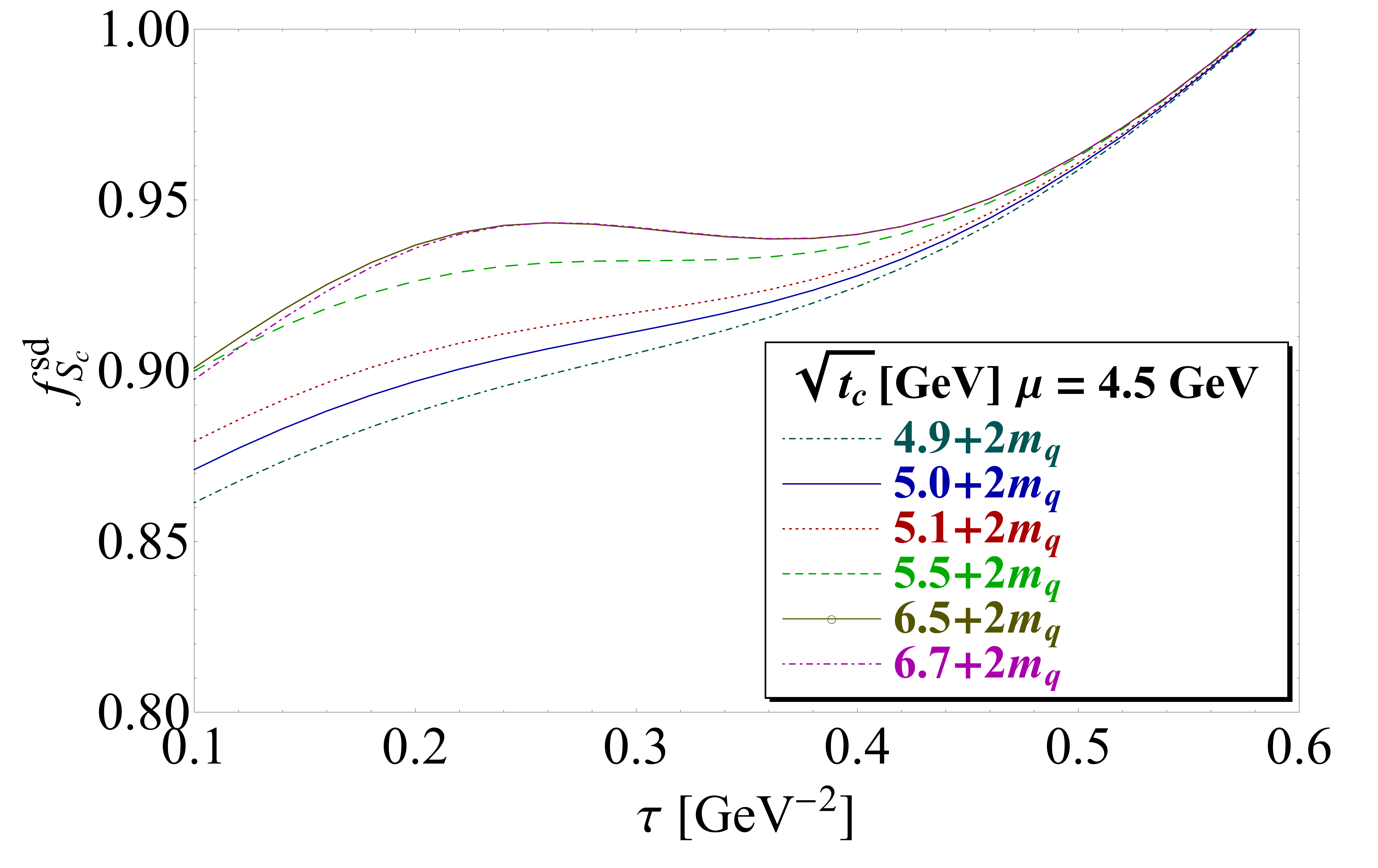}}
{\includegraphics[width=6.29cm  ]{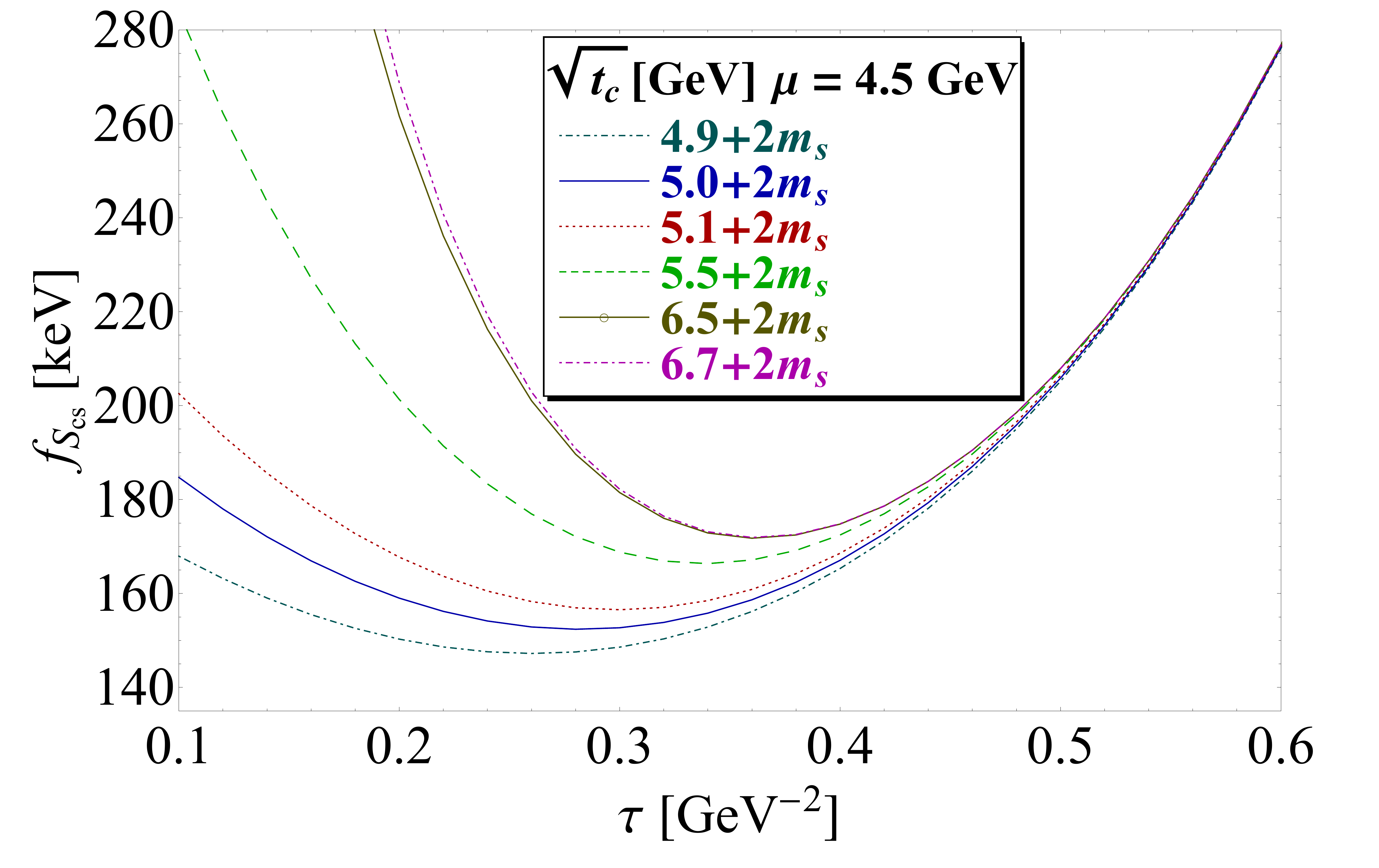}}
\centerline {\hspace*{-3cm} a)\hspace*{6cm} b) }
\caption{
\scriptsize 
{\bf a)} SU3 ratio of couplings $f^{sd}_{S_c}$ at NLO  as function of $\tau$ for different values of $t_c$, for $\mu=4.5$ GeV  and for the QCD parameters in Tables\,\ref{tab:param} and \ref{tab:alfa}; {\bf b)} The same as a) but for the couplings $f_{S_{sc}}$.
}
\label{fig:scf} 
\end{center}
\end{figure}

\begin{figure}[hbt] 
\begin{center}
{\includegraphics[width=6.29cm  ]{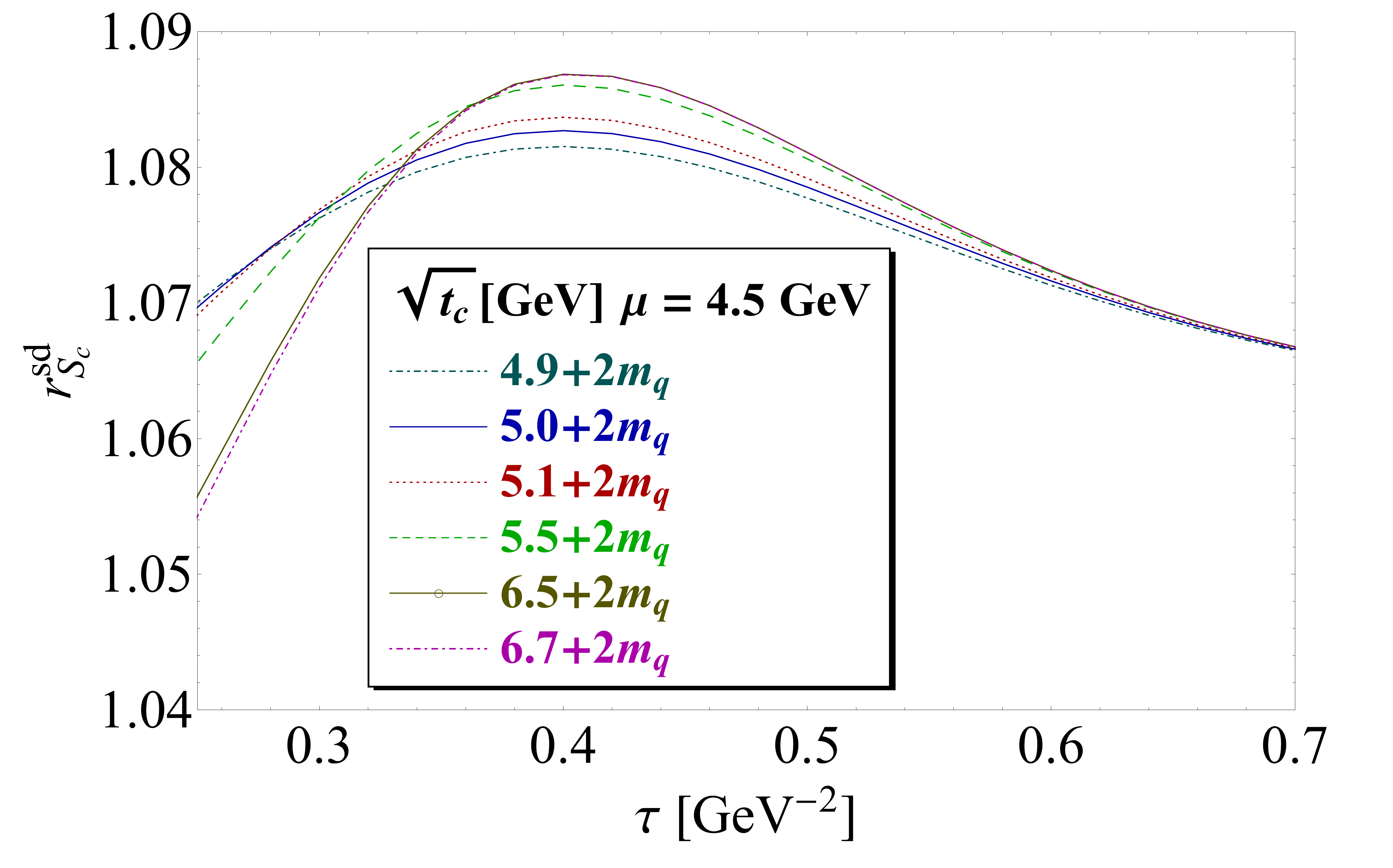}}
\caption{
\scriptsize 
 SU3 ratio of masses $r^{sd}_{S_c}$ at NLO  as function of $\tau$ for different values of $t_c$, for $\mu=4.5$ GeV  and for the QCD parameters in Tables\,\ref{tab:param} and \ref{tab:alfa}.
}
\label{fig:scrsd} 
\end{center}
\end{figure}

The behaviours of the corresponding curves are very similar to some of the previous molecule ones. They are shown in Figs.\,\ref{fig:scf} and \,\ref{fig:scrsd}.
\nin
\nin
The SU3 ratio of coupling presents $\tau$-stabilities from $\sqrt{t_c}=5+2\overline m_q$ to $\sqrt{t_c}=6.5+2\overline m_q$  GeV and for $\tau$=0.28 (resp. 0.36) GeV$^{-2}$.
 Within these range of $t_c$-values, the SU3 ratio of masses stabilizes for $\tau\simeq$  0.4(resp. 0.4) GeV$^{-2}$ and the coupling for $\tau\simeq$  0.28 (resp. 0.36) GeV$^{-2}$. 
We obtain:
\beq
f^{sd}_{S_{c}}\simeq 0.924(15)_{t_c}(3)_\tau\cdots,~r^{sd}_{S_{c}}\simeq 1.085(2)_{t_c}(0)_{\tau}\cdots,
~f_{S_{sc}} \simeq 162(10)_{t_c}(1)_{\tau}\cdots~{\rm keV}~,
\eeq   
 Using the NLO values: $M_{S_{c}}=3901(0.2)$ MeV, $ f_{S_c}=$184(9) keV from\,\cite{ChXYZ} and $f^{sd}_{S_{c}}$, we can deduce:
\beq
M_{S_{sc}}\simeq 4233(0.2)_M(7)_{t_c}(0)_\tau\cdots~{\rm MeV},
f_{S_{sc}}\simeq 170(8)_f(3)_{t_c}(1)_{\tau}\cdots~{\rm keV}~.
\eeq   
Taking the mean of $f_{S_{sc}}$ and re-using $ f_{S_c}\simeq$ 184(9) keV, we deduce the final estimate:
\beq
f_{S_{sc}}\simeq 166(7)\cdots~{\rm keV}~\lrar~  f^{sd}_{S_c}\simeq 0.902(17)\cdots,
\eeq
where the error of $f^{sd}_{S_c}$ comes from the direct determination. 
  \subsection{The $A_{sc} (1^{+})$ Charm Axial-Vector Four-Quark State}
The behaviours of the corresponding curves are very similar to the previous ones.
The SU3 ratio of coupling presents $\tau$-stabilities from $\sqrt{t_c}=5.1+2\overline m_q$ to $\sqrt{t_c}=6.5+2\overline m_q$ GeV and for $\tau$=0.26 (resp. 0.36) GeV$^{-2}$.
 Within these range of $t_c$-values, the SU3 ratio of masses stabilizes for $\tau\simeq$  0.34(resp. 0.34) GeV$^{-2}$ and the coupling for $\tau\simeq$  0.28 (resp. 0.34) GeV$^{-2}$. 
We obtain:
\beq
f^{sd}_{A_{c}}\simeq 0.834(17)_{t_c}(2)_\tau\cdots,~r^{sd}_{A_{c}}\simeq 1.081(4)_{t_c}(1)_{\tau}\cdots,
~f_{A_{sc}} \simeq 137(8)_{t_c}(5)_{\tau}\cdots~{\rm keV}~,
\eeq   
 Using the NLO values: $M_{A_{c}}=3890(27)$ MeV and $ f_{A_c}=$176(9) keV from\,\cite{ChXYZ}, we can deduce:
\beq
M_{A_{sc}}\simeq 4205(29)_M(16)_{t_c}(4)_\tau\cdots~{\rm MeV},
f_{A_{sc}}\simeq 147(8)_f(30)_{t_c}(4)_{\tau}\cdots~{\rm keV}~.
\eeq   
Taking the mean of $f_{A_{sc}}$ and re-using $ f_{A_c}\simeq$ 176(9) keV, we deduce the final estimate:
\beq
f_{A_{sc}}\simeq 141(6)\cdots~{\rm keV}~\lrar~  f^{sd}_{A_c}\simeq 0.80(3)\cdots,
\eeq
where the error of $f^{sd}_{A_c}$ comes from the direct determination. 

  \subsection{The  $\pi_{sc}(0^{-})$ Charm Pseudoscalar State}
The coupling presents $\tau$-stabilities from $\sqrt{t_c}=6.0+2\overline m_q$ to $\sqrt{t_c}=7.3+2\overline m_q$ GeV and for $\tau$=0.15(resp. 0.22) GeV$^{-2}$.
 Within these range of $t_c$-values, the mass stabilizes for $\tau\simeq$  0.20 (resp. 0.24) GeV$^{-2}$ and the ratio of couplings for $\tau\simeq$  0.23 (resp. 0.24) GeV$^{-2}$. 
We obtain:
\bea
f_{\pi_{sc}}&\simeq& 249(21)_{t_c}(6)_{\tau}\cdots~{\rm keV}~,\nnb\\
f^{sd}_{\pi_{c}}&\simeq&0.90(3)_{t_c}(4)_\tau\cdots,~M_{\pi_{sc}}\simeq 5671(159)_{t_c}(4)_{\tau}\cdots~{\rm MeV}.
\eea   
 Using the values: $M_{\pi_{c}}=5872(101)$ MeV and $ f_{\pi_c}=$ 292(5.7) keV from\,\cite{ChXYZ}, we can deduce:
\beq
r^{sd}_{\pi_c}\simeq 0.97(3)\cdots,~
f_{\pi_{sc}}\simeq 263(5)_f(9)_{t_c}(3)_{\tau}\cdots~{\rm keV}~.
\eeq   
Taking the mean of $f_{\pi_{sc}}$ and re-using $ f_{\pi_c}=$ 292(5.7) keV, we deduce the final estimate:
\beq
f_{\pi_{sc}}\simeq 256(9)\cdots~{\rm keV}~\lrar~ f^{sd}_{\pi_c}\simeq 0.88(3)\cdots
\eeq
  \subsection{The  $V_{sc}(1^{-})$ Charm Vector State}
The behaviours of the corresponding curves are very similar to the previous ones.
The coupling presents $\tau$-stabilities from $\sqrt{t_c}=6.0+2\overline m_q$ to $\sqrt{t_c}=7.3+2\overline m_q$ GeV and for $\tau$=0.11-0.15 (resp. 0.24) GeV$^{-2}$.
 Within these range of $t_c$-values, the mass stabilizes for $\tau\simeq$  0.19 (resp. 0.24) GeV$^{-2}$ and the ratio of couplings for $\tau\simeq$  0.23 (resp. 0.24) GeV$^{-2}$. 
We obtain:
\bea
f_{V_{sc}}&\simeq& 235(21)_{t_c}(5)_{\tau}\cdots~{\rm keV}~,\nnb\\
f^{sd}_{V_{c}}&\simeq&0.94(4)_{t_c}(1)_\tau\cdots,~M_{V_{sc}}\simeq 5654(222)_{t_c}(8)_{\tau}\cdots~{\rm MeV}.
\eea   
 Using the NLO values: $M_{V_{c}}=5904(90)$ MeV and $ f_{V_c}=$ 268(14) keV from\,\cite{ChXYZ}, we can deduce:
\beq
r^{sd}_{V_c}\simeq 0.96(4)\cdots,~
f_{V_{sc}}\simeq 252(13)_f(11)_{t_c}(3)_{\tau}\cdots~{\rm keV}~.
\eeq   
Taking the mean of $f_{V_{sc}}$ and re-using $ f_{V_c}=$ 268(14) keV, we deduce the final estimate:
\beq
f_{V_{sc}}\simeq 245(14)\cdots~{\rm keV}~\lrar~  f^{sd}_{V_c}\simeq 0.91(4)\cdots,
\eeq
where the error of $f^{sd}_{V_c}$ comes from the direct determination. 
{\scriptsize
\begin{table}[hbt]
\setlength{\tabcolsep}{0.5pc}
\tbl{Different sources of errors for the estimate of the four-quarks $[cs\bar c \bar s]$  pseudo (scalar)  $\pi_{sc}$ ($S_{sc}$) and  axial (vector) $A_{sc}$ ($V_{sc}$) masses (in units of MeV) and couplings (in units of keV). We use $\mu=4.5(5)$ GeV.}
   {\scriptsize
 {\begin{tabular}{@{}llllllllll@{}} 
&\\
\hline
\hline
\bf Inputs &$\Delta M_{S_{sc}}$&$\Delta f_{S_{sc}}$&$\Delta M_{A_{sc}}$&$\Delta f_{A_{sc}}$&$\Delta M_{\pi_{sc}}$&$\Delta f_{\pi_{sc}}$&$\Delta M_{V_{sc}}$&$\Delta f_{V_{sc}}$\\
\hline 
{\it LSR parameters}&\\
$(t_c,\tau)$&7&8.6&33.4&5.5&159.1&9&222.1&14\\
$\mu$&24.77&8.03&29.12&7.90&8.40&4.69&7.10&5.01 \\
{\it QCD inputs}&\\
$\bar M_Q$&11.87&5.20&10.88&4.09&27.18&6.49&27.01&6.17\\
$\alpha_s$&15.69&4.20&16.61&3.81&5.39&2.46&3.99&2.26\\
$N3LO$&  0.00  &  1.82 &  0.28 &  1.33 &10.29&0.77&8.05&0.91 \\
$\la\bar qq\ra$   &15.75&2.33&20.45&2.47& 6.17&0.58&5.47&0.46\\
$\la\alpha_s G^2\ra$&0.56&0.56&0.54&0.16&9.17&0.62&1.70&0.21\\
$M_0^2$&14.94&1.60&18.30&2.17& 0.3&0.0&5.17&1.37\\
$\la\bar qq\ra^2$&12.09&9.32&33.02&8.99&80.31&26.20&81.12&25.82\\
$\la g^3G^3\ra$&0.54&0.15&0.24&0.14&2.36&0.13& 2.19&0.12\\
$d\geq8$&45.09&0.91&91.51&1.30& 9.7&2.70&22.5&7.26\\
{\it Total errors}&60.83&16.79&112.14&14.79&181.43&28.83&239.44&30.50\\
\hline\hline
\end{tabular}}
\label{tab:4q-errorc}  
}
\end{table}
}
{\scriptsize
\begin{table}[hbt]
\setlength{\tabcolsep}{1.2pc}
\tbl{Different sources of errors for the direct estimate of the four-quarks $[cs\bar c \bar s]$  pseudo (scalar)  $\pi_{sc}$ ($S_{sc}$) and  axial (vector) $A_{sc}$ ($V_{sc}$) SU(3) ratio of masses $r^{sd}_{M}$ and of couplings $f^{sd}_{M}$. We use $\mu=4.5(5)$ GeV.}
   {\scriptsize
 {\begin{tabular}{@{}llllllllll@{}} 
&\\
\hline
\hline
\bf Inputs &$\Delta r^{sd}_{S_c}$&$\Delta f^{sd}_{S_c}$&$\Delta r^{sd}_{A_c}$&$\Delta f^{sd}_{A_c}$&$\Delta f^{sd}_{\pi_c}$&$\Delta f^{sd}_{V_c}$\\
\hline 
{\it LSR parameters}&\\
$(t_c,\tau)$&0.002&0.017&0.004&0.03&0.03&0.04\\
$\mu$&0.0&0.005&0.0&0.007&0.002&0.003  \\
{\it QCD inputs}&\\
$\bar M_Q$&0.0&0.004&0.0&0.004&0.002 & 0.002\\
$\alpha_s$&0.001&0.002&0.001&0.003&0.0 & 0.0\\
$N3LO$&0.0&  0.005  & 0.0  & 0.005  & 0.001&0.003  \\
$\la\bar qq\ra$   &0.002&0.002&0.003&0.003&0.002&0.001   \\
$\la\alpha_s G^2\ra$&0.0&0.001&0.0&0.0&0.0 & 0.0 \\
$M_0^2$&0.002&0.001&0.003&0.003& 0.0 & 0.002  \\
$\la\bar qq\ra^2$&0.003&0.029&0.004&0.024& 0.045  & 0.049\\
$\la g^3G^3\ra$&0.0&0.0&0.0&0.0& 0.0 & 0.0  \\
$d\geq8$ &0.010&0.005&0.016&0.009&0.049 & 0.072  \\
{\it Total errors}&0.011&0.035&0.018&0.041&0.073&0.096\\
\hline\hline
\end{tabular}}
\label{tab:error-rap-4c}
}
\end{table}
} 

\section{The  Heavy-light Beauty  Molecule States}
We extend the previous analysis to the case of beauty molecule states. The strategy for obtaining the results 
is very similar to the one of the charm. The different sources of errors are given in Tables \,\ref{tab:errorbplus} to \,\ref{tab:error-rap-bmoins} .
\subsection{The  $(0^{++})$ Beauty  Scalar Molecule States}
   \subsection*{\b The $\bar B_sB_s$ molecule state}
\begin{figure}[hbt] 
\begin{center}
{\includegraphics[width=6.29cm  ]{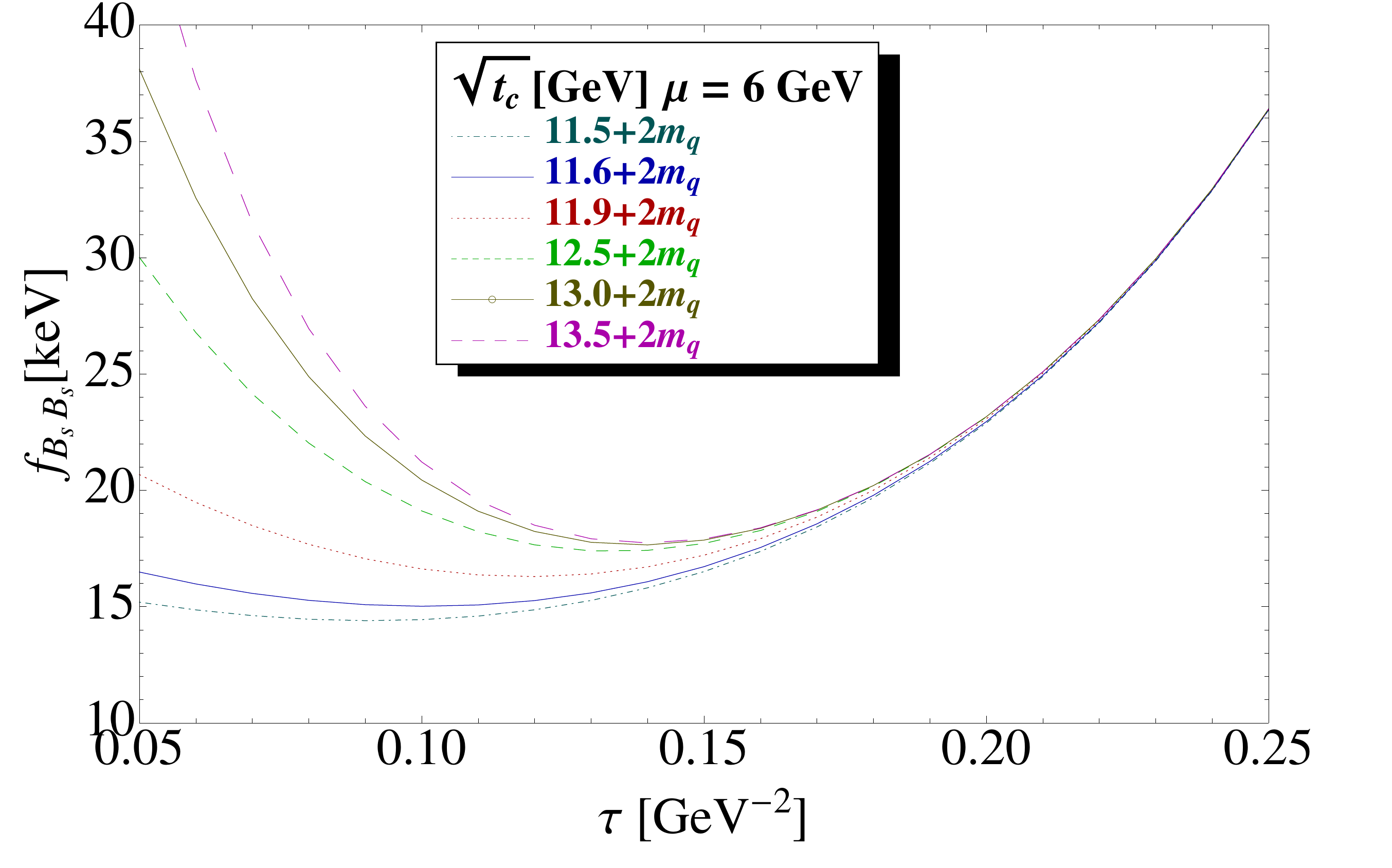}}
{\includegraphics[width=6.29cm  ]{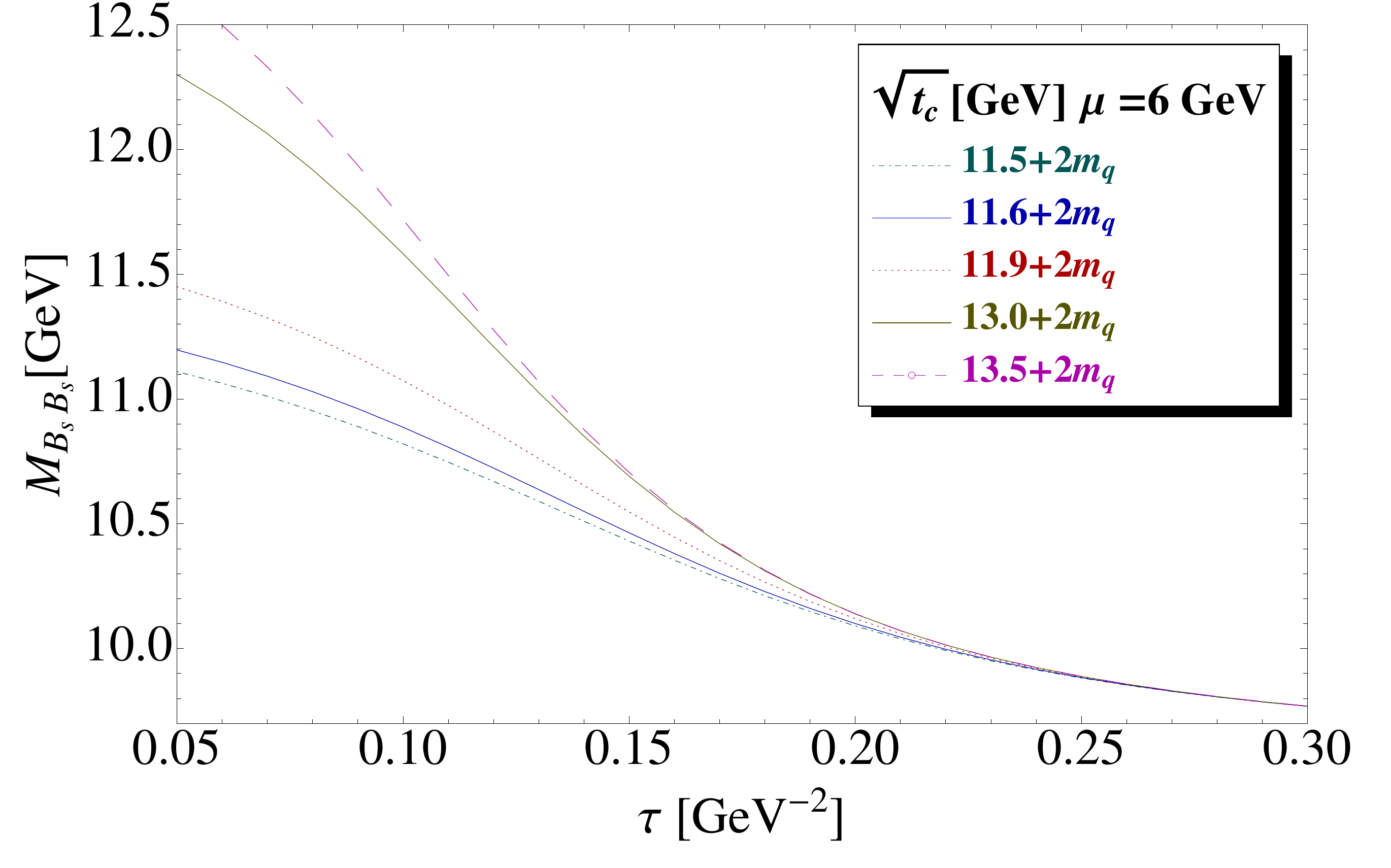}}
\centerline {\hspace*{-3cm} a)\hspace*{6cm} b) }
\caption{
\scriptsize 
{\bf a)} The coupling $f_{B_sB_s}$ at NLO  as function of $\tau$ for different values of $t_c$, for $\mu=6$ GeV  and for the QCD parameters in Tables\,\ref{tab:param} and \ref{tab:alfa}; {\bf b)} The same as a) but for the mass $M_{B_sB_s}$.
}
\label{fig:bsbs} 
\end{center}
\end{figure}

\begin{figure}[hbt] 
\begin{center}
{\includegraphics[width=6.29cm  ]{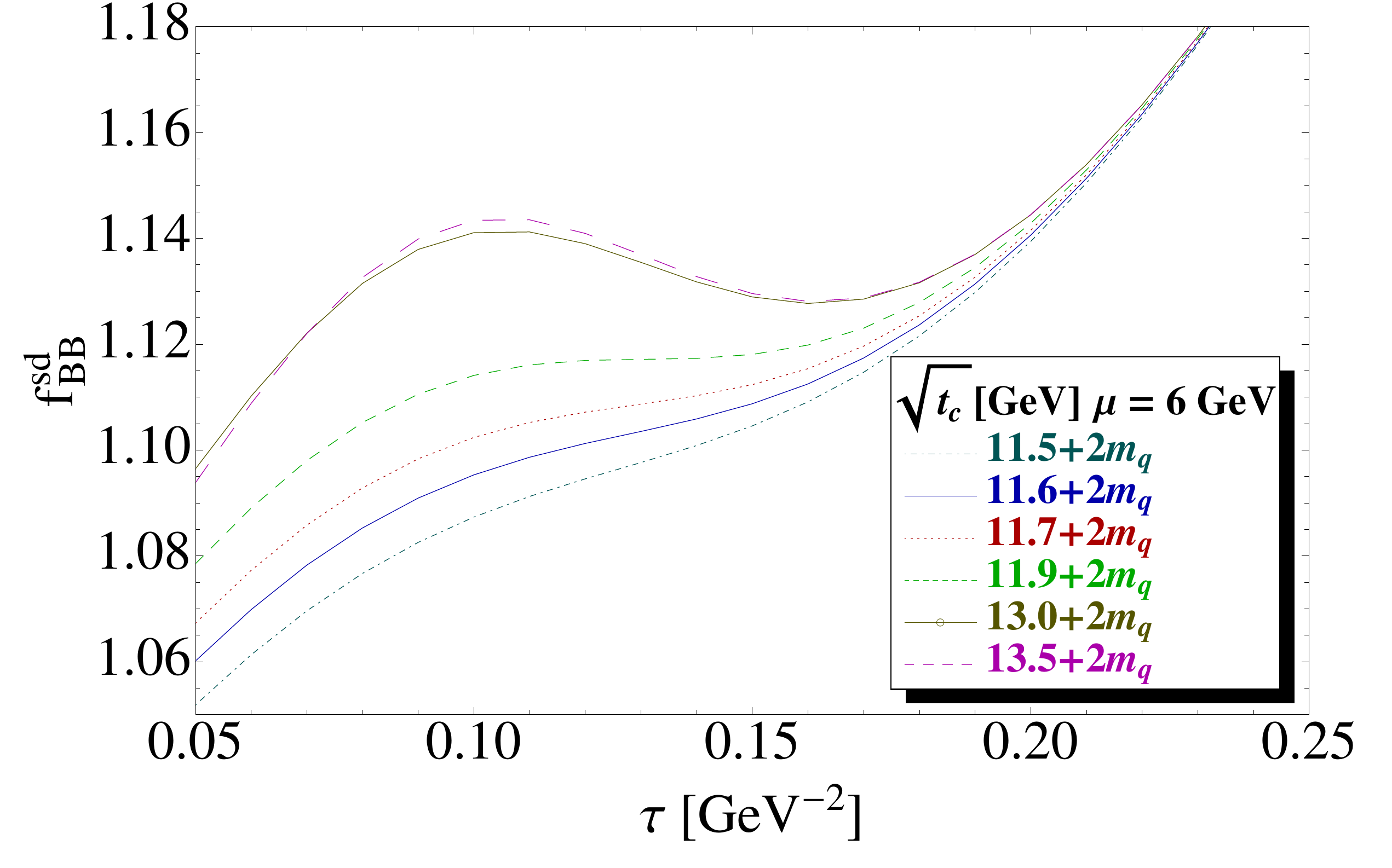}}
{\includegraphics[width=6.29cm  ]{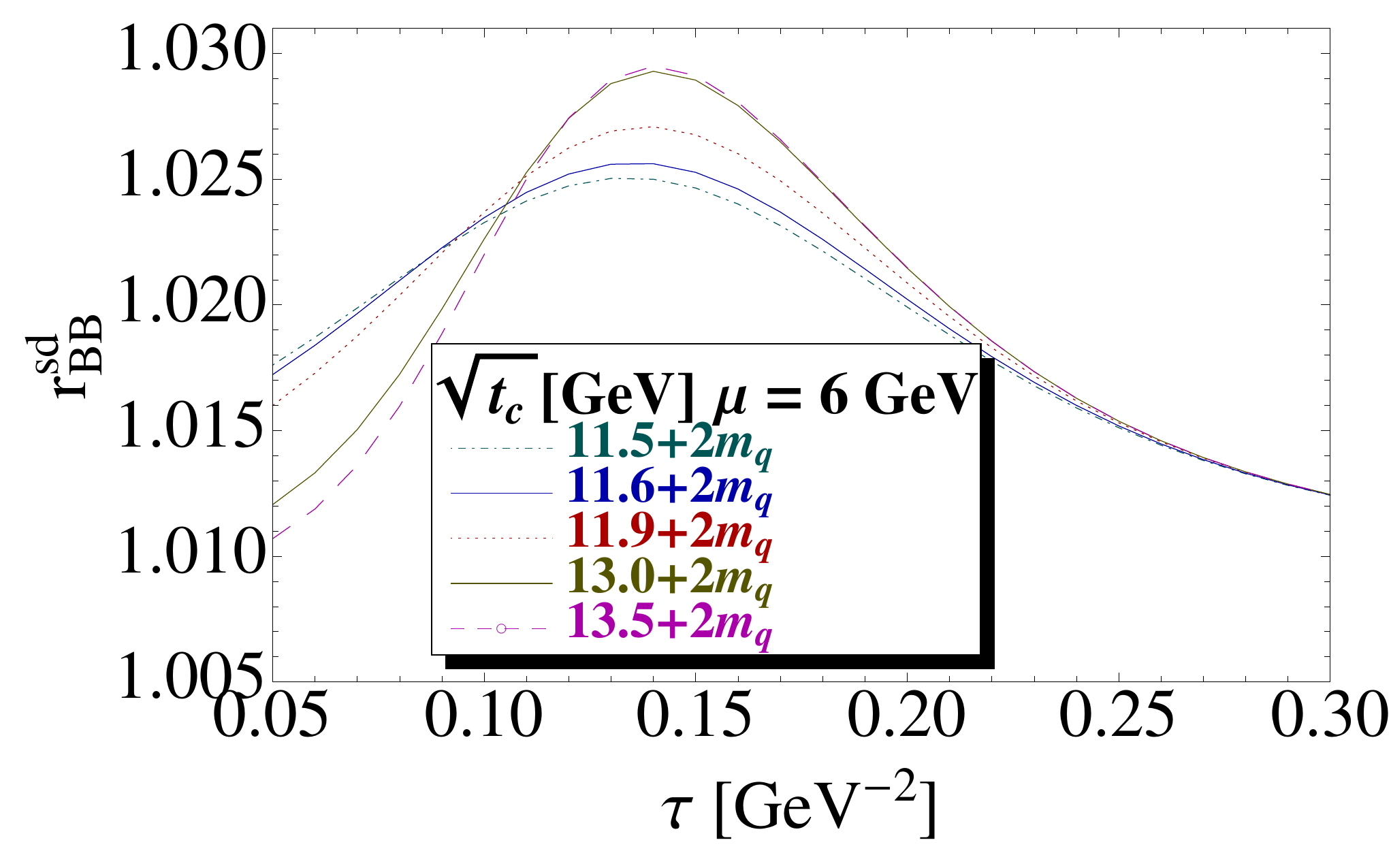}}
\centerline {\hspace*{-3cm} a)\hspace*{6cm} b) }
\caption{
\scriptsize 
{\bf a)} SU3 ratio of couplings $f^{sd}_{BB}$ at NLO  as function of $\tau$ for different values of $t_c$, for $\mu=6$ GeV  and for the QCD parameters in Tables\,\ref{tab:param} and \ref{tab:alfa}; {\bf b)} The same as a) but for the SU3 ratio of masses $r^{sd}_{BB}$.
}
\label{fig:bbs} 
\end{center}
\end{figure}

We shall illustrate the analysis by showing  the different figures (Figs.\,\ref{fig:bsbs} and \,\ref{fig:bbs}) in this channel. The subtraction point is taken at $\mu=6$ GeV, where $\mu$-stability has been obtained in\,\cite{ChXYZ}  for the non-strange quark case. 
\nin
\nin
From these figures, we obtain extrema or inflexion points from $\sqrt{t_c}\simeq 11.6$ +2$m_q$ to 13 +2$m_q$ GeV. The $\tau$-stabilties occur at 0.10 (resp. 0.14), about 0.13--0.15 (resp. 0.16) and  0.13 (resp. 0.14) GeV$^{-2}$ for the coupling, mass, SU3 ratio of couplings and of masses. We deduce the optimal results:
\beq
f_{B_sB_s}(6)\simeq 16.3(1.3)_{t_c}(0.2)_\tau~{\rm keV},~ f^{sd}_{BB}\simeq 1.116(13)_{t_c}(0)_\tau, ~r^{sd}_{BB}\simeq 1.027(2)_{t_c}(1)_\tau,
\eeq
where the QCD corrections are given in Table\,\ref{tab:errorbplus}.
We have not considered the value of the mass from the figure but combine the accurate ratio with the value $M_{BB}=10598(54)$ MeV (without QCD corrections) obtained in\,\cite{ChXYZ} from which we obtain:
\beq
M_{B_sB_s}=10884(56)_M(21)_{t_c}(10)_\tau\cdots~{\rm MeV},
\eeq
where $\cdots$ denotes QCD corrections given in Table\,\ref{tab:errorbplus}.
Combining the SU3 ratio of couplings with the NLO  value $f_{BB}(6)\simeq 16.2(1.6)~{\rm keV}$ from\,\cite{ChXYZ} runned at $\mu=6$ GeV, one deduces:
\beq
f_{B_sB_s}(6)\simeq 18.1(1.8)_f(0.2)_{t_c}(0)_\tau\cdots~{\rm keV}~.
\eeq
Taking the mean, we deduce:
\beq
f_{B_sB_s}(6)\simeq 16.9(1.1)\cdots~{\rm keV}~\lrar f^{sd}_{BB}\simeq 1.043(12)\cdots,
\eeq

   \subsection*{\b The $\bar B^*_sB^*_s$ molecule state}
In the same way as before, the extrema or inflexion points occur from $\sqrt{t_c}\simeq 11.6$+2 $m_q$  to 13 +2 $m_q$ GeV. The $\tau$-stabilties are at 0.08 (resp. 0.13), about 0.12-- 0.16 (resp. 0.14) and  0.11 (resp. 0.13) GeV$^{-2}$ for the coupling, mass, SU3 ratio of couplings and of masses. We deduce the optimal results:
\beq
f_{B^*_sB^*_s}(6)\simeq 30.2(34)_{t_c}(1)_\tau~{\rm keV},~ f^{sd}_{B^*B^*}\simeq 1.105(20)_{t_c}(6)_\tau, ~r^{sd}_{B^*B^*}\simeq 1.028(2)_{t_c}(0)_\tau,
\eeq
where the QCD corrections are given in Table\,\ref{tab:errorbplus}. Using $M_{B^*B^*}=10646(102)$ MeV from\,\cite{ChXYZ}, we deduce:
\beq
M_{B^*_sB^*_s}=10944(105)_M(21)_{t_c}(0)_\tau~{\rm MeV}.
\eeq
Combining the SU3 ratio of couplings with the NLO  value $f_{B^*B^*}(6)\simeq 31(5)~{\rm keV}$ from\,\cite{ChXYZ} runned at $\mu=6$ GeV, one deduces:
\beq
f_{B^*_sB^*_s}(6)\simeq 32.6(5.3)_f(0.7)_{t_c}(0)_\tau~{\rm keV}~.
\eeq
Taking the mean value of the coupling, we obtain:
\beq
f_{B^*_sB^*_s}(6)\simeq 30.9(2.9)\cdots~{\rm keV}~\lrar f^{sd}_{B^*B^*}\simeq 1.00(2)\cdots.
\eeq
   \subsection*{\b The $\bar B^*_{s0}B^*_{s0}$ molecule state}
 The different sum rules stabilize in the same range of $t_c$ as in the previous $0^{++}$ cases.  The $\tau$-stabilities are 
 at 0.08 (resp. 0.13), about 0.12--0.16 (resp. 0.14) and  0.11 (resp. 0.13) GeV$^{-2}$ for the coupling, mass, SU3 ratio of couplings and of masses. We deduce the optimal results:
\bea
f_{B^*_{s0}B^*_{s0}}(6)&\simeq& 11.7(3.7)_{t_c}(0.6)_\tau~{\rm keV},~ f^{sd}_{B^*_0B^*_0}\simeq 1.258(73)_{t_c}(10)_\tau, \nnb\\
r^{sd}_{B^*_0B^*_0}&\simeq& 1.050(7)_{t_c}(1)_\tau.
\eea
  Using $M_{B^*_0B^*_0}=10649(113)$ MeV from\,\cite{ChXYZ}, we deduce:
\beq
M_{B^*_{s0}B^*_{s0}}=11182(119)_M(74)_{t_c}(11)_\tau~{\rm MeV}.
\eeq
Combining the SU3 ratio of couplings with the NLO  value $f_{B^*_0B^*_0}(6)\simeq 11.7(3.3)~{\rm keV}$ from\,\cite{ChXYZ} runned at $\mu=6$ GeV, one deduces:
\beq
f_{B^*_{s0}B^*_{s0}}(6)\simeq 14.7(4.2)_f(0.9)_{t_c}(0.1)_\tau~{\rm keV}~.
\eeq
Taking the mean value of the coupling, we obtain:
\beq
f_{B^*_{s0}B^*_{s0}}(6)\simeq 13.0(2.9)\cdots~{\rm keV}~\lrar f^{sd}_{B^*_0B^*_0}\simeq 1.11(1)\cdots.
\eeq
   \subsection*{\b The $\bar B_{s1}B_{s1}$ and $\bar B_1B_1$ molecule states}
We perform a similar analysis. The behaviours of the different curves are very similar to the case of the $\bar B_sB_s$ molecule states. They present stabilites for $\sqrt{t_c}\simeq$ 11.6+$2\overline m_s$ to  13.0+$2\overline m_s$ GeV for $\tau\simeq 0.07-0.13$ (resp. 0.09--0.12) (resp. 0.09--0.12) GeV$^{-2}$ for the coupling $f_{B_{s1}B_{s1}}$ (resp. SU3 ratio of couplings $f^{sd}_{B_1B_1}$) (resp. SU3 ratio of masses  $r^{sd}_{B_1B_1}$) leading to the values at NLO:
\bea
f_{B_{s1}B_{s1}}\simeq 24(4)\cdots~{\rm keV},~f^{sd}_{B_1B_1}\simeq 1.197(41)\cdots~,r^{sd}_{B_1B_1}\simeq 1.040(1)\cdots~,
\eea
where the quoted errors come from the correlated values of $(t_c,\tau)$ and $\cdots$ are QCD corrections given in Table\,\ref{tab:errorcplus}. The mass presents an inflexion which is difficult to localize.  To fix the $\tau$-values, we take the  range where the SU3 ratio of masses optimizes, which corresponds to $\tau\simeq 0.09-0.12$ GeV$^{-2}$. In this way, we obtain:
\beq
M_{B_{s1}B_{s1}}\simeq 10935(155)\cdots~{\rm MeV}~.
\eeq
Using the previous values of the SU3 ratios, we can deduce for the $\bar B_1B_1$ molecule at NLO:
\beq
f_{B_{1}B_{1}}\simeq 20(3)\cdots~{\rm keV}~,~~~~M_{B_{1}B_{1}}\simeq 10514(149)\cdots~{\rm MeV}~.
\eeq
\subsection{The  $(1^{+\pm})$ Beauty Axial-Vector Molecule States}				     %
Here, within our choice of interpolating currents, the $(1^{++})$ and $(1^{+-})$ are degenerate in masses like in the cases of charmonium and pseusoscalar channels and have the same couplings. 
   \subsection*{\b The $\bar B^*_sB_{s}$  molecule state}
The different sum rules stabilize in the same range of $t_c$ as in the previous cases.  The $\tau$-stabilities are 
 at 0.09 (resp. 0.135), 0.12 (resp. 0.15) and  0.13 (resp. 0.145) GeV$^{-2}$for the coupling, SU3 ratio of couplings and of masses. We deduce the optimal results:
\beq
f_{B^*_{s}B_{s}}(6)\simeq 16.6(1.6)_{t_c}(0.1)_\tau~{\rm keV},~ f^{sd}_{B^*B}\simeq 1.114(17)_{t_c}(1)_\tau, ~r^{sd}_{B^*B}\simeq 1.028(3)_{t_c}(1)_\tau.
\eeq
  Using $M_{B^*B}=10673(150)$ MeV from\,\cite{ChXYZ}, we deduce:
\beq
M_{B^*_{s}B_{s}}=10972(154)_M(32)_{t_c}(11)_\tau~{\rm MeV}.
\eeq
Combining the SU3 ratio of couplings with the NLO  value $f_{B^*B}(6)\simeq 16.5(5)~{\rm keV}$ from\,\cite{ChXYZ} runned at $\mu=6$ GeV, one deduces:
\beq
f_{B^*_{s}B_{s}}(6)\simeq 18.4(5.6)_f(0.3)_{t_c}(0)_\tau~{\rm keV}~.
\eeq
Taking the mean value of the coupling, we obtain:
\beq
f_{B^*_{s}B_{s}}(6)\simeq 16.7(1.5)\cdots~{\rm keV}~\lrar f^{sd}_{B^*B}\simeq 1.01(1)\cdots.
\eeq
   \subsection*{\b The $\bar B^*_{s0}B_{s1}$ molecule state}
 In this channel,  only the coupling and the SU3 ratio of masses present net stabilities. The others present inflexion points which cannot be accurately localized. The $\tau$-stabilities are 
 at 0.04 (resp. 0.125), and 0.06 (resp. 0.115) GeV$^{-2}$ for  $\sqrt{t_c}\simeq 11.6$ +2 $m_q$ (resp.  13 +2 $m_q$) GeV. We deduce the optimal results:
\beq
f_{B^*_{s0}B_{s1}}(6)\simeq 9.1(10)_{t_c}(14)_\tau~{\rm keV},~ r^{sd}_{B^*_0B_1}\simeq 1.052(8)_{t_c}(3)_\tau.
\eeq
  Using $M_{B^*_0B_1}=10679(132)$ MeV from\,\cite{ChXYZ}, we deduce:
\beq
M_{B^*_{s0}B_{s1}}=11234(139)_M(85)_{t_c}(32)_\tau~{\rm MeV}.
\eeq
Using the NLO  value $f_{B^*_0B_1}(6)\simeq 11.3(1.6)~{\rm keV}$ from\,\cite{ChXYZ} runned at $\mu=6$ GeV, one deduces:
\beq
f^{sd}_{B^*_0B_1}\simeq 0.80(11)\cdots.
\eeq

{\scriptsize
\begin{table}[hbt]
\setlength{\tabcolsep}{0.3pc}
\tbl{Different sources of errors for the estimate of the $0^{+}$ and $1^{+}$ $\bar B_sB_s$-like molecule masses (in units of MeV) and couplings $f_{M_sM_s}(\mu)$ (in units of keV). We use $\mu=6.0(5)$ GeV.}
   {\scriptsize
 {\begin{tabular}{@{}llllllllllllllllll@{}} 
&\\
\hline
\hline
 &\multicolumn{2}{c}{$B_sB_s$}
					&\multicolumn{2}{c}{$B^{*}_{s}B^{*}_{s}$}
					&\multicolumn{2}{c}{$B^{*}_{s0}B^{*}_{s0}$}
					&\multicolumn{2}{c}{$B_{s1}B_{s1}$}
					&\multicolumn{2}{c}{$B^{*}_{s}B_{s}$}
					&\multicolumn{2}{c}{$B^{*}_{s0}B_{s1}$}
					\\
\cline{2-3} \cline{4-5}\cline{6-7}\cline{8-9}\cline{10-11}\cline{12-13}
                 & \multicolumn{1}{l}{$\Delta M$} 
                 & \multicolumn{1}{l }{$\Delta f$} 
                 & \multicolumn{1}{l}{$\Delta M$}
                 & \multicolumn{1}{l }{$\Delta f$} 
                 & \multicolumn{1}{l}{$\Delta M$}
                 & \multicolumn{1}{l}{$\Delta f$} 
		    & \multicolumn{1}{l}{$\Delta M$} 
                 & \multicolumn{1}{l}{$\Delta f$} 
                 & \multicolumn{1}{l}{$\Delta M$}
                 & \multicolumn{1}{l}{$\Delta f$} 
		    & \multicolumn{1}{l}{$\Delta M$} 
                 & \multicolumn{1}{l}{$\Delta f$} 
                  \\
\hline

{\bf Inputs}&\\
{\it LSR parameters}&\\
$(t_c,\tau)$&60.63&1.81&107.08&2.90&140.56&2.90&155&3.88&157.67&1.50&166.04&1.72&\\
$\mu$&5.14&0.0&7.30 &0.56&3.46&0.04&9.30&1.31&7.29&0.03&6.20&0.0\\
{\it QCD inputs}&\\
$\bar M_Q$&2.14&0.10&2.32&0.17&2.27&0.07&1.56&0.14&2.93&0.35&3.42&0.10\\
$\alpha_s$&10.79&0.35&11.53&0.63&19.04&0.23&10.55&0.49&13.02&0.10&17.64&0.24\\
$N3LO$&1.54&0.21&0.84&0.35&11.76&0.27&3.71&0.35&0.00&0.23&14.91&0.11\\
$\la\bar qq\ra$&14.95&0.17&6.40&0.20&25.01&0.34&20.20&0.28&8.66&0.18&21.37&0.36\\
$\la\alpha_s G^2\ra$&0.51&0.02&0.70&0.02&3.52&0.05&0.55&0.04&0.03&0.0&1.70&0.0\\
$M_0^2$&11.63&0.15&11.67&0.23&22.24&0.20&10.22&0.23&7.12&0.18&22.75&0.25\\
$\la\bar qq\ra^2$&27.03&0.95&19.24&2.12&23.34&1.49&51.30&2.49&23.52&0.98&21.38&1.55\\
$\la g^3G^3\ra$&0.02&0.0&0.08&0.0&0.06&0.0&0.06&0.0&0.02&0.0&1.71&0.0\\
$d\geq8$&22.0&0.83&76.0&2.78&171.5&0.57&42.9&1.09&110&0.22&116.5&0.10\\
{\it Total errors}&73.49&2.26&134.10&4.64&226.64&3.13&169.80&4.96&194.62&1.87&207.77&2.21\\
\hline\hline
\end{tabular}}
\label{tab:errorbplus}
}
\end{table}
} 
{\scriptsize
\begin{table}[hbt]
\setlength{\tabcolsep}{0.25pc}
\tbl{Different sources of errors for the estimate of the $0^{+}$ and $1^{+}$  $\bar BB$-like molecule SU(3) ratios of masses $r^{sd}_{MM}$ and SU(3) ratios of couplngs $f^{sd}_{MM}$. We use $\mu=6.0(5)$ GeV.}
   {\scriptsize
 {\begin{tabular}{@{}llllllllllllllllllll@{}} 
&\\
\hline
\hline
 &\multicolumn{2}{c}{$B_sB_s$}
					&\multicolumn{2}{c}{$B^{*}_{s}B^{*}_{s}$}
					&\multicolumn{2}{c}{$B^{*}_{s0}B^{*}_{s0}$}
					&\multicolumn{2}{c}{$B_{s1}B_{s1}$}
					&\multicolumn{2}{c}{$B^{*}_{s}B_{s}$}
					&\multicolumn{2}{c}{$B^{*}_{s0}B_{s1}$}
					\\
\cline{2-3} \cline{4-5}\cline{6-7}\cline{8-9}\cline{10-11}\cline{12-13}
                 & \multicolumn{1}{l}{$r^{sd}$} 
                 & \multicolumn{1}{l }{$f^{sd}$} 
                 & \multicolumn{1}{l}{$r^{sd}$}
                 & \multicolumn{1}{l }{$f^{sd}$} 
                 & \multicolumn{1}{l}{$r^{sd}$}
                 & \multicolumn{1}{l}{$f^{sd}$} 
		    & \multicolumn{1}{l}{$r^{sd}$} 
                 & \multicolumn{1}{l}{$f^{sd}$} 
                 & \multicolumn{1}{l}{$r^{sd}$}
                 & \multicolumn{1}{l}{$f^{sd}$} 
		    & \multicolumn{1}{l}{$r^{sd}$} 
                 & \multicolumn{1}{l}{$f^{sd}$} 
                  \\
\hline

\hline 
{\bf Inputs}\\
{\it LSR parameters}&\\
$(t_c,\tau)$&0.002&0.012&0.002&0.02&0.007&0.01&0.001&0.050&0.003&0.01&0.009&0.11\\
$\mu$&0.0&0.003&0.0&0.004&0.0&0.004&0.0&0.003&0.0&0.004&0.0&0.003\\
{\it QCD inputs}&\\
$\bar M_Q$&0.0&0.0&0.0&0.0&0.0&0.0&0.0&0.0&0.0&0.0&0.0&0.003\\
$\alpha_s$&0.0&0.0&0.0&0.0&0.0&0.0&0.0&0.001&0.0&0.0&0.0&0.003\\
$N3LO$&0.0&0.01&0.0&0.01&0.001&0.003&0.0&0.001&0.0&0.01&0.002&0.001\\
$\la\bar qq\ra$&0.001&0.001&0.0&0.001&0.002&0.002&0.001&0.0&0.0&0.001&0.002&0.002\\
$\la\alpha_s G^2\ra$&0.0&0.0&0.0&0.0&0.0&0.001&0.0&0.001&0.0&0.0&0.0&0.0\\
$M_0^2$ &0.001&0.001&0.0&0.001&0.001&0.001&0.001&0.0&0.001&0.001&0.001&0.0\\
$\la\bar qq\ra^2$&0.001&0.032&0.001&0.029&0.002&0.047&0.001&0.049&0.001&0.030&0.002&0.032\\
$\la g^3G^3\ra$&0.0&0.0&0.0&0.0&0.0&0.0&0.0&0.0&0.0&0.0&0.0&0.0\\
$d\geq8$&0.002&0.020&0.004&0.008&0.007&0.020&0.0&0.019&0.001&0.010&0.011&0.008\\
{\it Total errors}&0.004&0.041&0.005&0.033&0.011&0.051&0.002&0.073&0.004&0.032&0.014&0.036\\
\hline\hline
\end{tabular}}
\label{tab:error-rap-bplus}
}
\end{table}
} 
\subsection{ The  $(0^{-\pm})$ Beauty Pseudoscalar Molecule States}				     %
Here, within our choice of interpolating currents, the $(0^{--})$ and $(0^{-+})$ are degenerate in masses and have the same
couplings. Here, we choose $\mu=5.5$ GeV where inflexion point has been obtained for the non-strange channel\,\cite{ChXYZ}. 

   \subsection*{\b The $\bar B^*_{s0}B_{s}$  molecule state}
\begin{figure}[hbt] 
\begin{center}
{\includegraphics[width=6.29cm  ]{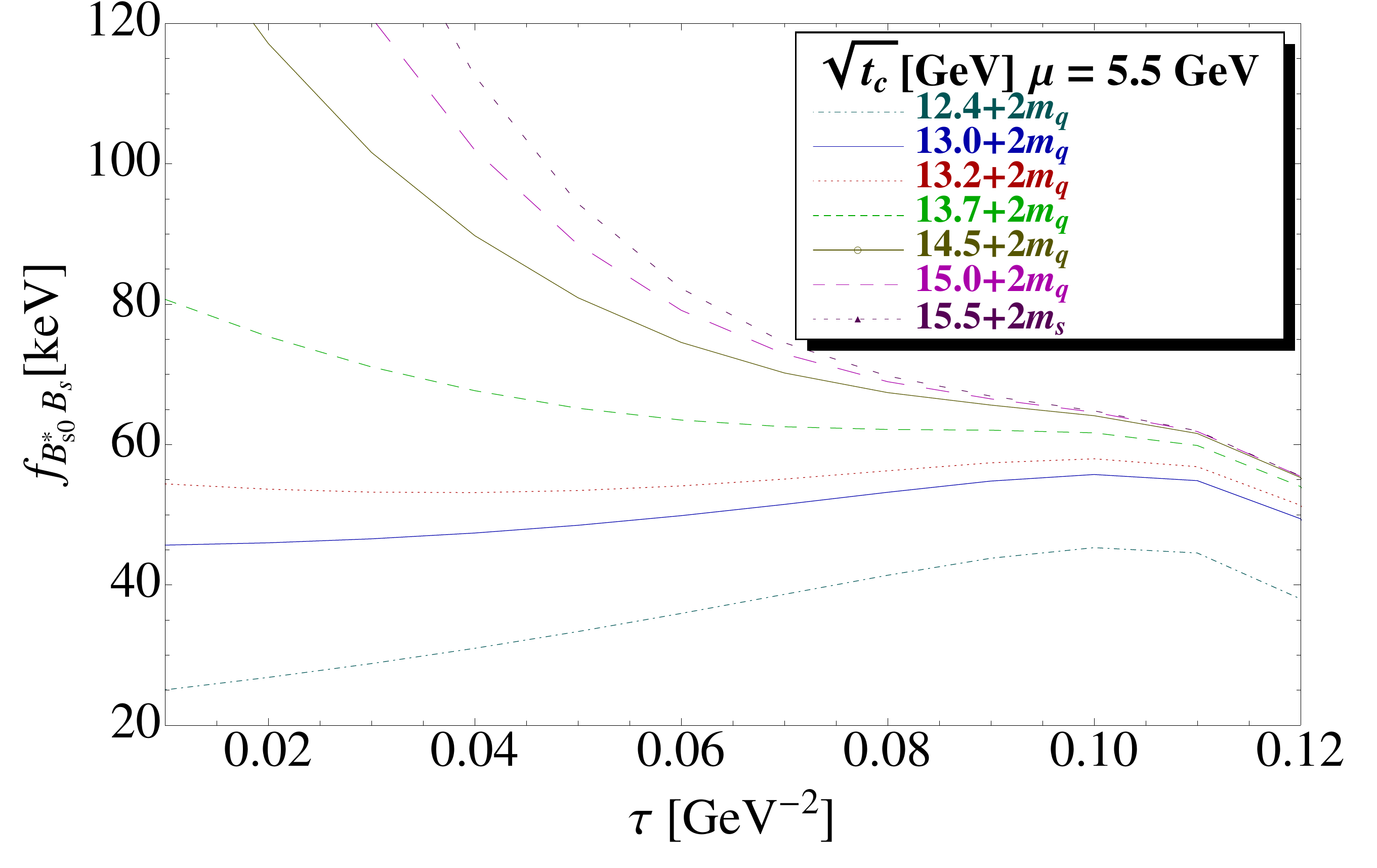}}
{\includegraphics[width=6.29cm  ]{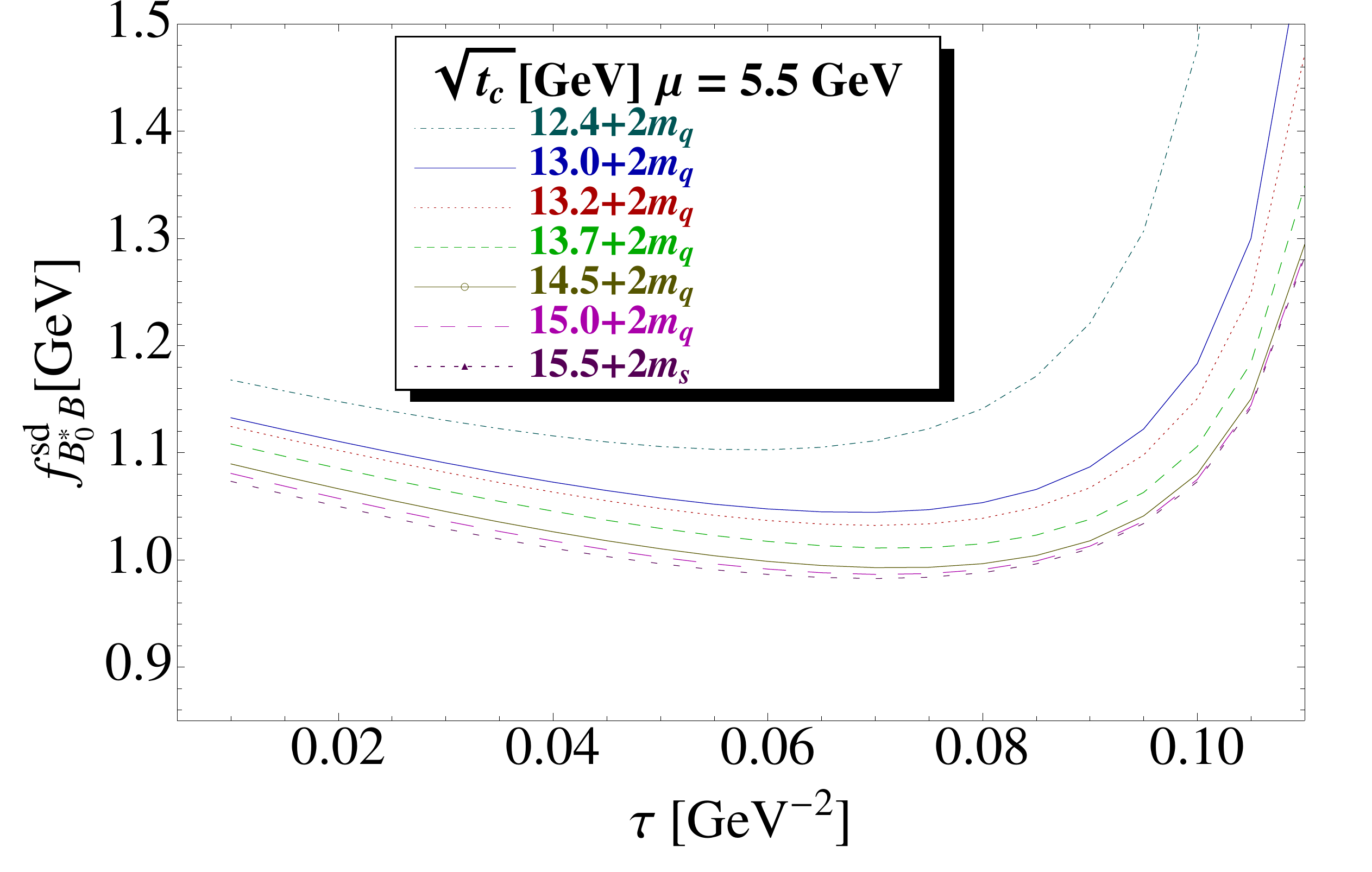}}
\centerline {\hspace*{-3cm} a)\hspace*{6cm} b) }
\caption{
\scriptsize 
{\bf a)} The coupling $f_{B^*_{s0}B_s}$ at NLO  as function of $\tau$ for different values of $t_c$, for $\mu=5.5$ GeV  and for the QCD parameters in Tables\,\ref{tab:param} and \ref{tab:alfa}; {\bf b)} The same as a) but for the SU3 ratio of couplings $f^{sd}_{B^*_{s0}B_s}$.
}
\label{fig:bst0b} 
\end{center}
\end{figure}

\begin{figure}[hbt] 
\begin{center}
{\includegraphics[width=6.29cm  ]{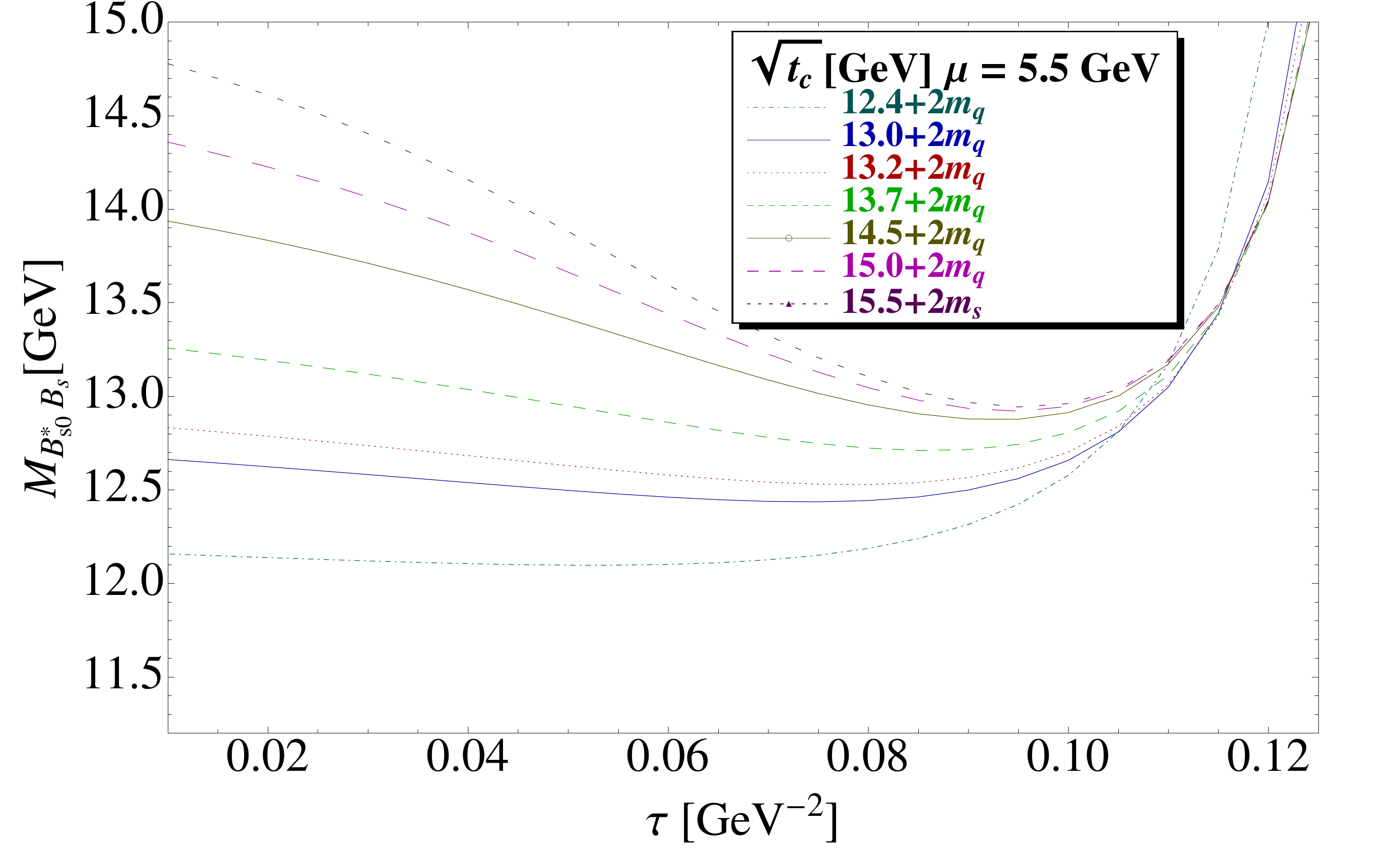}}
\centerline {\hspace*{-3cm} a)\hspace*{6cm} b) }
\caption{
\scriptsize 
$M_{B^*_{s0}B_s}$ at NLO  as function of $\tau$ for different values of $t_c$, for $\mu=5.5$ GeV  and for the QCD parameters in Tables\,\ref{tab:param} and \ref{tab:alfa}.
}
\label{fig:bst0bm} 
\end{center}
\end{figure}

In this channel, the curves present new features where the coupling, its SU3 ratio and the mass present $\tau$-minima as shown in Figs\,\ref{fig:bst0b} and \,\ref{fig:bst0bm}.
The results are similar to the one of $\bar B^*_{s0}B_{s}$. Stabilities are obtained from $\sqrt{t_c}\simeq 13.2$ +2 $m_q$ to 15 +2 $m_q$ GeV. The $\tau$-stabilties are at 0.04(resp. 0.09), 0.07 (resp. 0.07) and  0.07 (resp. 0.095) GeV$^{-2}$ for the coupling, SU3 ratio of couplings and the mass. We deduce the optimal results:
\bea
f_{B^*_{s0}B_{s}}(5.5)&\simeq& 59.8(6.7)_{t_c}(2.1)_\tau~{\rm keV},~ f^{sd}_{B^*_0B}\simeq 1.009(23)_{t_c}(5)_\tau~,\nnb\\
M_{B^*_{s0}B_{s}}&\simeq&12725(197)_{t_c}(37)_\tau~{\rm MeV}.
\eea
Using the NLO  value $f_{B^*_0B}(5.5)\simeq 55(9)~{\rm keV}$ and  $M_{B^*_0B}\simeq$12737(254) MeV from\,\cite{ChXYZ}, one deduces from $f^{sd}_{B^*_0B}$:
\beq
f_{B^*_{s0}B_s}\simeq55.5(8.8)_f(1.3)_{t_c}(0.3)_\tau\cdots~{\rm keV}~, r^{sd}_{B^*_0B}\simeq 1.00(2)\cdots.
\eeq
Taking the mean value of the coupling and re-using $f_{B^*_0B}(5.5)\simeq 55(9)~{\rm keV}$, we deduce the final estimate:
\beq
f_{B^*_{s0}B_s}\simeq58.2(5.5) \cdots ~{\rm keV}~\lrar ~f^{sd}_{B^*_0B}\simeq 1.058(23)\cdots.
\eeq
\nin
\nin
   \subsection*{\b The $\bar B^*_{s}B_{s1}$  molecule state}
The results are similar to the one of $\bar B^*_{s0}B_{s}$. Stabilities are obtained from $\sqrt{t_c}\simeq 13$ +2 $m_q$ to 15 +2 $m_q$ GeV The $\tau$-stabilties are at 0.04(resp. 0.09), 0.07 (resp. 0.07) and  0.07 (resp. 0.09) GeV$^{-2}$ for the coupling, SU3 ratio of couplings and the mass. We deduce the optimal results:
\bea
f_{B^*_{s}B_{s1}}(5.5)&\simeq& 96.7(10.1)_{t_c}(4.3)_\tau~{\rm keV},~ f^{sd}_{B^*B_1}\simeq 1.014(31)_{t_c}(10)_\tau,~\nnb\\
M_{B^*_{s}B_{s1}}&\simeq&12726(264){t_c}(5)_\tau~{\rm MeV}.
\eea
Using the NLO  value $f_{B^*B_1}(5.5)\simeq 105(15)~{\rm keV}$ and $M_{B^*B_1}\simeq$ 12794(228) MeV from\,\cite{ChXYZ}, one deduces:
\beq
f_{B^*_{s}B_{s1}}(5.5)\simeq 106.5(15.2)_f(3.3)_{t_c}(1.1)_\tau~{\rm keV},~r^{sd}_{B^*B_1}\simeq 1.00(3)
\eeq
Taking the mean of the couplings, we obtain:
\beq
f_{B^*_{s}B_{s1}}(5.5)\simeq 100.2(9.0)\cdots~{\rm keV}~\lrar ~ f^{sd}_{B^*B_1}\simeq 0.96(3)\cdots
\eeq
\subsection{ The  $(1^{--})$ Beauty Vector Molecule States}				     %
   \subsection*{\b The $(1^{--})~\bar B^*_{s0}B^*_{s}$ molecule state}
The behaviours of different curves are the same as in the case of the pseudoscalar ($0^{-\pm}$) molecules and will not be shown here.
$\tau$ and $t_c$-stabilities are obtained about the same values as for the $B^*_{s}B_{s1}~(0^{-\pm})$ state at which we deduce  the optimal results:
\bea
f_{B^*_{s0}B^*_{s}}(5.5)&\simeq& 49.8(5.8)_{t_c}(1.8)_\tau~{\rm keV},~ f^{sd}_{B^*_0B^*}\simeq 1.024(28)_{t_c}(7)_\tau~,\nnb\\
M_{B^*_{s0}B^*_{s}}&\simeq&12715(258)_{t_c}(37)_\tau~{\rm MeV}.
\eea
Using the NLO  value $f_{B^*_0B^*}(5.5)\simeq 54(9)~{\rm keV}$ and $M_{B^*_0B^*}\simeq$12756(261) MeV from\,\cite{ChXYZ}, one deduces from $f^{sd}_{B^*_0B^*}$:
\beq
f_{B^*_{s0}B^*_s}\simeq55.3(9.2)_f(1.5)_{t_c}(0.4)_\tau\cdots~{\rm keV},~r^{sd}_{B^*_0B^*}\simeq 1.00(3)~.
\eeq
Taking the mean value of the coupling and re-using $f_{B^*_0B^*}(5.5)\simeq 54(9)~{\rm keV}$, we deduce the final estimate:
\beq
f_{B^*_{s0}B^*_s}\simeq51.4(5.1) \cdots ~{\rm keV}~\lrar ~f^{sd}_{B^*_0B^*}\simeq 0.95(3)\cdots.
\eeq
   \subsection*{\b The $(1^{--})~\bar B_{s}B_{s1}$ molecule state}
The behaviours of different curves are the same as in the case of the pseudoscalar ($0^{-\pm}$) molecules and will not be shown here.
$\tau$-stabilities are obtained at 0.055(resp. 0.09), 0.07 (resp. 0.075) and  0.08 (resp. 0.095) GeV$^{-2}$ for the coupling, SU3 ratio of couplings and the mass
for $\sqrt{t_c}\simeq 13$ +2 $m_q$ to 15+2 $m_q$  GeV. We deduce  the optimal results:
\bea
f_{B_{s}B_{s1}}(5.5)&\simeq& 45.1(1.0)_{t_c}(0.2)_\tau~{\rm keV},~ f^{sd}_{BB_1}\simeq 0.997(33)_{t_c}(8)_\tau~,\nnb\\
M_{B_{s}B_{s1}}&\simeq&12615(221)_{t_c}(42)_\tau~{\rm MeV}.
\eea
Using the NLO  value $f_{BB_1}(5.5)\simeq 54(10.6)~{\rm keV}$ and $M_{BB_{1}}\simeq$ 12734(249) MeV from\,\cite{ChXYZ}, one deduces:
\beq
f_{B_{s}B_{s1}}\simeq53.8(10.6)_f(1.8)_{t_c}(0.4)_\tau\cdots~{\rm keV},~r^{sd}_{BB_{1}}\simeq 0.99(3)\cdots~.
\eeq
Taking the mean value of the coupling and re-using $f_{BB_1}(5.5)\simeq 54(10.6)~{\rm keV}$, we deduce the final estimate:
\beq
f_{B_{s}B_{s1}}\simeq 45.1(1.0) \cdots ~{\rm keV}~\lrar ~f^{sd}_{BB_1}\simeq 0.83(3)\cdots.
\eeq
\subsection{ The  $(1^{-+})$ Beauty Vector Molecule States}				     %
   \subsection*{\b The $(1^{-+})~\bar B^*_{s0}B^*_{s}$ molecule state}
The behaviours of different curves are the same as in the previous case of the pseudoscalar ($0^{-\pm}$) and $1^{--}$ vector  molecules and will not be shown here.
$\tau$ stabilities are obtained  are obtained at 0.07--0.08(resp. 0.09), 0.07 (resp. 0.07) and  0.09 (resp. 0.095) GeV$^{-2}$ for the coupling, the SU3 ratio of couplings and the mass for $\sqrt{t_c}\simeq 13.2$ +2 $m_q$ to 15 +2 $m_q$ GeV.We obtain  the optimal results:
\bea
f_{B^*_{s0}B^*_{s}}(5.5)&\simeq& 50.3(3.4)_{t_c}(2.2)_\tau~{\rm keV},~ f^{sd}_{B^*_0B^*}\simeq0.99(1.3)_{t_c}(3)_\tau~,\nnb\\
M_{B^*_{s0}B^*_{s}}&\simeq&12734(239)_{t_c}(92)_\tau~{\rm MeV}.
\eea
Using the NLO  value $f_{B^*_0B^*}(5.5)\simeq 54(9)~{\rm keV}$ and $M_{B^*_0B^*}\simeq$12774(261) MeV from\,\cite{ChXYZ}, one deduces from $f^{sd}_{B^*_0B^*}$:
\beq
f_{B^*_{s0}B^*_s}\simeq 53.5(8.9)_f(0.7)_{t_c}(1.6)_\tau\cdots~{\rm keV},~r^{sd}_{B^*_0B^*}\simeq 1.00(3)~.
\eeq
Taking the mean value of the coupling and re-using $f_{B^*_0B^*}(5.5)\simeq 54(9)~{\rm keV}$, we deduce the final estimate:
\beq
f_{B^*_{s0}B^*_s}\simeq50.8(3.7) \cdots ~{\rm keV}~\lrar ~f^{sd}_{B^*_0B^*}\simeq 0.94(3)\cdots.
\eeq
   \subsection*{\b The $(1^{-+})~\bar B_{s}B_{s1}$ molecule state}

The behaviours of different curves are the same as in the case of the pseudoscalar ($0^{-\pm}$) molecules and will not be shown here.
$\tau$-stabilities are obtained at 0.055(resp. 0.09), 0.07 (resp. 0.075) and  0.08 (resp. 0.095) GeV$^{-2}$ for the coupling, SU3 ratio of couplings and the mass
for $\sqrt{t_c}\simeq 13$+2 $m_q$ to 15 +2 $m_q$ GeV. We deduce  the optimal results:
\bea
f_{B_{s}B_{s1}}(5.5)&\simeq& 47.4(4.5)_{t_c}(1)_\tau~{\rm keV},~ f^{sd}_{BB_1}\simeq 1.005(44)_{t_c}(12)_\tau~,\nnb\\
M_{B_{s}B_{s1}}&\simeq&12602(234)_{t_c}(32)_\tau~{\rm MeV}.
\eea
Using the NLO  value $f_{BB_1}(5.5)\simeq 53(10.6)~{\rm keV}$ and $M_{BB_{1}}\simeq$ 12790(249) MeV from\,\cite{ChXYZ}, one deduces:
\beq
f_{B_{s}B_{s1}}\simeq53.3(10.6)_f(2.3)_{t_c}(0.6)_\tau\cdots~{\rm keV},~r^{sd}_{BB_{1}}\simeq 0.985(27)~.
\eeq
Taking the mean value of the coupling and re-using $f_{BB_1}(5.5)\simeq 54(10.6)~{\rm keV}$, we deduce the final estimate:
\beq
f_{B_{s}B_{s1}}\simeq 48.3(4.2) \cdots ~{\rm keV}~\lrar ~f^{sd}_{BB_1}\simeq 0.894(41)\cdots.
\eeq

{\scriptsize
\begin{table}[hbt]
\setlength{\tabcolsep}{0.12pc}
\tbl{Different sources of errors for the estimate of the $0^{- }$ and $1^{- }$ $\bar B_sB_s$-like molecule masses (in units of MeV) and couplings $f_{M_sM_s}$ (in units of keV). We use $\mu=5.5(5)$ GeV.}
   {\scriptsize
 {\begin{tabular}{@{}llllllllll@{}} 
&\\
\hline
\hline
\bf Inputs &$\Delta M_{B^*_{s0}B_s}$&$\Delta f_{B^*_{s0}B_s}$&$\Delta M_{B^*_sB_{s1}}$&$\Delta f_{B^*_{s}B_{s1}}$&$\Delta M_{B^*_{s0}B^*_s}$&$\Delta f_{B^*_{s0}B^*_s}$&$\Delta M_{B_sB_{s1}}$&$\Delta f_{B_{s}B_{s1}}$\\
\hline 
{\it LSR parameters}&\\
$(t_c,\tau)$&200.4&5.5&264.1&9.1&260.6&5.1&225.0&1.0\\
$\mu$&11 &1.13&12&2.15&11.6&1.39&12.4&1.04\\
{\it QCD inputs}&\\
$\bar M_Q$&3.70 &0.18&3.84&0.34&3.88&0.17&4.0&0.18\\
$\alpha_s$& 5.20&0.50 &4.82&0.94& 5.46&0.48&4.96&0.54  \\
$N3LO$&15.12&0.70&10.71&1.26&14.21&0.56&13.23&0.35   \\
$\la\bar qq\ra$   &5.10&0.05&5.73&0.11&7.23&0.03&0.0&0.0\\
$\la\alpha_s G^2\ra$&5.10&0.05&2.66&0.05&0.27&0.0&0.0&0.0 \\
$M_0^2$ &1.75&0.0&3.53&0.0&1.13&0.03&2.06&0.07\\
$\la\bar qq\ra^2$&35.65&1.51&33.8&3.18&32.15&1.11&42.95&1.51\\
$\la g^3G^3\ra$& 0.15&0.0&0.17&0.0& 0.15&0.0&0.16&0.0\\
$d\geq8$&71.80&0.82&125&8.09&41.6&1.81&54.6&2.00\\
{\it Total errors}&216.80&3.38&294.73&11.03&266.66&3.86&236.26&2.97\\
\hline\hline
\end{tabular}}
\label{tab:errorb}
}
\end{table}
} 
{\scriptsize
\begin{table}[hbt]
\setlength{\tabcolsep}{1.6pc}
\tbl{Different sources of errors for the direct estimate of the $0^{- }$ and $1^{- }$ $\bar B_sB_s$-like molecule SU(3) ratio of couplings $f^{sd}_{MM}$. We use $\mu=5.5(5)$ GeV.}
   {\scriptsize
 {\begin{tabular}{@{}llllllllll@{}} 
&\\
\hline
\hline
\bf Inputs&$\Delta f^{sd}_{B^*_{0}B}$&$\Delta f^{sd}_{B^*B_1}$&$\Delta f^{sd}_{B^*_{0}B^*}$&$\Delta f^{sd}_{BB_1}$\\
\hline 
{\it LSR parameters}&\\
$(t_c,\tau)$&0.023&0.03&0.03&0.03\\
$\mu$&0.001&0.001&0.006&0.002\\
{\it QCD inputs}&\\
$\bar M_Q$&0.0&0.0&0.0&0.0\\
$\alpha_s$&0.0&0.0&0.0&0.0  \\
$N3LO$&0.003&0.001&0.004&0.004  \\
$\la\bar qq\ra$   &0.001&0.001&0.001&0.0\\
$\la\alpha_s G^2\ra$&0.0&0.0&0.0&0.0  \\
$M_0^2$&0.0& 0.0&0.001& 0.001\\
$\la\bar qq\ra^2$&0.016&0.020&0.012&0.017\\
$\la g^3G^3\ra$&0.0&0.0&0.0&0.0\\
$d\geq8$&0.001&0.0&0.002&0.002\\
{\it Total errors}&0.028&0.036&0.033&0.035\\
\hline\hline
\end{tabular}}
\label{tab:error-rap-bmoins}
}
\end{table}
} 
\section{The  Heavy-light Beauty  Four-Quark States}
We do a similar analysis. The different sources of errors are given in Tables \,\ref{tab:error-4b} and \,\ref{tab:error-rap-4b}.
{\scriptsize
\begin{table}[hbt]
\setlength{\tabcolsep}{0.42pc}
\tbl{Different sources of errors for the estimate of the four-quarks $[bs\bar b \bar s]$  pseudo (scalar)  $\pi_{sb}$ ($S_{sb}$) and  axial (vector) $A_{sb}$ ($V_{sb}$) masses (in units of MeV) and couplings (in units of keV). We use $\mu=5.5(5)$ GeV.}
   {\scriptsize
 {\begin{tabular}{@{}llllllllll@{}} 
&\\
\hline
\hline
\bf Inputs &$\Delta M_{S_{sb}}$&$\Delta f_{S_{sb}}$&$\Delta M_{A_{sb}}$&$\Delta f_{A_{sb}}$&$\Delta M_{\pi_{sb}}$&$\Delta f_{\pi_{sb}}$&$\Delta M_{V_{sb}}$&$\Delta f_{V_{sb}}$\\
\hline 
{\it LSR parameters}&\\
$(t_c,\tau)$&21&1.4&24&0.9&203&6&264&6\\
$\mu$&3.02&1.23&4.27&1.20&13.3&1.54&12.2&1.36\\
{\it QCD inputs}&\\
$\bar M_Q$&5.53&0.13&2.70&0.12& 3.79&0.24& 3.90&0.20\\
$\alpha_s$&13.81&0.46&12.80&0.44& 5.67&0.66& 4.83&0.54\\
$N3LO$&  0.77  & 0.28  & 1.54&0.28&24.92&0.35 &21.35&0.28   \\
$\la\bar qq\ra$  &18.37&0.25&18.33&0.26& 5.00&0.08 &5.91&0.08\\
$\la\alpha_s G^2\ra$&0.42&0.01&0.13&0.007&1.48&0.02& 1.18&0.01\\
$M_0^2$&18.00&0.28&18.47&0.26& 0.46&0.0& 1.75&0.10\\
$\la\bar qq\ra^2$&6.16&0.89&12.26&0.82& 35.35&5.68& 35.34&3.36  \\
$\la g^3G^3\ra$&0.06&0.0&0.0&0.001&0.14&0.0& 0.15&0.0 \\
$d\geq8$&144.56&4.80&167.72&3.21&53.4&8.76& 50.5&4.49 \\
{\it Total errors}&149.23&5.27&172.41&3.69&214.78&11.80&272.36&7.56\\
\hline\hline
\end{tabular}}
\label{tab:error-4b}
}
\end{table}
} 
{\scriptsize
\begin{table}[hbt]
\setlength{\tabcolsep}{1.05pc}
\tbl{Different sources of errors for the direct estimate of the four-quarks $[bs\bar b \bar s]$  pseudo (scalar)  $\pi_{sb}$ ($S_{sb}$) and  axial (vector) $A_{sb}$ ($V_{sb}$) SU(3) ratio of masses $r^{sd}_{M}$ and  of couplings $f^{sd}_{M}$. We use $\mu=5.5(5)$ GeV.}
   {\scriptsize
 {\begin{tabular}{@{}llllllllll@{}} 
&\\
\hline
\hline
\bf Inputs &$\Delta r^{sd}_{S_b}$&$\Delta f^{sd}_{S_b}$&$\Delta r^{sd}_{A_b}$&$\Delta f^{sd}_{A_b}$&$\Delta f^{sd}_{\pi_b}$&$\Delta f^{sd}_{V_b}$\\
\hline 
{\it LSR parameters}&\\
$(t_c,\tau)$&0.002&0.016&0.002&0.02&0.03&0.04\\
$\mu$&0.0&0.003&0.0&0.004&0.002 &0.002  \\
{\it QCD inputs}&\\
$\bar M_Q$&0.0&0.0&0.0&0.0&0.0  & 0.0 \\
$\alpha_s$&0.0&0.007&0.0&0.001&0.0 & 0.0  \\
$N3LO$&0.003&0.004&0.003&0.004&0.003& 0.005  \\
$\la\bar qq\ra$   &0.001&0.002&0.001&0.003&0.001  &  0.001 \\
$\la\alpha_s G^2\ra$&0.0&0.001&0.0&0.0& 0.0& 0.0 \\
$M_0^2$&0.001&0.002&0.001&0.001& 0.0 & 0.001  \\
$\la\bar qq\ra^2$&0.001&0.012&0.001&0.019& 0.056 & 0.044   \\
$\la g^3G^3\ra$&0.0&0.0&0.0&0.0& 0.0 & 0.0 \\
$d\geq8$&0.003&0.013&0.004&0.015&0.016&0.006\\
{\it Total errors}&0.0050&0.026&0.006&0.032&0.065&0.060\\
\hline\hline
\end{tabular}}
\label{tab:error-rap-4b}
}
\end{table}
} 
  \subsection{The  $S_{sb}(0^{+})$ Beauty Scalar State}
In this case, the coupling stabilizes at $\tau\simeq$ 0.11(resp. 0.14) GeV$^{-2}$ from $\sqrt{t_c}=11.9+2\overline m_q$  to $\sqrt{t_c}=15+2\overline m_q$
GeV while the SU3 ratio of masses stabilizes at $\tau\simeq$ 0.17(resp. 0.17) GeV$^{-2}$ for the same range of $t_c$-values. The SU3 ratio of couplings stabilizes for $\tau\simeq$ 0.09(resp. 0.14) GeV$^{-2}$ from $\sqrt{t_c}=12.9+2\overline m_q$  to $\sqrt{t_c}=15+2\overline m_q$
GeV. We obtain the optimal results at NLO:
\bea
f_{S_{sb}}&\simeq& 21.7(1.4)_{t_c}(0.1)_{\tau}\cdots~{\rm keV}~, ~~~~~f^{sd}_{S_{b}}\simeq 0.919(17)_{t_c}(1)_\tau\cdots~,\nnb\\
r^{sd}_{S_{b}}&\simeq& 1.044(2)_{t_c}(0.3)_\tau\cdots~.
\eea
Using $f_{S_{b}}\simeq 17(0.14)$ keV and  $M_{S_{b}}\simeq 10653(0.1)$ MeV from\,\cite{ChXYZ}, wecan also  deduce:
\beq
f^{sd}_{S_{b}}\simeq 0.78(4)\cdots~,~~~~~M_{S_{sb}}\simeq 11122(21)\cdots~{\rm  MeV}~.
\eeq
Taking the mean of the SU3 ratio of couplings, we obtain our final estimate:
\beq
f^{sd}_{S_{b}}\simeq 0.898(16)\cdots~~\lrar~~f_{S_{sb}}\simeq 15.27(0.30)\cdots~{\rm keV}~
\eeq
  \subsection{The  $A_b(1^{+})$ Beauty  Axial-Vector  State}
Here the coupling stabilizes at $\tau\simeq$ 0.10(resp. 0.14) GeV$^{-2}$ from $\sqrt{t_c}=11.9+2\overline m_q$ to $\sqrt{t_c}=15+2\overline m_q$
GeV while the SU3 ratio of masses stabilizes at $\tau\simeq$ 0.18(resp. 0.16) GeV$^{-2}$ for the same range of $t_c$-values. We obtain the optimal results at NLO:
\beq
f_{A_{sb}}\simeq 21.9(1.2)_{t_c}(0.1)_{\tau}\cdots~{\rm keV}~, ~~~~~r^{sd}_{A_{b}}\simeq 1.042(2)_{t_c}(0.3)_\tau\cdots~.
\eeq
Using $f_{A_{b}}\simeq 18(0.9)$ keV and  $M_{A_{b}}\simeq 10701(9)$ MeV from\,\cite{ChXYZ}, we deduce:
\beq
f^{sd}_{A_{b}}\simeq 0.82(6)\cdots~,~~~~~M_{A_{sb}}\simeq 11150(24)\cdots~{\rm  MeV}~.
\eeq
Taking the mean of the previous SU3 ratio of couplings with the one from the direct determination obtained at $\tau\simeq$ 0.09(resp. 0.14) GeV$^{-2}$ from $\sqrt{t_c}=12.9+2\overline m_q$ GeV to $\sqrt{t_c}=15+2\overline m_q$ GeV:
\beq
f^{sd}_{A_{b}}\simeq 0.92(2)_{t_c}(1)_\tau\cdots~,
\eeq
we deduce:
\beq
f^{sd}_{A_{b}}\simeq 0.91(2)\cdots~~\lrar~~f_{A_{sb}}\simeq 16.4(0.9)\cdots~{\rm keV}~
\eeq
where $f_{A_{b}}$ has been used for deriving the last equation. 
  \subsection{The  $\pi_{sb}(0^{-})$  Beauty  Pseudoscalar State}
The coupling presents $\tau$-stabilities from $\sqrt{t_c}=13.2+2\overline m_q$ GeV to $\sqrt{t_c}=15+2\overline m_q$ GeV and for $\tau$=0.045(resp. 0.09) GeV$^{-2}$.
 Within these range of $t_c$-values, the mass stabilizes for $\tau\simeq$  0.08 (resp. 0.095) GeV$^{-2}$ and the ratio of couplings for $\tau\simeq$  0.08 (resp. 0.09) GeV$^{-2}$. 
We obtain:
\bea
f_{\pi_{sb}}&\simeq& 60(7)_{t_c}(2)_{\tau}\cdots~{\rm keV}~,\nnb\\
f^{sd}_{\pi_{b}}&\simeq&0.91(2)_{t_c}(3)_\tau\cdots,~M_{\pi_{sb}}\simeq 12730(197)_{t_c}(48)_{\tau}~{\rm MeV}.
\eea   
 Using the values: $M_{\pi_{b}}=12920(235)$ MeV and $ f_{\pi_b}=$83(9) keV from\,\cite{ChXYZ}, we can deduce:
\beq
r^{sd}_{\pi_b}\simeq 0.985(24)\cdots,~
f_{\pi_{sb}}\simeq 75.5(82)_f(17)_{t_c}(25)_{\tau}\cdots~{\rm keV}~.
\eeq   
Taking the mean of $f_{\pi_{sb}}$ and re-using $ f_{\pi_b}=$ 83(9) keV, we deduce the final estimate:
\beq
f_{\pi_{sb}}\simeq 66(6)\cdots~{\rm keV}~\lrar~ f^{sd}_{\pi_b}\simeq 0.80(3)\cdots,
\eeq
where the error comes from the direct determination of the SU3 ratio. 
  \subsection{The  $V_{sb}(1^{-})$ Beauty  Vector  State}
The behaviours of the corresponding curves are very similar to the previous ones.
The coupling presents $\tau$-stabilities from $\sqrt{t_c}=13.0+2\overline m_q$ to $\sqrt{t_c}=15.0+2\overline m_q$ GeV and for $\tau$=0.04(resp. 0.09) GeV$^{-2}$.
 Within these range of $t_c$-values, the mass stabilizes for $\tau\simeq$  0.07 (resp. 0.09) GeV$^{-2}$ and the ratio of couplings for $\tau\simeq$  0.06 (resp. 0.07) GeV$^{-2}$. 
We obtain:
\bea
f_{V_{sb}}&\simeq& 57(7)_{t_c}(2)_{\tau}\cdots~{\rm keV}~,\nnb\\
f^{sd}_{V_{b}}&\simeq&1.03(3)_{t_c}(2)_\tau\cdots,~M_{V_{sb}}\simeq 12716(260)_{t_c}(48)_{\tau}~{\rm MeV}.
\eea   
 Using the NLO values: $M_{V_{b}}=12770(214)$ MeV and $ f_{V_b}\simeq$ 62(9) keV from\,\cite{ChXYZ}, we can deduce:
\beq
r^{sd}_{V_b}\simeq 1.00(3)\cdots,~
f_{V_{sb}}\simeq 64(9)_f(2)_{t_c}(1)_\tau\cdots~{\rm keV}~.
\eeq   
Taking the mean of $f_{V_{sb}}$ and re-using $ f_{V_b}\simeq$ 62(9) keV, we deduce the final estimate:
\beq
f_{V_{sb}}\simeq 60(6)\cdots~{\rm keV}~\lrar~  f^{sd}_{V_b}\simeq 0.97(4)\cdots,
\eeq
where the error of $f^{sd}_{V_b}$ comes from the direct determination. 
\section{Summary Tables}\label{sec:summary}
  Our different results for the masses, couplings and their SU3 ratios are summarized in the Tables below. The SU3 ratios have been obtained either from a direct determination or/and by taking the ratio of masses (couplings) from this paper and the ones in the chiral limit from\,\cite{ChXYZ}. We complete Table\,\ref{tab:resultc} by the revised values of the $\bar D^*_0D^*_0$ and $\bar D^*_0D_1$ masses and couplings and by the new value of the $\bar D_1D_1$ ones.
  \subsection{Charm States}  
    \subsection*{\b  Molecules }
{\scriptsize
\begin{table}[H]
\setlength{\tabcolsep}{0.3pc}
 \tbl{$\bar{D}D$-like molecules couplings, masses and their corresponding SU3 ratios  from LSR within stability criteria at NLO to N2LO of PT. We include revised estimates of the $ \bar D^{*}_{0}D^{*}_{0}$,  $ \bar D^{*}_{0}D_{1}$ couplings and masses and new one for $ \bar D_{1}D_{1}$ . The errors are the quadratic sum of the ones in Tables
\,\ref{tab:errorcplus} to \ref{tab:error-rap-cmoins}.
}  
{\scriptsize{
\begin{tabular}{@{}ll   ll  ll  ll l@{}}
\hline
\hline
                \bf Channels &\multicolumn{2}{c}{$f^{sd}_M\equiv f_{M_s}/f_{M}$}
					&\multicolumn{2}{c}{$f_{M_s}$\bf [keV]}
					&\multicolumn{2}{c}{$r^{sd}_M\equiv M_{M_s}/M_{M}$}
					&\multicolumn{2}{c}{$M_{M_s}$ \bf [MeV]}\\
\cline{2-3} \cline{4-5}\cline{6-7}\cline{8-9}
                 & \multicolumn{1}{l}{{NLO}}
                 & \multicolumn{1}{l }{N2LO} 
                 & \multicolumn{1}{l}{NLO} 
                 & \multicolumn{1}{l }{N2LO} 
                 & \multicolumn{1}{l}{NLO} 
                 & \multicolumn{1}{l}{N2LO}
		    & \multicolumn{1}{l}{NLO} 
                 & \multicolumn{1}{l}{N2LO} 
                  \\
\hline
 \bf{Scalar($0^{++}$)}&&&&&&&&\\
$\bar D_sD_s$&$0.95(3)$&0.98(4)&156(17)&167(18)&1.069(4)&1.070(4)&4169(48)&4169(48)\\
$\bar D^{*}_{s}D^{*}_{s}$&0.93(3)&0.95(3)&265(31)&284(34)&1.069(3)&1.075(3)&4192(200)&4196(200)\\
$\bar D^{*}_{s0}D^{*}_{s0}$&0.88(6)&0.89(6)&85(12)&102(14)&1.069(69)&1.058(68)&4277(134)&4225(132)\\
$\bar D_{s1}D_{s1}$&0.906(33)&0.930(34)&209(28)&229(31)&1.097(7)&1.090(7)&4187(62)&4124(61) \\
\\
$\bar D^{*}_{0}D^{*}_{0}$&--&--&97(15)&114(18)&--&--&4003(227)&3954(224)\\
$\bar D_{1}D_{1}$&--&--&236(32)&274(37)&--&--&3838(57)&3784(56) \\
\\
\bf {Axial($1^{+\pm}$)}&&&&&&&&\\
$\bar D^{*}_{s}D_{s}$&0.93(3)&0.97(3)&143(16)&156(17)&1.070(4)&1.073(4)&4174(67)&4188(67)\\
$\bar D^{*}_{s0}D_{s1}$&0.90(1)&0.82(1)&87(14)&110(18)&1.119(24)&1.100(24)&4269(205)&4275(206)\\
\\
$\bar D^{*}_{0}D_{1}$&--&--&96(15)&112(17)&--&--&3849(182)&3854(182)\\
\\

\bf  {Pseudo($0^{-\pm}$)}&&&&&&&&\\
$\bar D^{*}_{s0}D_{s}$&0.94(5)&0.90(4)&225(24)&232(25)&0.970(50)&0.946(40)&5604(223)&5385(214)\\
$\bar D^{*}_{s}D_{s1}$&0.93(4)&0.90(4)&455(34)&508(38)&0.970(50)&0.972(34)&5724(195)&5632(192)\\
\\
\bf {Vector($1^{--}$)} &&&&&&&&\\
$\bar D^{*}_{s0}D^{*}_{s}$&0.87(4)&0.86(4)&208(11)&216(11)&0.980(33)&0.956(32)&5708(184)&5571(180)\\
$\bar D_{s}D_{s1}$&0.97(3)&0.93(3)&202(12)&213(13)&0.970(33)&0.951(31)&5459(122)&5272(120)\\
\\
\bf { Vector($1^{-+}$)} &&&&&&&&\\
$\bar D^{*}_{s0}D^{*}_{s}$&0.98(5)&0.92(5)&219(17)&231(18)&0.963(32)&0.948(32)&5699(184)&5528(179)\\
$\bar D_{s}D_{s1}$&0.92(3)&0.88(3)&195(13)&212(14)&0.959(34)&0.955(34)&5599(155)&5487(152)\\
\hline
\hline
\end{tabular}
}}
\label{tab:resultc}
\end{table}
}

\subsection*{\b Four-quark}
{\scriptsize
\begin{table}[H]
\setlength{\tabcolsep}{0.4pc}
 \tbl{4-quark couplings, masses and their corresponding SU3 ratios  from LSR within stability criteria at NLO and N2LO of PT. The errors are the quadratic sum of the ones in Tables
\,\ref{tab:4q-errorc} and\,\ref{tab:error-rap-4c}. The * indicates that the value does not come from a direct determination. 
}  
{\scriptsize{
\begin{tabular}{@{}ll   ll  ll  ll l@{}}
\hline
\hline
                \bf Channels &\multicolumn{2}{c}{$f^{sd}_M\equiv f_{M_s}/f_{M}$}
					&\multicolumn{2}{c}{$f_{M_s}$\bf [keV]}
					&\multicolumn{2}{c}{$r^{sd}_M\equiv M_{M_s}/M_{M}$}
					&\multicolumn{2}{c}{$M_{M_s}$ \bf [MeV]}\\
\cline{2-3} \cline{4-5}\cline{6-7}\cline{8-9}
                 & \multicolumn{1}{l}{{NLO}}
                 & \multicolumn{1}{l }{N2LO} 
                 & \multicolumn{1}{l}{NLO} 
                 & \multicolumn{1}{l }{N2LO} 
                 & \multicolumn{1}{l}{NLO} 
                 & \multicolumn{1}{l}{N2LO}
		    & \multicolumn{1}{l}{NLO} 
                 & \multicolumn{1}{l}{N2LO} 
                  \\
\hline
 \bf{c-quark}&&&&&&&&\\
$S_{sc}(0^+)$&0.91(4)&0.98(4)&161(17)&187(19)&1.085(11)&1.086(11)&4233(61)&4233(61)\\
$A_{sc}(1^+)$&0.80(4)&0.87(4)&141(15)&160(17)&1.081(4)&1.082(4)&4205(112)&4209(112)\\
$\pi_{sc}(0^-)$&0.88(7)&0.86(7)&256(29)&267(30)&0.97(3)*&0.96(3)*&5671(181)&5524(176)\\
$V_{sc}(1^-)$&0.91(10)&0.87(10)&245(31)&258(33)&0.96(4)*&0.96(4)*&5654(239)&5539(234)\\
\hline
\hline
\end{tabular}
}}
\label{tab:4q-resultc}
\end{table}
}
    \subsection{  Beauty States}
    \subsection*{\b  Molecules }
{\scriptsize
\begin{table}[H]
\setlength{\tabcolsep}{0.3pc}
 \tbl{$\bar{B}B$-like molecules couplings, masses and their corresponding SU3 ratios  from LSR within stability criteria at NLO to N2LO of PT. The errors are the quadratic sum of the ones in Tables
\,\ref{tab:errorbplus} to \,\ref{tab:error-rap-bmoins}.  The * indicates that the value does not come from a direct determination. 
}
{\scriptsize{
\begin{tabular}{@{}ll   ll  ll  ll l@{}}
\hline
\hline
                \bf Channels &\multicolumn{2}{c}{$f^{sd}_M\equiv f_{M_s}/f_{M}$}
					&\multicolumn{2}{c}{$f_{M_s}$\bf [keV]}
					&\multicolumn{2}{c}{$r^{sd}_M\equiv M_{M_s}/M_{M}$}
					&\multicolumn{2}{c}{$M_{M_s}$ \bf [MeV]}\\
\cline{2-3} \cline{4-5}\cline{6-7}\cline{8-9}
                 & \multicolumn{1}{l}{NLO} 
                 & \multicolumn{1}{l }{N2LO} 
                 & \multicolumn{1}{l}{NLO} 
                 & \multicolumn{1}{l }{N2LO} 
                 & \multicolumn{1}{l}{NLO} 
                 & \multicolumn{1}{l}{N2LO}
		    & \multicolumn{1}{l}{NLO} 
                 & \multicolumn{1}{l}{N2LO} 
                  \\
\hline
 \bf{Scalar($0^{++}$)}&&&&&&&&\\
$\bar B_sB_s$&1.04(4)&1.15(4)&17(2)&20(2)&1.027(4)&1.029(4)&10884(74)&10906(74)\\
$\bar B^{*}_{s}B^{*}_{s}$&1.00(3)&1.12(3)&31(5)&36(6)&1.028(5)&1.029(5)&10944(134)&10956(134)\\
$\bar B^{*}_{s0}B^{*}_{s0}$&1.11(5)&1.07(5)&13(3)&17(4)&1.050(11)&1.034(11)&11182(227)&11014(224)\\
$\bar B_{s1}B_{s1}$&1.197(73)&1.214(74)&24(5)&29(6)&1.040(2)&1.035(2)&10935(170)&10882(169) \\
 \\
$\bar B_{1}B_{1}$&--&--&20(3)&28.6(4)&--&--&10514(149)&10514(149) \\
\\
\bf {Axial($1^{+\pm}$)}&&&&&&&&\\
$\bar B^{*}_{s}B_{s}$&1.01(3)&1.18(4)&17(2)&20(2)&1.028(4)&1.030(4)&10972(195)&10972(195)\\
$\bar B^{*}_{s0}B_{s1}$&0.80(4)&0.79(4)&9(2)&11(3)&1.052(14)&1.031(14)&11234(208)&11021(204)\\
\\

\bf  {Pseudo($0^{-\pm}$)}&&&&&&&&\\
$\bar B^{*}_{s0}B_{s}$&1.06(3)&1.02(3)&58(3)&68(4)&1.00(3)*&1.00(3)*&12725(217)&12509(213)\\
$\bar B^{*}_{s}B_{s1}$&0.96(4)&0.95(4)&100(11)&118(13)&1.00(3)*&1.00(3)*&12726(295)&12573(292)\\
\\
\bf { Vector($1^{--}$)} &&&&&&&&\\
$\bar B^{*}_{s0}B^{*}_{s}$&0.95(3)&0.90(3)&51(4)&59(5)&1.00(3)*&0.99(3)*&12715(267)&12512(263)\\
$\bar B_{s}B_{s1}$&0.83(4)&0.77(3)&45(3)&50(3)&0.99(3)*&0.99(3)*&12615(236)&12426(233)\\
\\
\bf { Vector($1^{-+}$)} &&&&&&&&\\
$\bar B^{*}_{s0}B^{*}_{s}$&0.94(3)&0.92(3)&51(5)&59(6)&1.00(3)*&0.99(3)*&12734(262)&12479(257)\\
$\bar B_{s}B_{s1}$&0.89(4)&0.85(3)&48(5)&55(6)&0.99(3)*&0.98(3)*&12602(247)&12350(242)\\
\hline
\hline
\end{tabular}
}}
\label{tab:resultb}
\end{table}
}
\subsection*{\b Four-quark}
{\scriptsize
\begin{table}[H]
\setlength{\tabcolsep}{0.4pc}
 \tbl{4-quark couplings, masses and their corresponding SU3 ratios  from LSR within stability criteria at NLO and N2LO of PT. The errors are the quadratic sum of the ones in Tables
\,\ref{tab:4q-errorc} and\,\ref{tab:error-rap-4b}. 
}  
{\scriptsize{
\begin{tabular}{@{}ll   ll  ll  ll l@{}}
\hline
\hline
                \bf Channels &\multicolumn{2}{c}{$f^{sd}_M\equiv f_{M_s}/f_{M}$}
					&\multicolumn{2}{c}{$f_{M_s}$\bf [keV]}
					&\multicolumn{2}{c}{$r^{sd}_M\equiv M_{M_s}/M_{M}$}
					&\multicolumn{2}{c}{$M_{M_s}$ \bf [MeV]}\\
\cline{2-3} \cline{4-5}\cline{6-7}\cline{8-9}
                 & \multicolumn{1}{l}{{NLO}}
                 & \multicolumn{1}{l }{N2LO} 
                 & \multicolumn{1}{l}{NLO} 
                 & \multicolumn{1}{l }{N2LO} 
                 & \multicolumn{1}{l}{NLO} 
                 & \multicolumn{1}{l}{N2LO}
		    & \multicolumn{1}{l}{NLO} 
                 & \multicolumn{1}{l}{N2LO} 
                  \\
\hline
 \bf{b-quark}&&&&&&&&\\
$S_{sb}(0^+)$&0.78(3)&0.83(3)&22(5)&26(6)&1.044(4)&1.048(4)&11122(149)&11133((149)\\
$A_{sb}(1^+)$&0.92(3)&0.98(3)&22(4)&26(5)&1.042(6)&1.046(6)&11150(172)&11172(172)\\
$\pi_{sb}(0^-)$&0.80(7)&0.76(4)&66(12)&71(13)&0.985(2)*&0.975(2)*&12730(215)&12374(209)\\
$V_{sb}(1^-)$&0.97(6)&0.90(6)&64(8)&68(9)&0.996(3)*&0.984(30)*&12716(272)&12411(266)\\
\hline
\hline
\end{tabular}
}}
\label{tab:4q-resultb}
\end{table}
}
\section{Comments and Conclusions}
{\scriptsize
\begin{table}[hbt]
\setlength{\tabcolsep}{0.7pc}
\tbl{Comparison with some existing lowest order (LO) results for the $0^{++}$ and $1^+$ charm molecules and four-quark states with hidden strange quarks from QCD Laplace Sum Rules within the same choice of interpolating currents. The experimental candidates are listed in Table\,\ref{tab:exp}.}
   {\scriptsize
 {\begin{tabular}{@{}ll  ll   l   l@{}} 
\\
\hline
\hline
\bf Charm Sates&Mass[MeV]&\bf Beauty States&Mass[MeV]&PT Order&References\\
\hline 
\\
\footnotesize\bf{Scalar $(0^{++})$}\\
\bf Molecules\\
$\bar D_sD_s$ &  4169(48)&$\bar B_sB_s$&10906(74)&  N2LO& \it This work \\
&3910(100)&&&LO&\cite{ZHANG}\\
\\
$\bar D^*_sD^*_s$& 4196(200)&$\bar B^*_sB^*_s$&10956(134)& N2LO &\it This work \\
&4140(90)&&&LO&\cite{NIELSEN}\\
&4130(100)&&&LO&\cite{ZHANG}\\
&4480(170)&&11240(180)&LO&\cite{WANG2}\\
&4380(160)&&&LO&\cite{QIAO}\\
\\
$\bar D^*_{s0}D^*_{s0}$& 4225(132)&$\bar B^*_{s0}B^*_{s0}$&11014(224)& N2LO &\it This work \\
&4580(100)&&11350(90)&LO&\cite{ZHANG2}\\
\\
$\bar D_{s1}D_{s1}$&4124(61) &$\bar B_{s1}B_{s1}$& 10882(169)& N2LO &\it This work \\
&4660(120)&&11390(130)&LO&\cite{ZHANG2}\\
\\
\bf Four-quarks\\
$S_{sc}[\bar c\bar s cs]$& 4.233(61)&$S_{sb}[\bar b\bar s bs]$&11133(149)& N2LO &\it This work\\
&4180(190)&&10010(210)&LO&\cite{CHEN17}\\
\\
\footnotesize\bf{Axial $(1^{+})$}\\
\bf Molecules\\
$\bar D^*_sD_s$&4188(67)&$\bar B^*_sB_s$&10972(195)& N2LO&\it This work \\
&3980(150)&&&LO&\cite{QIAO}\\
&4010(100)&&10710(110)&LO&\cite{ZHANG}\\
\\
$\bar D^*_{s0}D_{s1}$& 4275(206)&$\bar B^*_{s0}B_{s1}$&11021(204)& N2LO&\it This work \\
&4640(100)&&11380(90)&LO&\cite{ZHANG2}\\
\\
\bf Four-quarks\\

$A_{sc}[\bar c\bar s cs]$&4209(112)&$A_{sb}[\bar b\bar s bs]$&11172(172)& N2LO & \it This work\\
&4240(100)&&10340(90)&LO&\cite{CHEN10}\\
&3950(90)&&&LO&\cite{WANG3}\\
&4183(115) &&&LO& \cite{AGAEV}\\
\hline
\hline
\end{tabular}}
\label{tab:theory}
}
\end{table}
} 
\subsection*{\b Comparison of the lowest order QCD expressions}\label{sec:confront3}
\begin{figure}[hbt] 
\begin{center}
{\includegraphics[width=6.29cm  ]{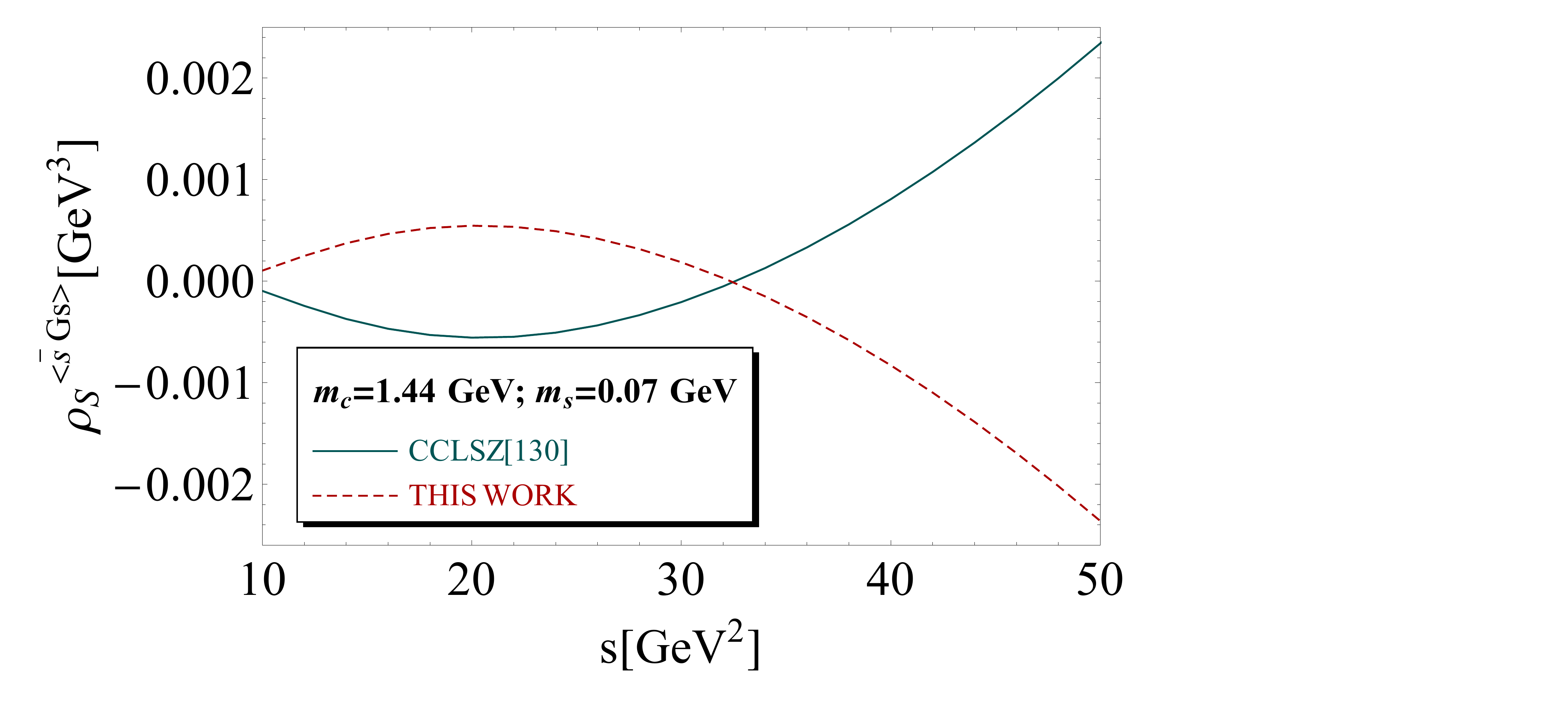}}
{\includegraphics[width=6.29cm  ]{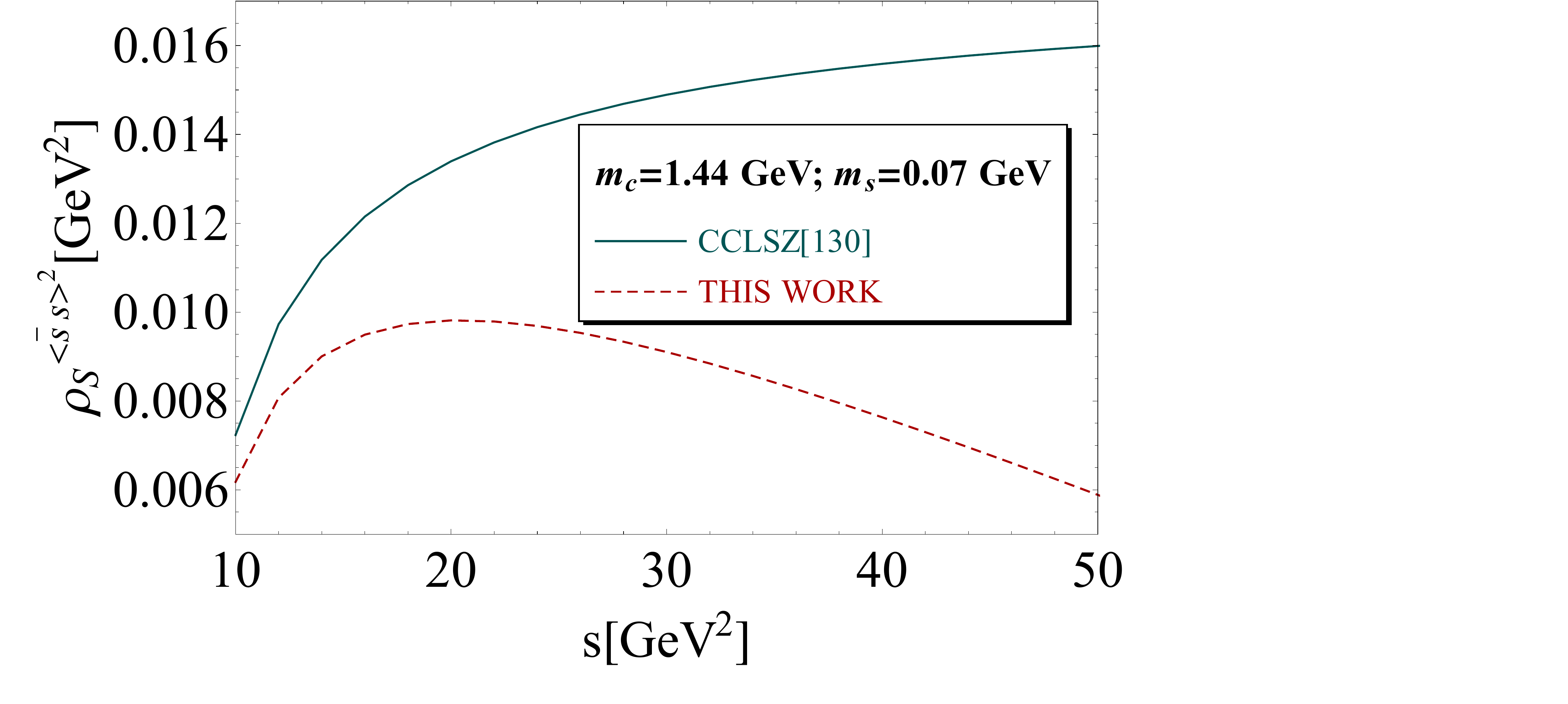}}
\centerline {\hspace*{-3cm} a)\hspace*{6cm} b) }
\caption{
\scriptsize 
Comparison of the Wilson coefficients of the $0^{++}$ four-quark spectral functions for different values of $s$ and for given values of $m_c$ and $m_s$: {\bf a)} mixed condensate; {\bf b)}  four-quark condensate.
}
\label{fig:comp0+} 
\end{center}
\end{figure}

\begin{figure}[hbt] 
\begin{center}
{\includegraphics[width=8.5cm]{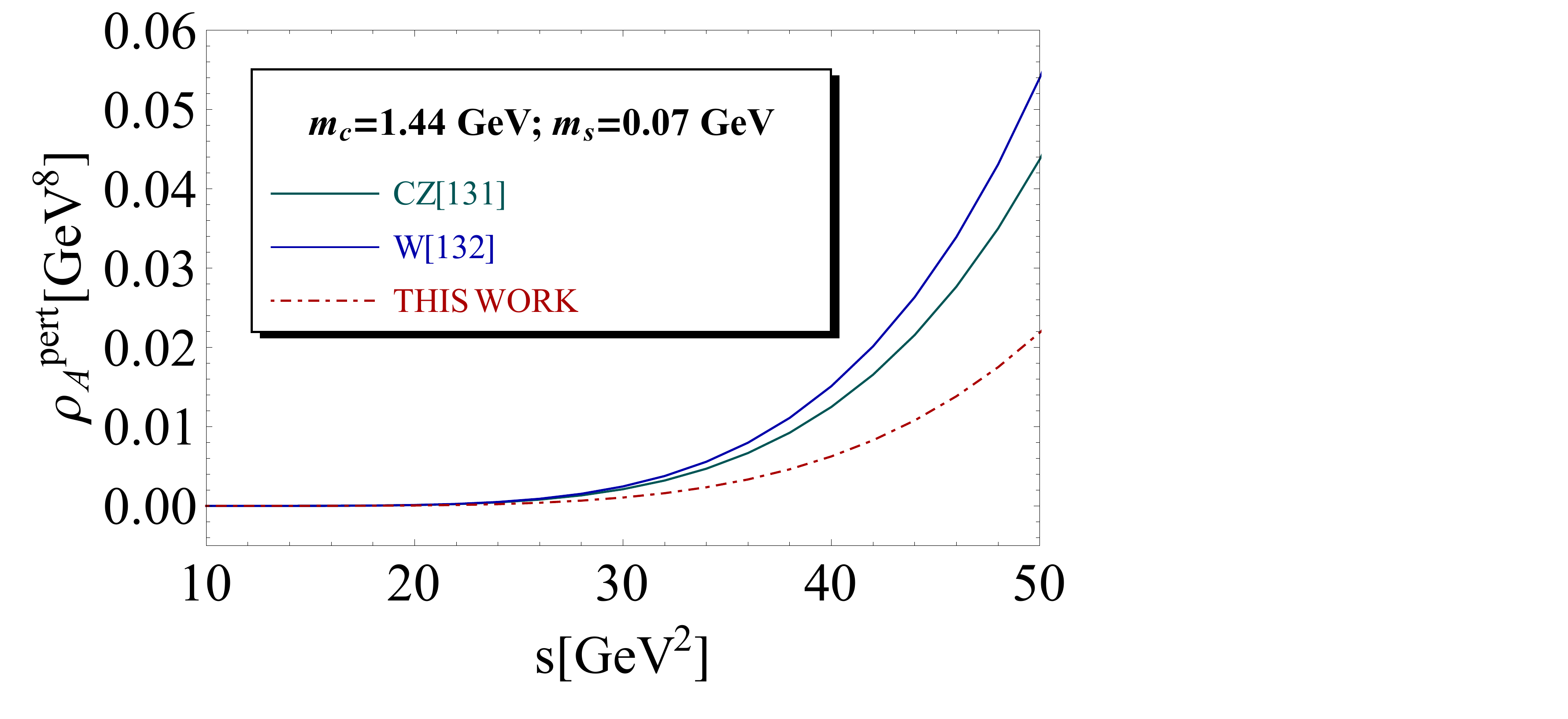}}
\caption{
\scriptsize 
Comparison of the perturbative expression of the $1^{++}$ four-quark spectral functions for different values of $s$ and for given values of $m_c$ and $m_s$.}
\label{fig:comp1+pert} 
\end{center}
\end{figure}

\begin{figure}[hbt] 
\begin{center}
{\includegraphics[width=6.29cm  ]{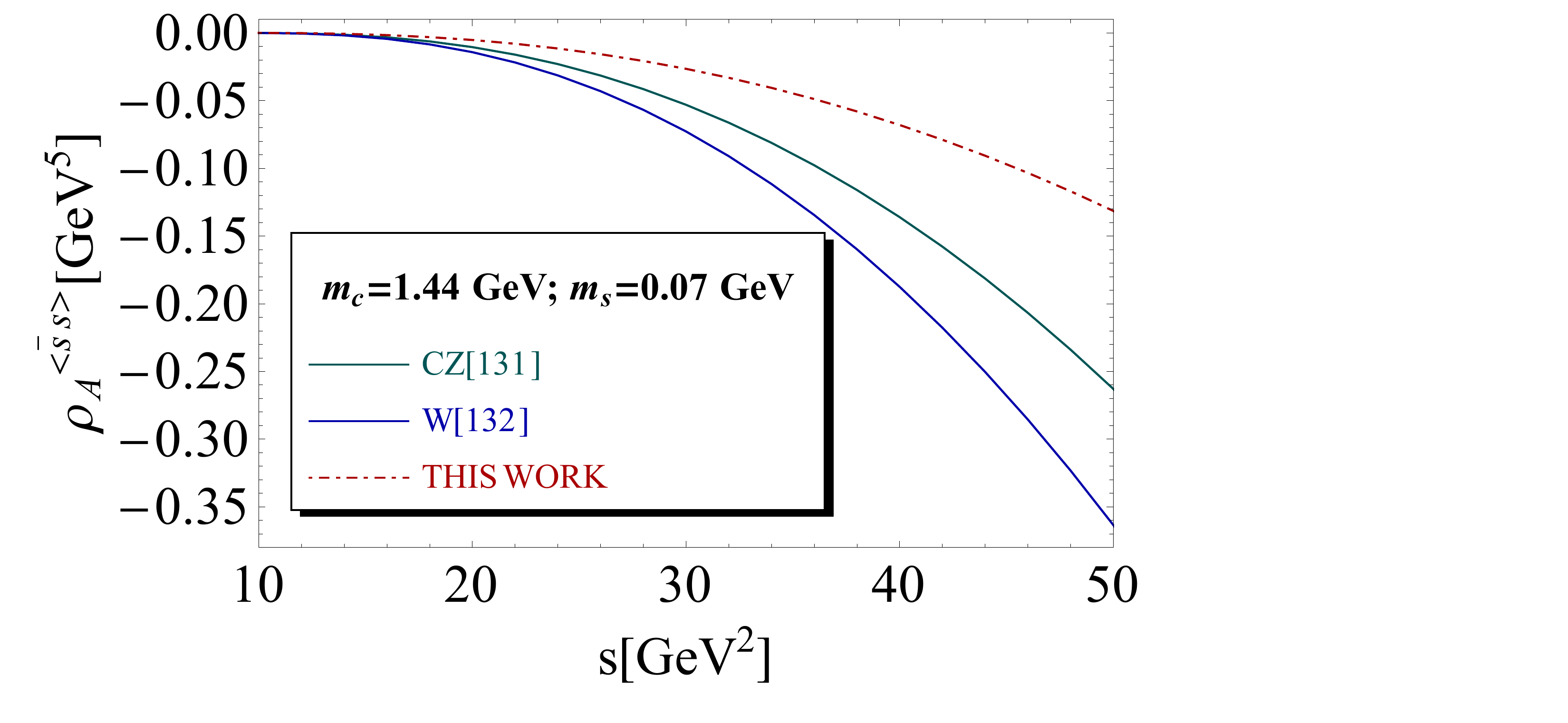}}
{\includegraphics[width=6.29cm  ]{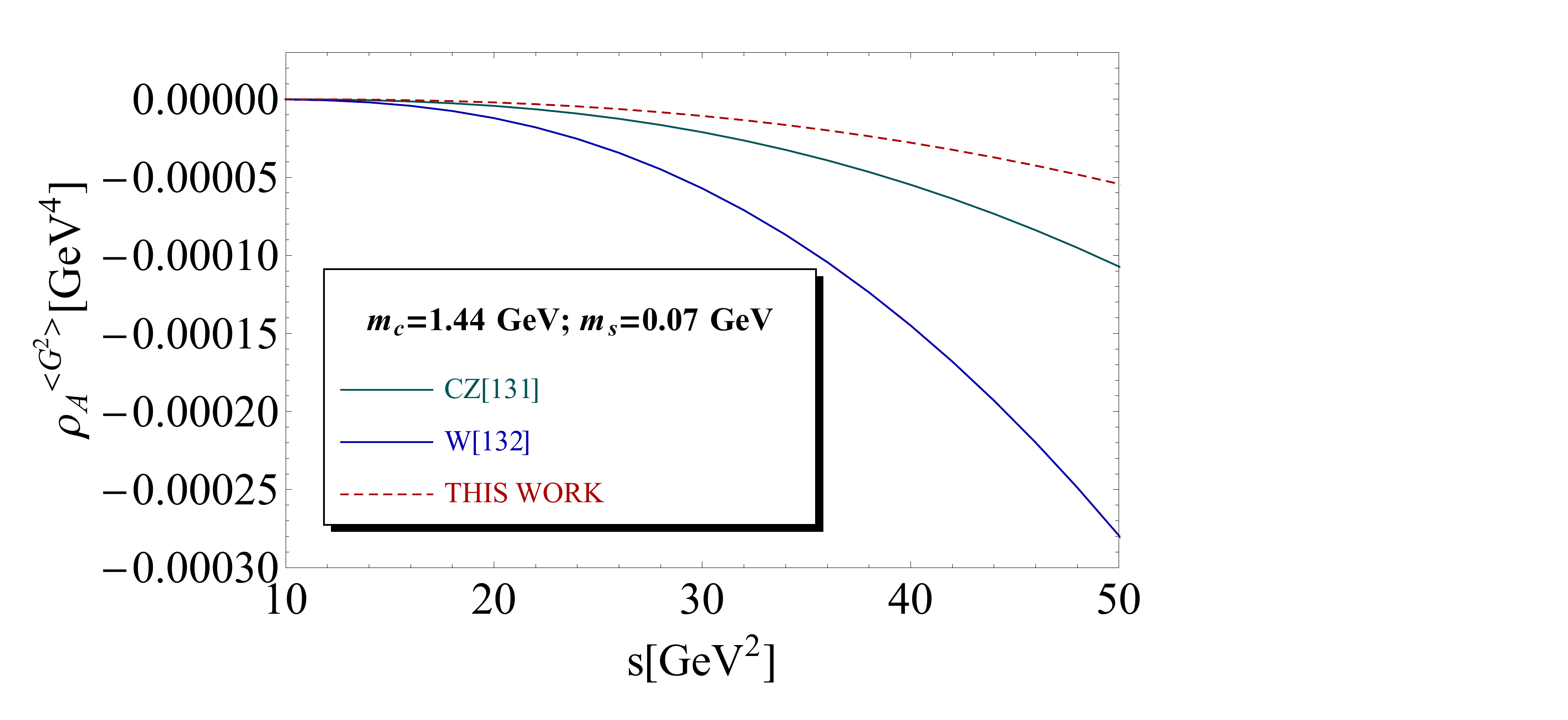}}
\centerline {\hspace*{-3cm} a)\hspace*{6cm} b) }
\caption{
\scriptsize 
Comparison of the Wilson coefficients of the $1^{++}$ four-quark spectral functions for different values of $s$ and for given values of $m_c$ and $m_s$: {\bf a)} $\la\bar ss\ra$ condensate; {\bf b)} $\alpha_s G^2\ra$ condensate.
}
\label{fig:comp1+4} 
\end{center}
\end{figure}

\begin{figure}[hbt] 
\begin{center}
{\includegraphics[width=6.29cm  ]{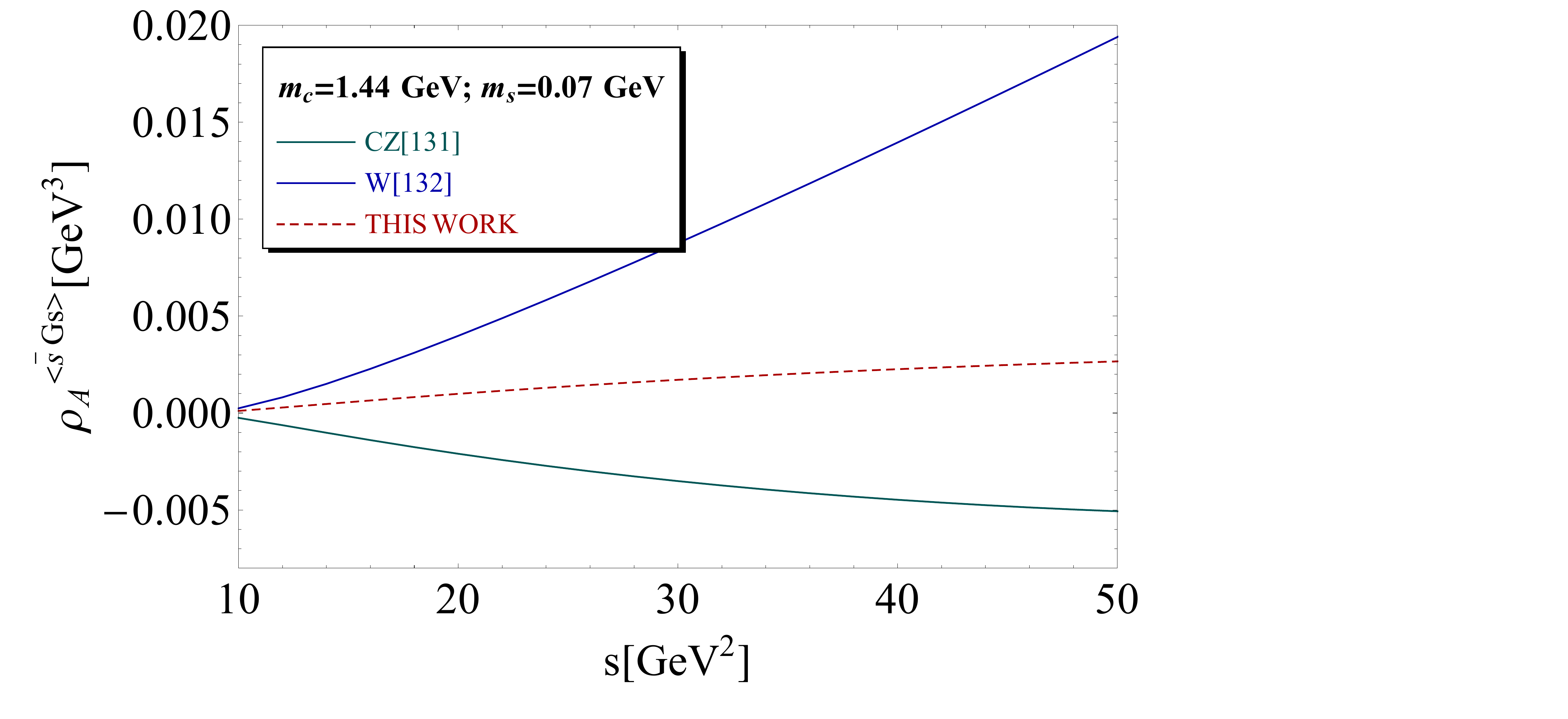}}
{\includegraphics[width=6.29cm  ]{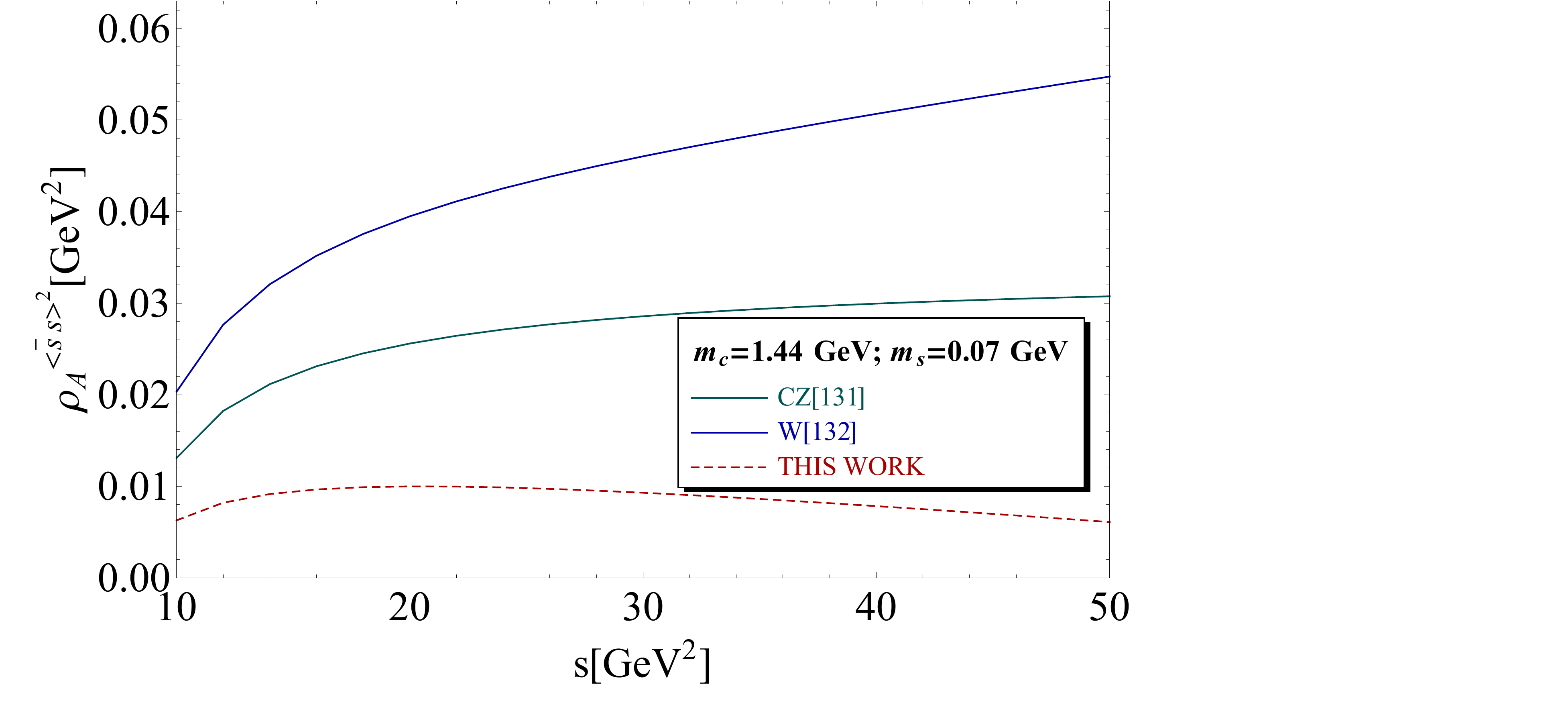}}
\centerline {\hspace*{-3cm} a)\hspace*{6cm} b) }
\caption{
\scriptsize 
Comparison of the Wilson coefficients of the $1^{++}$ four-quark spectral functions for different values of $s$ and for given values of $m_c$ and $m_s$: {\bf a)} mixed condensate; {\bf b)}  four-quark condensate.
}
\label{fig:comp1+6} 
\end{center}
\end{figure}

We compare numerically  the different LO QCD expressions of the spectral functions up to dimension-six from different authors. For definiteness, we consider  the examples of the $0^{++}$ and $1^{++}$ with some specific channels where the results from the QSSR analysis are listed inTable\,\ref{tab:theory}. \\
-- {\it $D^*_sD^*_s$ molecule states}\\ 
There is a complete agreement with our results and the ones in\,\cite{NIELSEN,WANG2}. For the four-quark condensates, we retain the linear $m_s$-corrections while $m_s^2$ corrections are included in \,\cite{NIELSEN,WANG2}.\\
-- {\it $0^{++}$ four-quarks states}\\ 
Our results are compared with the ones in \cite{CHEN17}.  There is a discrepancy at high $s$ as shown in Fig.\,\ref{fig:comp0+},
which originates from the fact that we only keep the linear $m_s$ corrections.\\
-- {\it $1^{++}$ four-quarks states}  \\
We compare in Figs. \ref{fig:comp1+pert} to \ref{fig:comp1+6} our results  with the ones from\,\cite{CHEN10,WANG3}. One can notice that the disagreement among different expressions occurs mainly at high vlaues of $s$. The disagreement for the four-quark condensate in Fig.\,\ref{fig:comp1+6} at low $s$ of our result with the one from\,\cite{WANG3} by a factor 2.  \\
However, due to the 
few informations given by the authors on the derivation of their QCD expressions, it is difficult to trace back the exact origin of such discrepancies. Hopefully, within  the accuracy of the approach, such discrepancies affect only slightly the final results listed in Table\,\ref{tab:theory} if the errors are taken properly. 
\nin	
\nin		
\nin		
\nin																 
\subsection*{\b Comparison with some previous lowest order QSSR results}\label{sec:confront2}
We list in Table\,\ref{tab:theory} some previous results for the $0^{++}$ and $1^{++}$ charm states obtained from QSSR at lowest order (LO) of perturbative QCD for the scalar and axial-vctor channels. The comparison is only informative as it is known that the LO results suffer from the ill-defined definition of the heavy quark mass used in the analysis at this order.  Most of the authors use the running mass value which is not justified when one implicitly uses the QCD expression obtained within the on-shell scheme. The difference between some results is also due to the way for extracting the optimal information from the analysis (different choices of $t_c$ and $\tau$). Here, we use well-defined based stability criteria verified from the example of the harmonic oscillator in quantum mechanics and from different well-known hadronic channels.  Another source of discrepancy in  the four-quark channels is the choice of the interpolating currents. We have taken the simplest choice of currents and used the optimal choice ($k=0$) determined in our earlier works\,\cite{X3A,X3B}. The results obtained in\,\cite{X3A,X3B} by matching the Laplace sum rules with Finite Energy moments  at N2LO will not be reported in the Table as this way of doing may lead to erroneous results due to the high-sensitivity of the Finite Energy moments on the continuum contribution. There, we also use the range of values spanned by the running and the pole mass (which one should do at LO) in the analysis. 
\subsection*{\b Confrontation with experiments}\label{sec:confront}
  We compare our results of the scalar $0^{++}$ and axial-vector $1^{++}$ charm states obtained by using the lowest dimension currents with the experimental $X$ candidates given in Table\,\ref{tab:exp}.  We conclude from the previous analysis that: \\
 -- The $0^{++}$ X(4700) experimental candidate might be identified with a $\bar D^*_{s0}D^*_{s0}$ molecule ground state. \\
 -- The interpretation of the $0^{++}$ candidates as pure 4-quark ground states is not favoured by our result. \\
 -- The masses of $1^{++}$ X(4147) and X(4273) 
  are compatible within the error with the one of the $\bar D^*_{s}D_{s}$ molecule state and with the one of the axial-vector $A_c$ 4-quark state. \\
  -- Our predictions suggest the presence of $0^{++}$ $\bar D_sD_s$ and $\bar D^*_sD^*_s$ molecule states in the range (4121$\sim$ 4396) MeV and a $D^*_{s0}D_{s1}$ state around 4841 MeV. \\
  -- We present new predictions for the $0^{-\pm}$, $1^{-\pm}$ and for different beauty states which can be tested in future experiments. \\
  --  Noting that the QCD continuum model smears all higher mass states, one may approximately expect  that their masses are in the vicinity of the value of the continuum threshold. In most case, the optimization region starts from 300(resp. 600) MeV above the lowest ground state mass. Then, one expects that the radial excitations might be visible in these regions if they couple strongly enough to the interpolating currents. \\
\subsection*{\b Theoretical Results and Perspectives}\label{sec:theory}
-- Our previous results show that the SU3 breakings are relatively small for the masses ($\leq 10$(rep. 3)\% for the charm (resp. bottom) channels while its can be large for the couplings ($\leq 20\%$). This can be understood as, in the ratios of sum rules, the corrections tend to cancel out.  \\
-- The approach cannot clearly separate (within the errors) some molecule states from the  four-quark ones of a given quantum number.\\
-- Like in the chiral limit case\,\cite{ChXYZ}, we also observe that the couplings behave as $1/m_b^{3/2}$ (resp. $1/m_b$) for the $1^+,0^+$ (resp. $1^-,0^-$) molecules and four-quark states which can be compared with $f_B\sim 1/m_b^{1/2}$ for open beauty mesons. These results which are important for further building of an effective theory for these exotic states can be tested by lattice calculations. \\
-- A natural extension of our analysis is the estimate of the meson widths. We plan to do this project in a future work.
\section*{Acknowledgements}
We thank A. Rabemananjara for participating at the early stage of this work. 
\appendix 
\section{SU3 Breakings to the Molecule Spectral Functions}\label{app.a}
They are defined from Eq.\,\ref{2po} as: $\frac{1}{\pi}{\rm Im}\Pi_{mol}^{(1)}(t)$ for spin 1 particles and $\frac{1}{\pi}{\rm Im}\psi_{mol}^{(s,p)}(t)$ from Eq.\,\ref{2po5} for spin 0 ones and normalized in the same way as the spectral functions in Ref.\,\cite{ChXYZ} . 
In the following, we shall give the SU3 breaking corrections (denoted by $\delta\rho$) to the spectral functions obtained in the chiral limit ($m_q=0$) \,\cite{ChXYZ}. 
We shall use the same notations and definitions:
\begin{table}[hbt]
\setlength{\tabcolsep}{0.6pc}
\footnotesize{
{\begin{tabular}{lllll}
$Q$&$\equiv$&$ c,\, b~$,  &$x ={M_Q^{2}}/{t} $~,&$ v =\sqrt{1-4x}\,~, $  \\
$\lv$&=& $\ln{\frac{(1+v)}{(1-v)}}~,$&$\lp=\lid\left(\frac{1+v}{2}\right)-\lid\left(\frac{1-v}{2}\right).$ \\
\end{tabular}
}
}
\end{table}
\vspace*{-1cm}
{\subsection{$(0^{++})$ $\bar D_s D_s,~\bar B_sB_s$ Molecules }}
\begin{eqnarray*}
  \delta\rho^{pert}_{m_s}(s) &=& -\frac{m_s M_Q^7}{2^{11} \:\pi^6} 
  \bigg[ v \Big( 60 + \frac{130}{x} - \frac{18}{x^2} - \frac{1}{x^3} \Big)
  +\nn\\
  &&12 {\cal L}_v \Big( 10x - 4 - 6\:\mbox{Log}(x) - \frac{6}{x} + \frac{1}{x^2} \Big)
  - 144 {\cal L}_+ \bigg] \\
  \delta \rho^{\qqs}_{m_s}(s) &=& -\frac{m_s M_Q^4 \qqs}{2^{8} \:\pi^4}
  \bigg[ v \Big( 24 + \frac{22}{x} - \frac{1}{x^2} \Big) 
 + 12{\cal L}_v \Big( 4x - 5 \Big) \bigg]  \\
 \delta \rho^{\sGs}_{m_s}(s) &=& \frac{m_s M_Q^2 \sGs}{ 2^{7} \:\pi^4} 
  \bigg[ v \Big( 10 + \frac{1}{x} \Big) + 3{\cal L}_v \Big( 4x - 3 \Big) \bigg] \\
   \delta\rho_{m_s}^{\qqs^2}(s) &=& -\frac{m_s M_Q \:\rho \qqs^2}{2^5 \:\pi^2} \:v (1+ s \tau) \nn\\
  \delta \rho_{m_s }^{\qqs \sGs}(s) &=& \frac{(m_s/M_Q) \qqs \sGs}{3 \cdot 2^7 \:\pi^2} \:v
  \Big( 6 + 6s\tau - 6s^2\tau^2 + 5s^3 \tau^3 x \Big) 
\end{eqnarray*}
\subsection{ $(0^{++})$ $D^\ast_s D^\ast_s$, $B^\ast_s B^\ast_s$ Molecules}
\vspace{-0.1cm}
\begin{eqnarray*}
  \delta\rho_{m_s}^{pert}(s) &=& -\frac{m_s M_Q^7}{2^{10} \:\pi^6} 
  \bigg[ v \Big( 60 + \frac{130}{x} - \frac{18}{x^2} - \frac{1}{x^3} \Big)
  +\nn\\
  &&12 {\cal L}_v \Big( 10x - 4 - 6\:\mbox{Log}(x) - \frac{6}{x} + \frac{1}{x^2} \Big)
  - 144 {\cal L}_+ \bigg] \\
  \delta \rho^{\qqs}_{m_s}(s) &=& -\frac{m_s M_Q^4 \qqs}{2^{6} \:\pi^4}
  \bigg[ v \Big( 24 + \frac{22}{x} - \frac{1}{x^2} \Big) 
  + 12{\cal L}_v \Big( 4x - 5 \Big) \bigg]  \\
 \delta \rho^{\sGs}_{m_s}(s) &=& \frac{m_s M_Q^2 \sGs}{ 2^{6} \:\pi^4} 
  \:v \Big( 8 - \frac{1}{x} \Big) \\
  \delta\rho_{m_s}^{\qqs^2}(s) &=& -\frac{m_s M_Q \:\rho \qqs^2}{2^4 \:\pi^2} \:v (1+ s \tau) \nn\\
  \rho^{m_s \cdot \qqs \sGs}(s) &=& \frac{(m_s/M_Q) \qqs \sGs}{3 \cdot 2^6 \:\pi^2} \:v \:x
  \Big( 12s^2\tau^2 + 5s^3 \tau^3 \Big) 
\end{eqnarray*}
\subsection{ $(0^{++})$ $D^\ast_{s0} D^\ast_{s0}$, $B^\ast_{s0} B^\ast_{s0}$ Molecules}
\vspace*{-0.45cm}
\begin{eqnarray*}
  \delta\rho_{m_s}^{pert}(s) &=& \frac{m_s M_Q^7}{2^{11} \:\pi^6} 
  \bigg[ v \Big( 60 + \frac{130}{x} - \frac{18}{x^2} - \frac{1}{x^3} \Big)
  +\nn\\
  &&12 {\cal L}_v \Big( 10x - 4 - 6\:\mbox{Log}(x) - \frac{6}{x} + \frac{1}{x^2} \Big)
  - 144 {\cal L}_+ \bigg] \\
  \delta \rho^{\qqs}_{m_s}(s) &=& -\frac{m_s M_Q^4 \qqs}{2^{8} \:\pi^4}
  \bigg[ v \Big( 24 + \frac{22}{x} - \frac{1}{x^2} \Big) 
  + 12{\cal L}_v \Big( 4x - 5 \Big) \bigg]  \\
 \delta \rho^{\sGs}_{m_s}(s) &=& \frac{m_s M_Q^2 \sGs}{ 2^{7} \:\pi^4} 
  \bigg[ v \Big( 10 + \frac{1}{x} \Big) + 3{\cal L}_v \Big( 4x - 3 \Big) \bigg] \\
  \delta\rho_{m_s}^{\qqs^2}(s) &=& \frac{m_s M_Q \:\rho \qqs^2}{2^5 \:\pi^2} \:v (1+ s \tau) \nn\\
 \delta \rho_{m_s }^{\qqs \sGs}(s) &=& -\frac{(m_s/M_Q) \qqs \sGs}{3 \cdot 2^7 \:\pi^2} \:v
  \Big( 6 + 6s\tau - 6s^2\tau^2 + 5s^3 \tau^3 x \Big) 
\end{eqnarray*} 
\vspace{0.5cm}

\subsection{ $(0^{++})$ $D_{s1} D_{s1}$, $B_{s1} B_{s1}$ Molecules}
\vspace*{-1.2cm}
\begin{eqnarray*}
  \rho^{pert} (s) &=& \frac{M_Q^8}{5 \cdot 2^{12} \:\pi^6} 
  \bigg[ v \Big( 480 + \frac{1460}{x} - \frac{274}{x^2} - \frac{38}{x^3} + \frac{1}{x^4} \Big)+\nnb\\
  &&120 {\cal L}_v \Big( 8x - 1 - 6\:\mbox{Log}(x) - \frac{8}{x} + \frac{2}{x^2} \Big)
  - 1440 {\cal L}_+ \bigg] \\
 \delta \rho^{pert}_{m_s} (s) &=& \frac{m_s M_Q^7}{2^{10} \:\pi^6} 
  \bigg[ v \Big( 60 + \frac{130}{x} - \frac{18}{x^2} - \frac{1}{x^3} \Big)
  +\nnb\\
  &&12 {\cal L}_v \Big( 10x - 4 - 6\:\mbox{Log}(x) - \frac{6}{x} + \frac{1}{x^2} \Big)
  - 144 {\cal L}_+ \bigg] \\
  \rho^{\qqs} (s) &=& -\frac{M_Q^5 \qqs}{2^{6} \:\pi^4}
  \bigg[ v \Big( 6 - \frac{5}{x} - \frac{1}{x^2} \Big) 
  + 6{\cal L}_v \Big( 2x - 2 + \frac{1}{x} \Big) \bigg]  \\
  \delta\rho^{  \qqs}_{m_s} (s) &=& -\frac{m_s M_Q^4 \qqs}{2^{6} \:\pi^4}
  \bigg[ v \Big( 24 + \frac{22}{x} - \frac{1}{x^2} \Big) 
  - 12{\cal L}_v \Big( 5 - 4x \Big) \bigg]  \\
  \rho^{\GG} (s) &=& \frac{M_Q^4 \gGG}{3 \cdot 2^{10} \:\pi^6} 
 \bigg[ v \Big( 6 - \frac{5}{x} - \frac{1}{x^2} \Big) 
  + 6{\cal L}_v \Big( 2x - 2 + \frac{1}{x} \Big) \bigg] \\
  %
  \rho^{\sGs} (s) &=& -\frac{3M_Q^3 \sGs}{ 2^{7} \:\pi^4} 
  \bigg[ \frac{v}{x} - 2 {\cal L}_v \bigg] \\
 \delta \rho^{\sGs}_{m_s} (s) &=& \frac{ m_s M_Q^2 \sGs}{ 2^{6} \:\pi^4} \:v
  \bigg[ 8 - \frac{1}{x} \bigg] \\
  \rho^{\qqs^2} (s) &=& \frac{M_Q^2 \:\rho \qqs^2 \:v}{4 \:\pi^2}  \nn\\
 \delta \rho^{\qqs^2}_{m_s } (s) &=& \frac{m_s M_Q \:\rho \qqs^2}{2^4 \:\pi^2} \:v(3 - s \:\tau)  \nn\\
  \rho^{\GGG} (s) &=& -\frac{M_Q^2 \gGGG}{ 3\cdot 2^{12} \:\pi^6} 
  \bigg[ v \Big( 6 - \frac{25}{x} + \frac{1}{x^2} \Big) + 
  6{\cal L}_v \Big( 2x + 2 + \frac{1}{x} \Big) \bigg] \\
  \rho^{\qqs\sGs} (s) &=& -\frac{M_Q^2 \qqs \sGs}{8 \:\pi^2} \:v \:s \:\tau^2\\
  \delta\rho^{ \qqs\sGs}_{m_s} (s) &=& \frac{m_s M_Q \qqs \sGs}{3 \cdot 2^6 \:\pi^2} \:v \:\tau
  \bigg[ 12 \:s \tau + 5 \:s^2 \tau^2 \bigg]
\end{eqnarray*}

\subsection{$(1^{+\pm})$ $D^\ast_s D_s$, $B^\ast_s B_s$ Molecules }
\begin{eqnarray*}
  \delta\rho_{m_s}^{pert}(s) &=& -\frac{m_s M_Q^7}{5 \cdot 2^{14} \:\pi^6} 
  \bigg[ v \Big( 420x + 1270 + \frac{4174}{x} - \frac{617}{x^2} - \frac{36}{x^3} \Big)+ \\ &&
  60 {\cal L}_v \Big( 14x^2 + 40x - 12 - 36\:\mbox{Log}(x) - \frac{40}{x} + \frac{7}{x^2} \Big)
  - 4320 {\cal L}_+ \bigg] \\
  \delta \rho^{\qqs}_{m_s}(s) &=& \frac{m_s M_Q^4 \qqs}{2^{10} \:\pi^4}
  \bigg[ v \Big( 12x - 94 - \frac{74}{x} + \frac{3}{x^2} \Big) 
  + 24{\cal L}_v \Big( x^2 - 8x + 9 \Big) \bigg]  \\
 \delta \rho^{\sGs}_{m_s}(s) &=& -\frac{m_s M_Q^2 \sGs}{ 3 \cdot 2^{8} \:\pi^4} 
  \bigg[ v \Big( 6x - 37 + \frac{1}{x} \Big) + 6{\cal L}_v \Big( 2x^2 - 6x + 3 \Big) \bigg] \\
  \delta\rho_{m_s}^{\qqs^2}(s) &=& -\frac{m_s M_Q \:\rho \qqs^2}{2^6 \:\pi^2} \:v (1+ 2 s \tau) \nn\\
 \delta \rho_{m_s }^{\qqs \sGs}(s) &=& -\frac{(m_s/M_Q) \qqs \sGs}{3 \cdot 2^8 \:\pi^2} \:v
  \bigg[ s\tau \Big( 13x - 12 \Big) + s^2 \tau^2 \Big( x + 6 \Big) - 10 s^3 \tau^3 x \bigg]  
\end{eqnarray*}

\subsection{$(1^{+\pm})$ $D^\ast_{s0} D_{s1}$, $B^\ast_{s0} B_{s1}$ Molecules }
\begin{eqnarray*}
  \delta\rho_{m_s}^{pert}(s) &=& \frac{m_s M_Q^7}{5 \cdot 2^{14} \:\pi^6} 
  \bigg[ v \Big( 420x + 1270 + \frac{4174}{x} - \frac{617}{x^2} - \frac{36}{x^3} \Big)+ \\ &&
  60 {\cal L}_v \Big( 14x^2 + 40x - 12 - 36\:\mbox{Log}(x) - \frac{40}{x} + \frac{7}{x^2} \Big)
  - 4320 {\cal L}_+ \bigg] \\
  \delta \rho^{\qqs}_{m_s}(s) &=& \frac{m_s M_Q^4 \qqs}{2^{10} \:\pi^4}
  \bigg[ v \Big( 12x - 94 - \frac{74}{x} + \frac{3}{x^2} \Big) 
  + 24{\cal L}_v \Big( x^2 - 8x + 9 \Big) \bigg]  \\
 \delta \rho^{\sGs}_{m_s}(s) &=& -\frac{m_s M_Q^2 \sGs}{ 3 \cdot 2^{8} \:\pi^4} 
  \bigg[ v \Big( 6x - 37 + \frac{1}{x} \Big) + 6{\cal L}_v \Big( 2x^2 - 6x + 3 \Big) \bigg] \\
  \delta\rho_{m_s}^{\qqs^2}(s) &=& \frac{m_s M_Q \:\rho \qqs^2}{2^6 \:\pi^2} \:v (1+ 2 s \tau) \nn\\
\delta \rho_{m_s }^{\qqs \sGs}(s)&=& \frac{(m_s/M_Q) \qqs \sGs}{3 \cdot 2^8 \:\pi^2} \:v
  \bigg[ s\tau \Big( 13x - 12 \Big) + s^2 \tau^2 \Big( x + 6 \Big) - 10 s^3 \tau^3 x \bigg]  
\end{eqnarray*}
\subsection{$(0^{-\pm})$ $D^\ast_s D_{s1}$, $B^\ast_s B_{s1}$ Molecules }
\begin{eqnarray*}
  \delta\rho_{m_s}^{pert}(s) &=& 0 \\
  \delta \rho^{\qqs}_{m_s}(s) &=& \frac{m_s M_Q^4 \qqs}{2^{6} \:\pi^4}
  \bigg[ v \Big( 24 + \frac{2}{x} + \frac{1}{x^2} \Big) 
  + 12{\cal L}_v \Big( 4x - 3 \Big) \bigg]  \\
 \delta \rho^{\sGs}_{m_s}(s) &=& -\frac{m_s M_Q^2 \sGs}{ 2^{6} \:\pi^4} 
  \:v \Big( 4 + \frac{1}{x} \Big) \\
   \delta\rho_{m_s}^{\qqs^2}(s) &=& 0\\
 \delta \rho_{m_s }^{\qqs \sGs}(s) &=& 0 
\end{eqnarray*} 
\subsection{$(0^{-\pm})$ $D^\ast_{s0} D_s$, $D^\ast_{s0} D_s$ Molecules }
\begin{eqnarray*}
  \delta\rho_{m_s}^{pert}(s) &=& 0 \\
  \delta \rho^{\qqs}_{m_s}(s) &=& \frac{m_s M_Q^4 \qqs}{2^{8} \:\pi^4}
  \bigg[ v \Big( 24 + \frac{2}{x} + \frac{1}{x^2} \Big) 
  + 12{\cal L}_v \Big( 4x - 3 \Big) \bigg]  \\
 \delta \rho^{\sGs}_{m_s}(s) &=& -\frac{m_s M_Q^2 \sGs}{ 2^{7} \:\pi^4} 
  \bigg[ v \Big( 8 - \frac{1}{x} \Big) + 3{\cal L}_v \Big( 4x - 1 \Big) \bigg] \\
   \delta\rho_{m_s}^{\qqs^2}(s) &=& 0\\
  \delta \rho_{m_s }^{\qqs \sGs}(s) &=& 0 
\end{eqnarray*}
\subsection{$(1^{--})$ $D^\ast_{s0} D^\ast_s$, $B^\ast_{s0} B^\ast_s$ Molecules}
\begin{eqnarray*}
  \delta\rho_{m_s}^{pert}(s) &=& 0 \\
  \delta \rho^{\qqs}_{m_s}(s) &=& \frac{m_s M_Q^4 \qqs}{2^{9} \:\pi^4}
  \bigg[ v \Big( 60 + \frac{20}{x} + \frac{1}{x^2} \Big) 
  + 12{\cal L}_v \Big( 10x - 9 \Big) \bigg]  \\
 \delta \rho^{\sGs}_{m_s}(s) &=& \frac{m_s M_Q^2 \sGs}{ 3 \cdot 2^{8} \:\pi^4} 
  \bigg[ v \Big( 6x - 50 - \frac{1}{x} \Big) + 3{\cal L}_v \Big( 4x^2 - 18x + 9 \Big) \bigg] \\
  \delta\rho_{m_s}^{\qqs^2}(s) &=& 0 \\
\delta \rho_{m_s }^{\qqs \sGs}(s) &=& -\frac{(m_s/M_Q) \qqs \sGs}{3 \cdot 2^7 \:\pi^2} \:v
  \bigg[ 3 - s\tau \Big( x - 3 \Big) + s^2 \tau^2 \Big( 14x - 3 \Big) \bigg]  
\end{eqnarray*} 
\subsection{$(1^{-+})$ $D^\ast_{s0} D^\ast_s$,  $B^\ast_{s0} B^\ast_s$ Molecules }
\begin{eqnarray*}
  \delta\rho_{m_s}^{pert}(s) &=& -\frac{m_s M_Q^7}{5 \cdot 2^{13} \:\pi^6} 
  \bigg[ v \Big( 420x - 1130 - \frac{1026}{x} + \frac{103}{x^2} + \frac{4}{x^3} \Big) +\\ &&
  60 {\cal L}_v \Big( 14x^2 - 40x + 20 + 12\:\mbox{Log}(x) + \frac{8}{x} - \frac{1}{x^2} \Big)
  + 1440 {\cal L}_+ \bigg] \\
  \delta \rho^{\qqs}_{m_s}(s) &=& -\frac{m_s M_Q^4 \qqs}{2^{8} \:\pi^4}
  \bigg[ v \Big( 6x + 19 + \frac{1}{x} + \frac{1}{x^2} \Big) 
  + 6{\cal L}_v \Big( 2x^2 + 6x - 5 \Big) \bigg]  \\
 \delta \rho^{\sGs}_{m_s}(s) &=& \frac{m_s M_Q^2 \sGs}{ 3 \cdot 2^{8} \:\pi^4} 
  \bigg[ v \Big( 18x + 20 + \frac{1}{x} \Big) + 9{\cal L}_v \Big( 4x^2 + 2x - 1 \Big) \bigg] \\
   \delta\rho_{m_s}^{\qqs^2}(s) &=& \frac{m_s M_Q \:\rho \qqs^2}{2^5 \:\pi^2} \:v \\
\delta \rho_{m_s }^{\qqs \sGs}(s)&=& -\frac{(m_s/M_Q) \qqs \sGs}{2^7 \:\pi^2} \:v
  \bigg[ 1 + s\tau \Big( 4x - 3 \Big) - s^2 \tau^2 \Big( 3x - 1 \Big) \bigg]  
\end{eqnarray*}

\subsection{$(1^{--})$ $D_s D_{s1}$,  $B_s B_{s1}$ Molecules }
\begin{eqnarray*}
  \delta\rho_{m_s}^{pert}(s) &=& 0 \\
  \delta \rho^{\qqs}_{m_s}(s) &=& 0  \\
 \delta \rho^{\sGs}_{m_s}(s) &=& \frac{3 m_s M_Q^2 \sGs}{ 2^{7} \:\pi^4} 
  \bigg[ v + {\cal L}_v \Big( 2x - 1 \Big) \bigg] \\
   \delta\rho_{m_s}^{\qqs^2}(s) &=& 0 \\
\delta \rho_{m_s }^{\qqs \sGs}(s)&=& -\frac{(m_s/M_Q) \qqs \sGs}{2^6 \:\pi^2} \:v
  \Big( s^2 \tau^2 x \Big)
\end{eqnarray*} 
\subsection{$(1^{-+})$ $D_s D_{s1}$,  $B_s B_{s1}$ Molecules }
\begin{eqnarray*}
  \delta\rho_{m_s}^{pert}(s) &=& \frac{m_s M_Q^7}{5 \cdot 2^{13} \:\pi^6} 
  \bigg[ v \Big( 420x - 1130 - \frac{1026}{x} + \frac{103}{x^2} + \frac{4}{x^3} \Big)+ \\ &&
  60 {\cal L}_v \Big( 14x^2 - 40x + 20 + 12\:\mbox{Log}(x) + \frac{8}{x} - \frac{1}{x^2} \Big)
  + 1440 {\cal L}_+ \bigg] \\
  \delta \rho^{\qqs}_{m_s}(s) &=& -\frac{m_s M_Q^4 \qqs}{2^{9} \:\pi^4}
  \bigg[ v \Big( 12x + 98 + \frac{22}{x} + \frac{3}{x^2} \Big) 
  + 24{\cal L}_v \Big( x^2 + 8x - 7 \Big) \bigg]  \\
 \delta \rho^{\sGs}_{m_s}(s) &=& \frac{m_s M_Q^2 \sGs}{ 3 \cdot 2^{7} \:\pi^4} 
  \bigg[ v \Big( 6x + 26 + \frac{1}{x} \Big) + 3{\cal L}_v \Big( 4x^2 + 6x - 3 \Big) \bigg] \\
   \delta\rho_{m_s}^{\qqs^2}(s) &=& -\frac{m_s M_Q \:\rho \qqs^2}{2^5 \:\pi^2} \:v \\
 \delta \rho_{m_s }^{\qqs \sGs}(s)&=& \frac{(m_s/M_Q) \qqs \sGs}{3 \cdot 2^7 \:\pi^2} \:v
  \bigg[ s\tau \Big( 13x - 12 \Big) - s^2 \tau^2 \Big( 17x - 6 \Big) \bigg]  
\end{eqnarray*}
\section{Four-Quark States Spectral Functions}\label{app.b}
The spectral functions corresponding to the four-quark interpolating currents given in Table\,\ref{tab:current4}
read:
\subsection{$(0^+)$ Scalar State}
\begin{eqnarray*}
\rho^{m_s}_S(s) &=& -\frac{(1-k^2) m_s M_c^7}{3\cdot 2^{9} \:\pi^6} \bigg[ 
	v \bigg( 60 + \frac{130}{x} - \frac{18}{x^2} - \frac{1}{x^3} \bigg)+ \nnb\\
	&& 12{\cal L}_v \bigg( 10x - 4 - 6 \log x - \frac{6}{x} + \frac{1}{x^2} \bigg)  - 144 \:{\cal L}_+
	\bigg] \hspace{1cm}\nnb
\\
%
\rho^{m_s \cdot \qqs}_S(s) &=& -\frac{(1+k^2) m_s M_c^4 \qqs}{3 \cdot 2^6 \:\pi^4} \bigg[ 
	v \bigg( 24 + \frac{22}{x} - \frac{1}{x^2} \bigg)
	+ 12{\cal L}_v \bigg( 4x - 5 \bigg)
	\bigg] \nnb
\\
%
\rho^{m_s \cdot \sGs}_S(s) &=& \frac{(1+k^2) m_s M_c^2 \sGs}{3\cdot 2^7 \:\pi^4} \bigg[
	v \bigg( 28 + \frac{1}{x} \bigg)
	+ 6{\cal L}_v \bigg( 4x - 3 \bigg)
	\bigg]\nnb
\\
%
\rho^{m_s \cdot \qqs^2}_S(s) &=& -\frac{(1-k^2) m_s M_c \,\rho\qqs^2}{3 \cdot 2^3 \:\pi^2} \:v (1 + s\tau)\nnb\\
%
\rho^{m_s \cdot \qqs \sGs}_S(s) &=& \frac{(1-k^2) (m_s/M_c) \qqs \sGs}{3^2 \cdot 2^{5} \:\pi^2} \:v \bigg[
	3 + 3s\tau - 3s^2 \tau^2 +5x s^3 \tau^3 \bigg]
\end{eqnarray*}

\subsection{$(1^+)$ Axial-vector State}
\begin{eqnarray*}
\rho^{m_s}_A(s) &=& -\frac{(1-k^2) m_s M_c^7}{5\cdot 3\cdot 2^{12} \:\pi^6} \bigg[ 
	v \bigg( 420x + 1270 + \frac{4174}{x} - \frac{617}{x^2} - \frac{36}{x^3} \bigg)+\nnb\\
	&& 60{\cal L}_v \bigg( 14x^2 + 40x - 12 - 36 \log x - \frac{40}{x} + \frac{7}{x^2} \bigg)  - 4320 \:{\cal L}_+
	\bigg] \hspace{1cm}\nnb
\\
%
\rho^{m_s \cdot \qqs}_A(s) &=& \frac{(1+k^2) m_s M_c^4 \qqs}{3 \cdot 2^8 \:\pi^4} \bigg[ 
	v \bigg( 12x - 94 - \frac{74}{x} + \frac{3}{x^2} \bigg)
	+ 24{\cal L}_v \bigg( x^2 - 8x + 9 \bigg)
	\bigg]\nnb
\\
%
\rho^{m_s \cdot \sGs}_A(s) &=& -\frac{(1+k^2) m_s M_c^2 \sGs}{3\cdot 2^7 \:\pi^4} \bigg[
	v \bigg( 2x - 19 + \frac{1}{x} \bigg)
	+ 2{\cal L}_v \bigg( 2x^2 - 6x + 3 \bigg)
	\bigg]\nnb
\\
%
\rho^{m_s \cdot \qqs^2}_A(s) &=& -\frac{(1-k^2) m_s M_c \,\rho\qqs^2}{3 \cdot 2^4 \:\pi^2} \:v (1 + 2s\tau)\nnb\\
%
\rho^{m_s \cdot \qqs \sGs}_A(s) &=& \frac{(1-k^2) (m_s/M_c) \qqs \sGs}{3^2 \cdot 2^{6} \:\pi^2} \:v \bigg[
	3s\tau (2 - 3x) - 3s^2 \tau^2 (1+x) + 10x \,s^3 \tau^3 \bigg]
\end{eqnarray*}

\subsection{$(0^-)$ Pseudoscalar State}
\begin{eqnarray*}
\rho^{m_s}_P(s) &=& 0 \hspace{1cm}\nnb
\\
%
\rho^{m_s \cdot \qqs}_P(s) &=& \frac{(1+k^2) m_s M_c^4 \qqs}{3 \cdot 2^6 \:\pi^4} \bigg[ 
	v \bigg( 24 + \frac{2}{x} + \frac{1}{x^2} \bigg)
	+ 12{\cal L}_v \bigg( 4x - 3 \bigg)
	\bigg]\nnb
\\
%
\rho^{m_s \cdot \sGs}_P(s) &=& -\frac{(1+k^2) m_s M_c^2 \sGs}{3\cdot 2^7 \:\pi^4} \bigg[
	v \bigg( 20 - \frac{1}{x} \bigg)
	+ 6{\cal L}_v \bigg( 4x - 1 \bigg)
	\bigg]\nnb
\\
%
\rho^{m_s \cdot \qqs^2}_P(s) &=& 0\nnb
\\
%
\rho^{m_s \cdot \qqs \sGs}_P(s) &=& 0
\end{eqnarray*}

\subsection{$(1^-)$ Vector State}
\begin{eqnarray*}
\rho^{m_s}_V(s) &=& -\frac{(1-k^2) m_s M_c^7}{5\cdot 3\cdot 2^{12} \:\pi^6} \bigg[ 
	v \bigg( 420x - 1130 - \frac{1026}{x} + \frac{103}{x^2} + \frac{4}{x^3} \bigg)+\nnb\\
	&& 60{\cal L}_v \bigg( 14x^2 - 40x + 20 + 12 \log x + \frac{8}{x} - \frac{1}{x^2} \bigg) + 1440 \:{\cal L}_+
	\bigg] \hspace{1cm}\nnb
\\
%
\rho^{m_s \cdot \qqs}_V(s) &=& \frac{(1+k^2) m_s M_c^4 \qqs}{3 \cdot 2^8 \:\pi^4} \bigg[ 
	v \bigg( 12x + 98 + \frac{22}{x} + \frac{3}{x^2} \bigg)
	+ 24{\cal L}_v \bigg( x^2 + 8x - 7 \bigg)
	\bigg]\nnb
\\
%
\rho^{m_s \cdot \sGs}_V(s) &=& -\frac{(1+k^2) m_s M_c^2 \sGs}{3\cdot 2^7 \:\pi^4} \bigg[
	v \bigg( 2x + 17 + \frac{1}{x} \bigg)
	+ 2{\cal L}_v \bigg( 2x^2 + 6x - 3 \bigg)
	\bigg]\nnb \\
%
\rho^{m_s \cdot \qqs^2}_V(s) &=& \frac{(1-k^2) m_s M_c \,\rho\qqs^2}{3 \cdot 2^4 \:\pi^2} \:v\nnb\\
%
\rho^{m_s \cdot \qqs \sGs}_V(s) &=& \frac{(1-k^2) (m_s/M_c) \qqs \sGs}{3 \cdot 2^{6} \:\pi^2} \:v \bigg[
	s\tau (2 - 3x) - s^2 \tau^2 (1 - 3x) \bigg]
\end{eqnarray*}


\end{document}